\documentclass[conference,compsoc]{IEEEtran}
\IEEEoverridecommandlockouts
\usepackage{cite}
\usepackage{amsmath,amssymb,amsfonts}
\usepackage{graphicx}
\usepackage{textcomp}
\usepackage{xcolor}
\usepackage{amsfonts}
\usepackage{mathrsfs}
\usepackage[english]{babel}
\usepackage{blindtext}
\usepackage[utf8]{inputenc}
\usepackage{breqn,xspace}
\usepackage{multirow}
\usepackage{threeparttable}
\usepackage{soul}
\usepackage{amsmath}
\usepackage{algorithm}
\usepackage{color}
\usepackage{booktabs}
\usepackage{enumitem}
\usepackage{bm}
\usepackage{url}
\usepackage{subfig}
\usepackage{wrapfig}
\usepackage{algorithmic}
\usepackage{float}
\usepackage{boondox-cal}
\usepackage{balance}

\usepackage{setspace}
\def\BibTeX{{\rm B\kern-.05em{\sc i\kern-.025em b}\kern-.08em
    T\kern-.1667em\lower.7ex\hbox{E}\kern-.125emX}}

\newcommand{\name}{\textsc{mimo}Crypt\xspace}
\newcommand{\sname}{\textsc{mimo}Crypt\xspace}
\ifodd 0
\newcommand{\rev}[1]{{\color{blue}#1}}      % revised by ZC
\newcommand{\newrev}[1]{{\color{blue}#1}}    % revised by JL
\else
\newcommand{\rev}[1]{#1} 
\newcommand{\newrev}[1]{#1} 
 
\fi 

\AtBeginDocument{%
  \providecommand\BibTeX{{%
    Bib\TeX}}}

\begin{document}

% The Major Revision criteria: 

% 1. Better describe and explore the multi-user case. 

% 2. Improved analysis and comparison to the literature 

% 3. Provide a security analysis and underpin security guarantees or describe limitations accordingly. Which (motivated) attacker can the system protect against, and which not? Describe how several factors affect the guarantees (tradeoffs), or why not (or maybe even provide a model to estimate their strength or lifetime). Consider the limited entropy of CSI, known patterns in the signal, and if or how an attacker could optimize against decoding performance to reverse the scrambling.

\title{{\large MIMO}Crypt: Multi-User Privacy-Preserving Wi-Fi Sensing via MIMO Encryption}
\ifodd 0
\author{\IEEEauthorblockN{Accepted Submission \#55 to IEEE S\&P 2024}
\IEEEauuthorblockA{\textcolor{white}{abc.edu.sg \country{Singapore}}}
\IEEEauthorblockA{\textcolor{white}{abc.edu \country{USA}}\\ \textcolor{white}{abc.com}}
}
\else
\author{\IEEEauthorblockN{
		Jun Luo\IEEEauthorrefmark{1}, Hangcheng Cao\IEEEauthorrefmark{2}, Hongbo Jiang\IEEEauthorrefmark{2}, Yanbing Yang\IEEEauthorrefmark{3}, Zhe Chen\IEEEauthorrefmark{4}
                        }
        \IEEEauthorblockA{\IEEEauthorrefmark{1}Nanyang Technological University, Singapore}
        \IEEEauthorblockA{\IEEEauthorrefmark{2}Hunan University, China \quad \IEEEauthorrefmark{3}Sichuan University, China \quad \IEEEauthorrefmark{4}Fudan University, China}
        \IEEEauthorblockA{Email: junluo@ntu.edu.sg, \{hangchengcao, hongbojiang\}@hnu.edu.cn, 
        % yangyanbing@scu.edu.cn, 
        zhechen@fudan.edu.cn}
    }
\fi	

\maketitle
\begin{abstract}
Wi-Fi signals may help realize low-cost and non-invasive human sensing, yet it can also be exploited by eavesdroppers to capture private information. Very few studies rise to handle this privacy concern so far; they either \textit{jam all sensing attempts} or rely on sophisticated technologies to support only a \textit{single sensing user}, rendering them impractical for multi-user scenarios. Moreover, these proposals 
% are incompatible with Wi-Fi design, especially its 
\newrev{all fail to exploit Wi-Fi's} multiple-in multiple-out (MIMO) capability. To this end, we propose \name, a privacy-preserving Wi-Fi sensing framework to support realistic \textit{multi-user} scenarios. To thwart unauthorized eavesdropping while retaining the sensing and communication capabilities for legitimate users, \sname innovates in exploiting \newrev{MIMO 
% and channel diversities 
to \textit{physically encrypt}} Wi-Fi channels, \newrev{treating the sensed human activities as \textit{physical plaintext}s.} The encryption scheme is further enhanced via an optimization framework, aiming to strike a balance among i) risk of eavesdropping, ii) sensing accuracy, and iii) communication quality, upon securely conveying \textit{\newrev{decryption} key}s to legitimate users. We implement a prototype of \sname\ on an SDR platform and perform extensive experiments to evaluate its effectiveness in common application scenarios, especially privacy-sensitive human gesture recognition.
\end{abstract}

% \begin{CCSXML}
% <ccs2012>
%  <concept>
%   <concept_id>10010520.10010553.10010562</concept_id>
%   <concept_desc>Computer systems organization~Embedded systems</concept_desc>
%   <concept_significance>500</concept_significance>
%  </concept>
%  <concept>
%   <concept_id>10010520.10010575.10010755</concept_id>
%   <concept_desc>Computer systems organization~Redundancy</concept_desc>
%   <concept_significance>300</concept_significance>
%  </concept>
%  <concept>
%   <concept_id>10010520.10010553.10010554</concept_id>
%   <concept_desc>Computer systems organization~Robotics</concept_desc>
%   <concept_significance>100</concept_significance>
%  </concept>
%  <concept>
%   <concept_id>10003033.10003083.10003095</concept_id>
%   <concept_desc>Networks~Network reliability</concept_desc>
%   <concept_significance>100</concept_significance>
%  </concept>
% </ccs2012>
% \end{CCSXML}

%\ccsdesc[500]{Computer systems organization~Embedded systems}
%\ccsdesc[300]{Computer systems organization~Redundancy}
%\ccsdesc{Computer systems organization~Robotics}
%\ccsdesc[100]{Networks~Network reliability}

%
\maketitle

\section{\MakeUppercase{Introduction}} \label{sec:intro}
Given the increasingly wide deployments of Wi-Fi infrastructure, an eminent \textit{threat} has arisen along with the ubiquitous wireless access. In fact, this threat has been explored as an \textit{opportunity} in the past two decades to sense our ambient environments in a ``sensor-free'' and ``contact-free'' manner~\cite{CThru-SIGCOMM13,WiHear-MobiCom14,iLocScan,WiDeo-NSDI15,VitalSign-MobiHoc15,WiKey-MobiCom15,LiFS-MobiCom16,SSNR-UbiComp22}; it is made possible since the disturbances brought to Wi-Fi signals (in particular \textit{channel state information}, or CSI~\cite{CSI-CCR11}) can be leveraged to infer the status and behaviors of their sources. Nonetheless, as this opportunity enables a common Wi-Fi user (under certain mild conditions) to get aware of others (not necessary Wi-Fi users) in terms of their locations~\cite{iLocScan,WiDeo-NSDI15,LiFS-MobiCom16}, activity/gesture~\cite{CThru-SIGCOMM13,WiHear-MobiCom14,WiKey-MobiCom15,Widar3-MobiSys19}, and even vital signs~\cite{VitalSign-MobiHoc15,SSNR-UbiComp22,zhang2022can}, it can potentially be exploited by malicious eavesdroppers and thus turned into a threat. For example, \newrev{someone typing a password can be ``overheard'' via Wi-Fi sensing~\cite{WiKey-MobiCom15, fang2018no, yang2022wink, WiKI-Eve-CCS23},} and the physical condition of an important person can be leaked via his/her vital signs~\cite{SSNR-UbiComp22,zhang2022can}. 
\begin{figure}[t]
    \setlength\abovecaptionskip{8pt}
	\centering
	\includegraphics[width = 0.98\columnwidth]{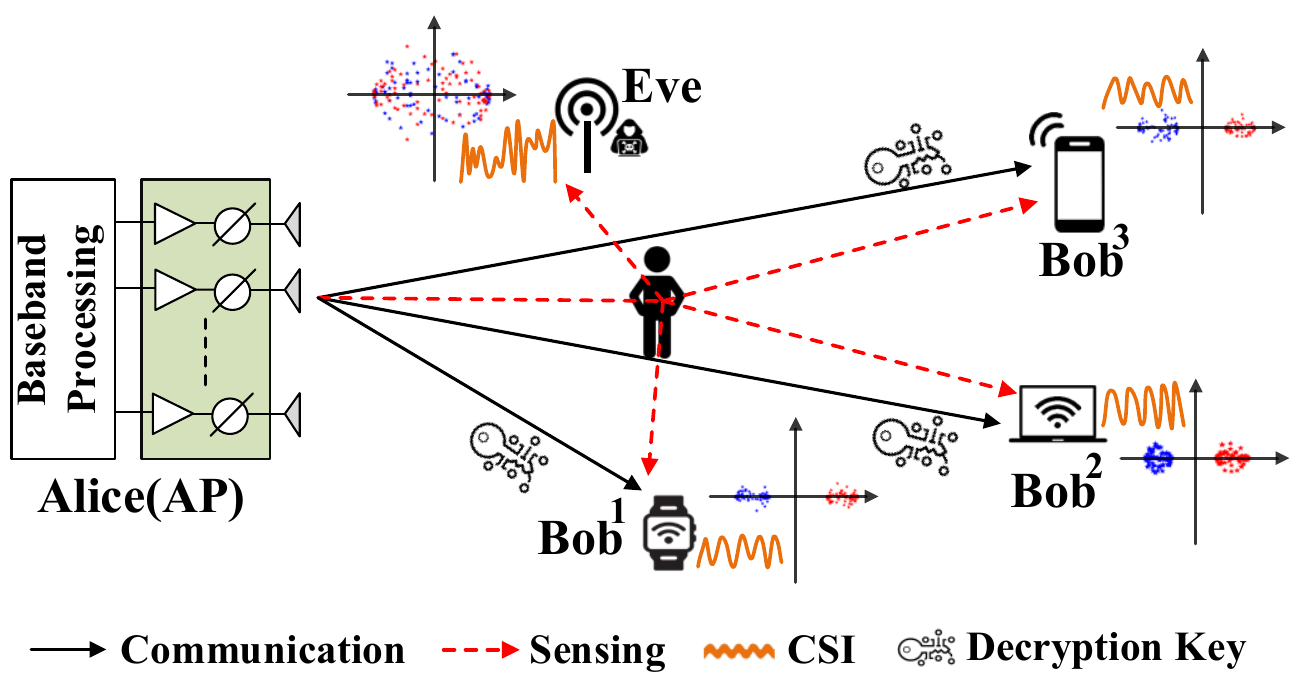}
	\caption{\sname leverages MIMO diversity at an AP (Alice) to encrypt the channel and thus thwart malicious eavesdroppers (Eve); it retains the sensing and communication ability of legitimate users (Bobs) by securely conveying a \newrev{decryption} key to them.} 
	\label{fig:teaser}
	\vspace{-1ex}
\end{figure}

Whereas this threat has been recognized for years~\cite{PhyCloak-NSDI16,RogueAP-CCS16}, the proposed solutions are still quite limited in their application scopes. PhyCloak~\cite{PhyCloak-NSDI16} presents a ``selfish'' countermeasure where only one user can perform sensing as it jams Wi-Fi signals after absorbing the clean ones. Similar \textit{signal obfuscation} technique is adopted again in~\cite{EtTuAlexa-NDSS20} but made omnipresent by initiating it from APs (access points): ``fake'' data traffic is injected to thwart all RSS (received signal strength) based sensing attempts. As the most recent improvement on signal obfuscation, IRShield~\cite{IRShield-SP22} leverages IRSs (intelligent reflecting surfaces) to reach a similar objective to~\cite{EtTuAlexa-NDSS20} but upon CSI. As a result, existing solutions, if adopted in practical scenarios, %can 
face three major issues: i) incapable of supporting multiple legitimate sensing users~\cite{MUSE-Fi-MobiCom23}, ii) jamming communications as the cost of defending rare sensing attacks, iii) \newrev{neglecting the important
% incompatible with commodity Wi-Fi, especially its 
\textit{multiple-in multiple-out} (MIMO) capability of Wi-Fi.}

To overcome all these issues, we propose \name as our answer to the threat imposed by the ubiquitous sensing capability of Wi-Fi. As illustrated in Figure~\ref{fig:teaser}, we realize signal obfuscation in a \textit{source-defined} manner by exploiting the MIMO capability to \newrev{\textit{physically encrypt} CSIs containing the to-be-sensed human activities as \textit{plaintexts}, rendering \sname compatible} with Wi-Fi standard without the need for extra components, such as FDR (full-duplex radio) in~\cite{PhyCloak-NSDI16} and IRS in~\cite{IRShield-SP22}. Apart from full compatible with commodity Wi-Fi, the source-defined obfuscation promoted by \sname is superior to existing solutions in two other aspects, given that the \newrev{decryption} keys can be securely conveyed to legitimate users: on one hand, \sname supports multi-user scenarios while thwarting unauthorized eavesdropping; on the other hand, the Wi-Fi communication quality can be largely retained for legitimate users.

Performing signal obfuscation always affects both communication and sensing for legitimate users, but this is a price to be paid for thwarting unauthorized eavesdropping.\footnote{Therefore, \sname is not meant for all application scenarios; it should be deployed only where privacy sensitivity is high.}  
Fortunately, \sname's source-defined channel encryption offers a full control on various parties, including both legitimate and unauthorized users, as well as users of both sensing and communication services. Consequently, the CSI encryption mechanism can be optimized so as to strike an adequate balance among i) risk of unauthorized eavesdropping, ii) legitimate sensing accuracy, and iii) communication quality. 
% As a result, \sname is able to spend the minimum amount of resources (only native Wi-Fi capabilities without adding extra components) to support a much larger user group in terms of both sensing and communication. 
%In addition, we propose an efficient descrambling procedure for sensing driven by a deep neural model, exploiting the distinction between sensing and communication in required signal granularity. 
%[zhe] Since we remove deep learning part from main body, do we revise this part?
In addition, we propose an efficient decryption procedure for relatively coarse-grained sensing tasks such as user gesture and activity recognition driven by a deep neural model, exploiting the distinction between sensing and communication in required signal granularity.
\footnote{Fine-grained sensing tasks such as vital signs monitoring~\cite{chen2021movifi} are not considered in our context, as they often require rather short distances potentially exposing the eavesdroppers.}
In summary, we make the following major contributions in \sname:
\begin{itemize}
    \item We realize the first secure Wi-Fi sensing system for multi-user scenarios; it reconciles the conflict between thwarting unauthorized eavesdropping and maintaining service qualities to legitimate users.
    \item We propose a source-defined channel encryption scheme, in order to maintain compatibility with Wi-Fi while delivering a flexible control to meet the demands from various parties under the multi-user scenario.
    \item We explore a multi-objective optimization framework for designing the channel encryption scheme, aiming to strike an adequate balance among functions with conflicting targets.
    \item We propose a novel decryption procedure for sensing tasks leveraging deep learning; it can be made more efficient than its counterpart for communications.
    % thanks to the distinction in the required signal granularity.
    %
    \item We implement a \sname prototype based on WARP~\cite{WARP-web} and conduct extensive experiments upon it to demonstrate the efficacy of \name.
\end{itemize}
The rest of the paper is organized as follows. Section~\ref{sec:background} sets up the background for the overall presentation before defining the attack model, discussing the related literature, and describing our brief solution. Section~\ref{sec:motiv} uses simple experiments to motivate the design of \sname, and Section~\ref{sec:sys} presents the design details. In these three sections, while describing the background, motivation, and workflow behind \sname, we gradually formalize the mathematical models for general Wi-Fi sensing and for specific techniques under \sname. Subsequently, Sections~\ref{sec:imple} and~\ref{sec:eval} respectively report the experiment setting and performance evaluation results of \sname. 
% Section~\ref{sec:eval} evaluates the performance of \sname. 
We then discuss the limitations in Section~\ref{sec:limfur} and finally conclude our paper in Section~\ref{sec:conclusion}. 
We postpone \newrev{extended security analysis,}
% an alternative decryption procedure driven by deep learning, 
several details on parameter settings, as well as future works to the Appendices.

\section{\MakeUppercase{Background and Literature}}
\label{sec:background}
\vspace{-1ex}
In this section, we provide basic knowledge on Wi-Fi sensing and define the attack model to characterize the potential risk of privacy leakage. We then discuss more on the literature based on explained basics and model. \newrev{Finally, we briefly introduce the key rationale behind our solution.}

\vspace{-1ex}
\subsection{Wi-Fi Sensing and CSI} \label{ssec:basic}
\vspace{-1ex}
%
% We hereby discuss how Wi-Fi sensing can be realized in general. 
During a multipath propagation process, Wi-Fi signals may experience diversified distortions determined by environment (hence channel) conditions. In particular, 
% Nevertheless, 
a moving object can substantially alter the signal propagation environments and thus the Wi-Fi channels. For instance, performing gestures between a pair of Wi-Fi transmitter and receiver can block certain signal paths while generating new ones, thereby causing the corresponding variations in Wi-Fi channels. In other words, the foundation of Wi-Fi sensing is the correlation between Wi-Fi channel variations and the states of moving objects.

While enabling sensing, Wi-Fi channel (i.e., CSI) estimation is also the default key to decode communication symbols. Therefore, existing Wi-Fi standards heavily rely on preambles (in particular the \textit{long training sequence}~(LTS)~\cite{biswaschannel} contained inside) for performing CSI estimations and in turn for correctly decoding the communication symbols. Wi-Fi LTS is a piece of \newrev{known yet unalterable information} without any encryption~\cite{yangdeath}, so anyone (regardless of whether registered to a Wi-Fi network or not) can potentially leverage a device equipped with Wi-Fi NICs (network interface cards) to overhear the LTSs and extract the CSI passively. 
% When a Wi-Fi transmitter (Tx) sends an LTS signal $\mathcal{s}_{m}$ (in time domain) of the $m$-th packet (cardinality $M$), the receiving (Rx) signals become $\mathcal{y}_{m} = \mathcal{h}_{m} * \mathcal{s}_{m} $ where the $\mathcal{h}_{m}$ represents the CSI of a Wi-Fi channel, and the $*$ means convolution operation. According to~\cite{MaZW19}, the CSI $\mathcal{h}_{m}$ can be modeled as:
% %
% \begin{align} \label{eq:csi_subc}
%     \mathcal{h}_{m} = \sum_{k = 1}^{K} \alpha_{m,k} e^{-j 2 \pi f_c  \tau_{m,k}  } 
% \end{align}
% %
% where $\alpha_{m,k}$ is the attenuation along the $k$-th path,  $f_c$ denotes the carrier frequency, and $\tau_{m,k} = \tau_k^{\mathrm{S}} + \tau_{m,k}^{\mathrm{D}}$ refers to the time delay: the static one $\tau_k^{\mathrm{S}}$ from an arbitrary reflector and the dynamic one $\tau_{m,k}^{\mathrm{D}}$ caused by its motions. 
When a Wi-Fi transmitter (Tx) sends an LTS signal $\mathcal{s}_{n,m}$ 
% (in frequency domain) 
at the $n$-th OFDM subcarrier (cardinality $N$) of the $m$-th packet (cardinality $M$), the receiving (Rx) signals become $\mathcal{y}_{n,m} = \mathcal{h}_{n,m} \mathcal{s}_{n,m} $ where the CSI $\mathcal{h}_{n,m}$
% of a Wi-Fi channel 
is obtained as
% To obtain the CSI, one often performs zero-forcing channel estimation algorithm~\cite{Argos-MobiCom12} to deduce 
$\mathcal{h}_{n,m} = \mathcal{y}_{n,m} / \mathcal{s}_{n,m}$. 
\newrev{To facility estimating $\mathcal{h}_{n,m}$, $\mathcal{s}_{n,m}$ actually remains constant in $m$.}
According to~\cite{MaZW19}, $\mathcal{h}_{n,m}$ can be modeled as:
\begin{align} \label{eq:csi_subc}
    \mathcal{h}_{n,m} = \textstyle{\sum_{k = 1}^{K}} \alpha_{n,m,k} e^{-j 2 \pi (f_c + n \Delta f ) \tau_{m,k}  } 
\end{align}
where $\alpha_{n,m,k}$ is the attenuation along the $k$-th path,  $f_c$ denotes the carrier frequency, $\Delta f$ represents the subcarrier spacing, and $\tau_{m,k} = \tau_k^{\mathrm{S}} + \tau_{m,k}^{\mathrm{D}}$ refers to the time delay: the static one $\tau_k^{\mathrm{S}}$ from an arbitrary reflector and the dynamic one $\tau_{m,k}^{\mathrm{D}}$ caused by its motions.

\begin{figure}[t]
	\setlength\abovecaptionskip{8pt}
        \setlength\belowcaptionskip{-3pt}
	\vspace{-2ex}
	\centering
	\subfloat[Amplitude of push-pull.]{
		\begin{minipage}[b]{0.47\linewidth}
			\centering
			\includegraphics[width = \textwidth]{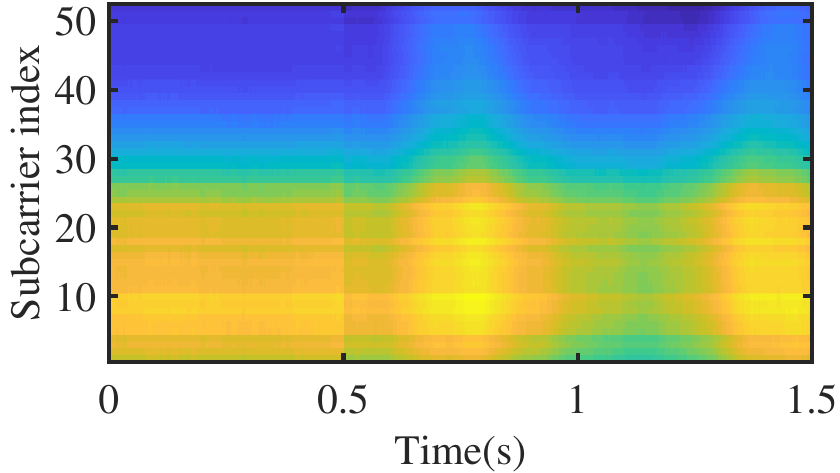}
			\label{sfig:apushpull}
			\vspace{-3ex}
		\end{minipage}
	}
	\subfloat[Phase of push-pull.]{
		\begin{minipage}[b]{0.47\linewidth}
			\centering
			\includegraphics[width = \textwidth]{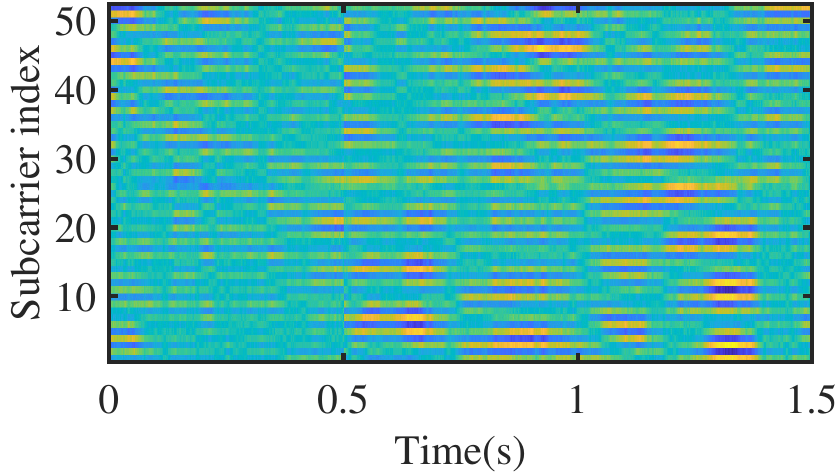}
			\label{sfig:ppushpull}
			\vspace{-3ex}
		\end{minipage}
	}
	\\
        \vspace{-1ex}
	\subfloat[Amplitude of zigzag.]{
		\begin{minipage}[b]{0.47\linewidth}
			\centering
			\includegraphics[width = \textwidth]{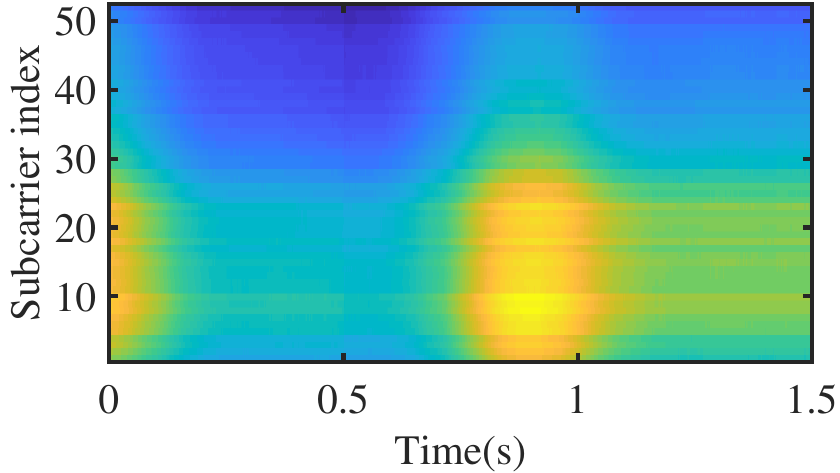}
			\label{sfig:azigzag}
			\vspace{-3ex}
		\end{minipage}
	}
	\subfloat[Phase of zigzag.]{
		\begin{minipage}[b]{0.47\linewidth}
			\centering
			\includegraphics[width = \textwidth]{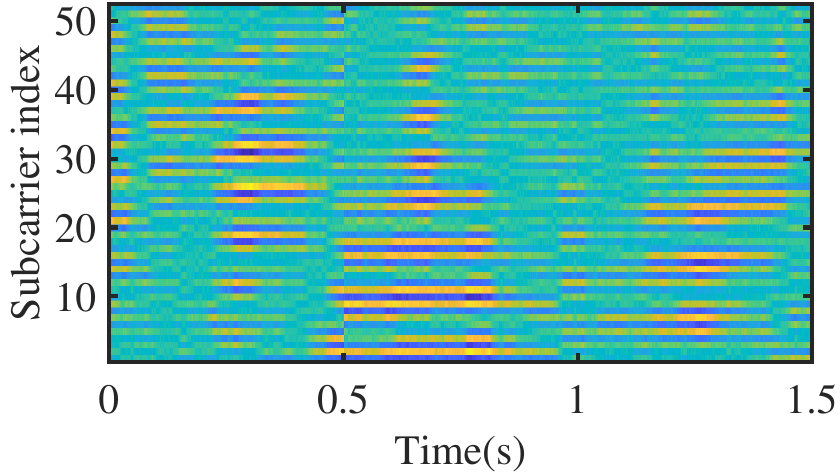}
			\label{sfig:pzigzag}
			\vspace{-3ex}
		\end{minipage}
	}
	\caption{\rev{CSI amplitudes and phases of two common gestures, while (a)-(b) of ``push-pull'' and (c)-(d) of ``zigzag''.}}
\label{fig:csiSensing}
\end{figure}

% \rev{
% Given a subcarrier CSI vector $\bm{h}_q = [\mathcal{h}_{m}]_q$ from the $q$-th Tx-Rx pair, the in total $Q$ ($Q = 8$ for the $4\times2$ Tx-Rx case) CSI vectors (a.k.a. \textit{CSI matrix})}
Given a CSI matrix $\bm{H}_q = [\mathcal{h}_{n,m}]_q$ from 
\newrev{the $q$-th Tx antenna, the in total $Q$ ($Q = 8$ for a 8 Tx antenna case)}
CSI matrices % (a.k.a. \textit{CSI tensor}) 
offers a substantial amount of information on the status and variations of the ambient environments; they have thus enabled a wide spectrum of Wi-Fi sensing applications, including activity tracking~\cite{WiDeo-NSDI15}, indoor localization~\cite{LiFS-MobiCom16}, and gesture recognition~\cite{Widar3-MobiSys19}. Although targeting distinct sensing goals, these proposals all rely on features extracted from CSI, i.e., amplitudes and phases of individual subcarriers, as well as their variations across time $m$ and frequency $n$. Since the aggregated amplitude of all subcarriers (a.k.a. RSSI) used by earlier proposals~\cite{EtTuAlexa-NDSS20,IRShield-SP22} is merely a low-end variant of CSI, we focus on preserving privacy for CSI-enabled Wi-Fi sensing throughout our paper.
%is . Nevertheless, the channel information embedded in RSS is a subset of CSI amplitude.
To further illustrate the correlations between CSIs and user motion states, we take gesture recognition as an example. Figure~\ref{fig:csiSensing} presents CSI amplitudes and phases of all subcarriers and across several LTSs upon performing two hand gestures: push-pull and drawing a zigzag. We can clearly observe the distinctions in CSI features 
% (especially in phases) 
between the two gestures, so these results indicate the feasibility of using CSI features to differentiate user motion states. \newrev{To simplify our design, we decide to protect all subcarriers as a whole by imposing distinct ``scrambling'' in time. Consequently, we omit the frequency dimension $n$ from now on.}

\vspace{-.5ex}
\subsection{Attack Model} \label{ssec:attackm}
\vspace{-1ex}
As mentioned earlier, \newrev{no encryption is applied to Wi-Fi LTS, hence} anyone with a Wi-Fi NIC can potentially overhear the CSIs from various APs around it, even without being registered (hence authenticated). \newrev{We consider a \textit{multi-user scenario} where an AP (Alice) can thwart multiple eavesdroppers (Eves, albeit only one is shown in Figure~\ref{fig:teaser}), while maintaining services to multiple legitimate sensing and communication users Bob$^{\mathrm{S}}$ and Bob$^{\mathrm{C}}$.
As an eavesdropper,
% In our commonly accepted attack model, an attacker 
Eve appears to be common users with Wi-Fi NICs in possession and bear in mind \textit{stealthiness} when launching attacks.}
% Therefore, in our commonly accepted attack model, an attacker Eve (as shown in Figure~\ref{fig:teaser}) is just a common user (hence stealthy) with a Wi-Fi NIC in possession.
To maintain the stealthiness, an Eve performs \textit{passive} Wi-Fi sensing by acquiring \newrev{CSI vector $\bm{\mathcal{h}} = [\mathcal{h}_{m}]$} from an AP (Alice) without being registered to it,\footnote{As both Eve and Bob$^{\mathrm{S}}$ act passively, only one instance needs to be considered in performace evaluation.} 
and thus eavesdrops on private information of close-by persons (\textit{victim}s who are not necessary Wi-Fi users), such as their password inputs and/or control gestures. 
To remain focusing on the defending scheme, only \textit{hand gesture recognition} (HGR) task is considered throughout the paper, yet our proposal is readily extensible to defend other sensing tasks. Moreover, Bobs and Eves adopt a publicly-known neural model (e.g., WiSee~\cite{WiSee-MobiCom13}) to perform HGR.
Finally, the AP (Alice) is always deemed as a trusted party, so security issues such as rogue APs~\cite{RogueAP-TMC,RogueAP-CCS16} are not our concern. Specifically, our attack model endows 
% for verifying the ability of \sname in privacy-preserving for Wi-Fi sensing, we enable 
Eve with the following capabilities stronger or at least equal to existing ones~\cite{PhyCloak-NSDI16}:

\emph{Eavesdropping Position.}
We do not impose any physical limitations on Eves' positions. In particular, Eves 
% can either share the same room with 
and Bobs do not have to be separated into different physical spaces; this differs significantly from existing proposals~\cite{Aegis-INFOCOM18,EtTuAlexa-NDSS20,IRShield-SP22} to be elaborated in Section~\ref{ssec:survey}. % In particular, we allow Eve to approach Bob in an identical room and adjacent positions as described in Section~\ref{sssec:scramblingPer}. 
% Even if without position limitations, Eve still cannot successfully eavesdrop.% due to lacking descrambling key.
Essentially, we allow Eves to choose any proper positions to maximize their chance of eavesdropping while still maintaining stealthiness. 
% Our target, as will be evaluated in Section~\ref{sssec:scramblingPer}, is to thwart such eavesdropping attempts regardless of Eves' physical positions.

\emph{Antenna Quantity.}
Considering the hardware configuration of common Wi-Fi NICs (e.g., those used by smartphones and laptops) and the stealthiness required by Eve, we assume that Eve is equipped with two antennas~\cite{AntennaNumber-TMC}. However, to demonstrate the strength of our proposal in thwarting more powerful eavesdropping capabilities, we use both security analysis in \newrev{Section~\ref{ssec:secana}} and experiments in Section~\ref{sssec:antqua} to confirm that increasing antenna numbers does not make the attack model stronger.
% study the proposed mechanism's ability in Wi-Fi based securing sensing, 
% we also evaluate its performance when Eve owns antennas larger than eight in Section~\ref{sssec:antqua}.

\emph{Multiple Attackers.} 
\newrev{We allow multiple Eves to collude with each other, so as to increase the number of antennas under their possession. However, the stealthiness of Eves forbid them to collude in the form of launching \textit{known-plaintext attack} (KPA). Since the information (hence the \textit{plaintext}) to be protected under our model is a \textit{physical process} of hand gesture, launching KPA requires the \textit{physical presence} of at least two Eves in the service range of the AP: one acts as the victim to generate physical plaintexts (hand gestures) and another observes the resulting CSI variations. Nonetheless, such an obvious attack would severely expose Eves and hence totally go against their stealthiness.}

\emph{HGR Model.} 
To follow the \textit{open design principle} that the security of a mechanism should not depend on the secrecy of its design or implementation, we let Bobs and Eves share the same HGR model readily retrievable from certain public domains. 
%
%rely on distinct models for gesture recognition, there exists a non-negligible deviation in performance evaluation since breaks the rule of controlling the single variable (i.e., our scrambling mechanism). Therefore, we requires them to utilize identical 
%
In our evaluations, common HGR models Widar3~\cite{Widar3-MobiSys19} and WiSee~\cite{WiSee-MobiCom13}) are adopted to quantify the effectiveness against malicious eavesdropping.

\iffalse
	We need to  clarify the fundamental difference between sensing security and (conventional) communication security in the concept of plaintext (i.e., contents to be secured): it is physical for sensing but digital for communications, hence making the known-plaintext attack (KPA) very different. 
	%	In particular, the plaintext is not LTS (All Reviewers); LTS is only the channel information carrier, so changing it, on one hand, is impossible as it is part of the communication protocol fixed in firmware, and on the other hand would compromise communications. Instead, 
	Plaintext for sensing is the physical phenomenon (hand gestures and the incurred channel variations in our context); such plaintext raises the bar for attacker, as they have to be physical yet digital defence may thwart physical attacks. Also, conventional security analysis may not be applicable to cases with both plaintext and attacks being physical. For example, launching KPA is possible but requires the collusion of at least two Eves: one performs hand gestures (and labels them) while another observing the scrambled CSIs. Nonetheless, such an obvious attack would totally expose Eves and hence go against their stealthiness (a realistic requirement specified in our attack model). This is in stark contrast to compromising the security of Wi-Fi data traffics that doesn’t require such an obvious presence of Eves.
\fi

%
% PS: the whole process: \\
% Tx: $ e^{j 2 \pi fc t} s(t) $ where $s(t) = IFFT (\bm{s}_j)$

\vspace{-.5ex}
\subsection{Existing Solutions} \label{ssec:survey}
\vspace{-1ex}
Given the attack model defined in Section~\ref{ssec:attackm}, Qiao \textit{et~al}. propose PhyCloak~\cite{PhyCloak-NSDI16} as one of the seminal countermeasure to thwart such eavesdropping attacks. Essentially, PhyCloak assumes an HGR setting similar to Figure~\ref{fig:teaser} but, instead of multiple legitimate users (Bobs) and unauthorized eavesdroppers (Eves), only one Bob is allowed and all others are deemed as Eves. Therefore, PhyCloak lets Bob to first perform sensing and then obfuscate the channel \rev{$\mathcal{h}_{m}$} so as to prevent Eves from doing the same. To perform these two tasks simultaneously (otherwise Eves can still perform sensing before the arrival of obfuscated signals), sophisticated yet immature FDR~\cite{FDR-SIGCOMM13} was adopted. While PhyCloak apparently cannot support multiple legitimate users, the need for FDR has substantially increased the system complexity and made PhyCloak incompatible with Wi-Fi standard. Though the experiments of PhyCloak show that it barely affects communication quality, one would suspect the results of being biased because, as far as Bob sufficiently scrambled \rev{$\mathcal{h}_{m}$}, communication quality is almost surely affected, albeit to a little less extent than Eves' sensing.

A similar idea of obfuscating the channel \rev{$\mathcal{h}_{m}$} has been further extended in Aegis~\cite{Aegis-INFOCOM18} and it also aims to defend HGR, so all the aforementioned drawbacks of PhyCloak are inherited by Aegis. The only difference is that Aegis involves a motorized platform so that obfuscating \rev{$\mathcal{h}_{m}$} is achieved by changing the speed and direction of Bob's antennas. 
%\rev{Their differences present in two aspects: Aegis involves a motorized platform so that obfuscating $\mathcal{h}_{n,m}$ is achieved by changing the speed and direction of Bob's antennas; it requires placing Bob in an indoor room while Eve's eavesdropping implemented in outdoor positions.}
Recently, IRshield~\cite{IRShield-SP22} considers the more aggressive scenario deeming everyone to be an Eve, which has significantly simplified the solution.
% \rev{moreover, this work similar to Aegis imposes the eavesdropping position limitation, greatly compromising the primordial attack ability of Eve.}
% of disabling Eves' eavesdropping while retaining the data communication capability of Bobs. For making this strategy practical, 
Basically, IRShield adopts customized IRSs~\cite{IRS-LSA14} installed beside an AP to obfuscate \rev{$\mathcal{h}_{m}$}. As a result, no one 
% around the AP 
can perform CSI-based sensing anymore, yet communication quality should also be affected, though the experiments (with only one link) seem to suggest a negligible level of variation. Clearly, \textit{direct channel obfuscation} is impractical since it lacks necessary controllability to differentiate between legitimate users (of both sensing and communications) and eavesdroppers.
%unauthorized eavesdroppers.

To better control the channel obfuscation, one may consider scrambling \rev{$\mathcal{s}_{m}$} so that the derived \rev{$\mathcal{h}_{m} = \mathcal{y}_{m} / \mathcal{s}_{m}$} gets obfuscated. As \rev{$\mathcal{s}_{m}$} can be fully controlled by Alice (an AP), differentiating ``good or evil'' in a multi-user scenario becomes possible. As a primitive version of this \textit{source-defined channel scrambling} approach, \cite{EtTuAlexa-NDSS20} dynamically adjusts the Tx power to scramble the amplitude of \rev{$\mathcal{h}_{m}$}. 
Since CSI phases $\tau_k$ are not affected, this solution only works for certain sensing tasks such as (rough) indoor localization. Jiao \textit{et~al}.~\cite{openwifi-WiSec21} leverages a secret switching sequence among a set of pre-defined LTSs to perform scrambling upon \rev{$\mathcal{s}_{m}$}. 
\newrev{Such a solution is surely impractical because LTS is fixed in the digital circuit of an Wi-Fi NIC; it can be altered in~\cite{openwifi-WiSec21} only thanks to the adoption of customized hardware. Also, the binary pattern of LTS offers very little scrambling space.}
%
% Although it is able to differentiate Bobs from Eves by sharing the LTS switching sequence with Bobs, this hard-coded approach is totally incompatible with commodity Wi-Fi. In practice, changing LTS is impossible, since it is part of Wi-Fi communication protocol fixed in firmware. Moreover, that hardcoded approach is  highly susceptible to Eve's enumeration attack. 

% to introduces large overhead to Wi-Fi communication.

\vspace{-1ex}
\subsection{Rationale behind Our Solution}
\vspace{-1ex}
\newrev{
According to the aforementioned discussions, a distinction between normal (communication) security and sensing security becomes evident: the former has \textit{digital} plaintexts while those of the latter are \textit{physical}. This has made existing sensing security solutions (wrongly) believe that arbitrary scrambling (sacrificing Bobs) could be the only solution to thwart eavesdropping attacks. Fortunately, 
% correctly recognizing the plaintexts in our context, 
we can formally encrypt them (albeit physically) instead, so as to thwarts Eves without harming Bobs.
%
%As briefly discussed physical plaintexts in Section~\ref{ssec:attackm},  Eves can infer them from CSI variations, but  Alice can perform physical encryption on physical plaintexts to thwart Eves.  There is a close analogy between our physical encryption and digital encryption. 
Recalling that digital encryption widely used for data protection, such as advanced encryption standard~(AES)~\cite{heron2009advanced}, essentially leverages cryptographic keys to digitally ``modulate'' plaintexts.
% AES leverages a block cipher to break time serials symbols~(a.k.a data) into fixed-size blocks and tackle each block individually. Multiple rounds of permutation, substitution and XOR operations are performed on each block  resulting in  difficultly cracking.    
%
The same idea of ``modulation for encryption'' can be applied to physical plaintexts, though it may require a combination of digital control and physical execution. In particular, though LTS cannot be altered, we may exploit (digital) Wi-Fi
%
% Considering $M$ Wi-Fi packets, similar to digital encryption, our physical encryption also treats each Wi-Fi packet as a fixed-size block, and encrypts signals of LTSs in each packet.  In physical world, permutation and substitution operations on signals do not exist, but XOR signals in the air can be implemented via 
%
MIMO beamformers~\cite{van2004implementation} to physically change the transmitted form of LTS: as far as the sequence of beamformers (equivalent to a cryptographic key) are kept secret, our physical plaintexts are successfully encrypted and thus hidden from Eves with no knowledge of the key.
It is worth noting that the idea of \textit{encrypting physical plaintext} differs fundamentally from the \textit{physical layer security} mechanisms meant to protect only \textit{digital plaintext} without a cryptographic key~\cite{Strobe, KPA-NDSS14, Robin}.
}

\section{\MakeUppercase{Motivations \& Feasibility Study}} \label{sec:motiv}
\vspace{-.5ex}
We hereby leverage simple experiments to further motivate the design of \sname, while verifying its feasibility.

\subsection{Thwarting Eavesdroppers} \label{ssec:thw-eve}
As discussed in Section~\ref{ssec:attackm}, an eavesdropper stealthily obtains the CSI vector from an AP, so as to extract gesture-dependent features and perform HGR for inferring the private information of another person. \newrev{Let us briefly demonstrate how 
% that exploiting 
MIMO beamformers can encrypt CSIs $[\mathcal{h}_{m}]$ (hence the gestures) to thwart the eavesdroppers. Essentially, 
% we can apply 
an \textit{encryption matrix} $\bm{\Psi}$ is applied to scramble an LTS vector $\bm{\mathcal{s}} = [\mathcal{s}_m]$: since $\mathcal{s}_m = 1,\forall m$ without loss of generality (see Section~\ref{ssec:basic}), the outcome $\bm{\mathcal{s}}' = \mathcal{E}(\bm{\mathcal{s}}, \bm{\Psi})$ leads to an encryption of $[\mathcal{h}_{m}]$ as $\mathcal{h}_{m} = \mathcal{y}_{m} / \mathcal{s}'_{m}$.
%
% have their OFDM (subcarrier) symbols pseudo-randomly assigned to different antennas. 
% Consequently, the actually CSI tensor $[\bm{H}_q] = [\mathcal{h}_{n,m,q}]$ gets scrambled as $[\mathcal{h}_{n,m,q}]\bm{\Psi}$ to eavesdroppers unaware of $\bm{\Psi}$, causing failure in successful 
%
Consequently, Eves would fail to eavesdrop HGRs without the knowledge of $\bm{\Psi}$. To ease the exposition of the encryption procedure $\mathcal{E}(\cdot)$, we hereby give a simple case study of transmitting an LTS vector $\bm{\mathcal{s}} = [\mathcal{s}_m]_{1 \le m \le M}$ with 8 Tx antennas.
%
% 64 subcarriers and $8\times1$ Tx-Rx antennas (hence $N=64, Q=8$). 
% and transmitted packet quantity $M$ equal to 1, 
Specifically, the component in $\bm{\Psi}$ for each time instance $m$ are 8 complex values (corresponding to the 8 antennas); they are multiplied to 8 $\mathcal{s}_m$'s (actually just 1's) individually and then get randomly and independently altered for $m+1$.
% $64\times64$ complex matrix; it aims to scramble the 64 subcarriers via modulation and reshuffle. 
% We then employ the $64\times64$ complex matrix of each Tx, to scramble the original LTS with 64 OFDM symbols. 
Finally, the 8 independently scrambled $[\mathcal{s}'_{m, 1 \le q \le 8}]$ are transmitted \textit{simultaneously} via 8 Tx antennas of Alice and then received by each Rx antenna of Bobs or Eves after these transmissions (via 8 independent channels) get mixed in the air.
% air-channel mixed versions.}
% scrambled version (containing $8\times64\times64$ elements mixed in the air. 
}

\begin{figure}[t]
	\setlength\abovecaptionskip{8pt}
        \setlength\belowcaptionskip{-2ex}
	\vspace{-1ex}
	\centering
	\subfloat[Amplitude pattern.]{
		\begin{minipage}[b]{0.47\linewidth}
			\centering
			\includegraphics[width = \textwidth]{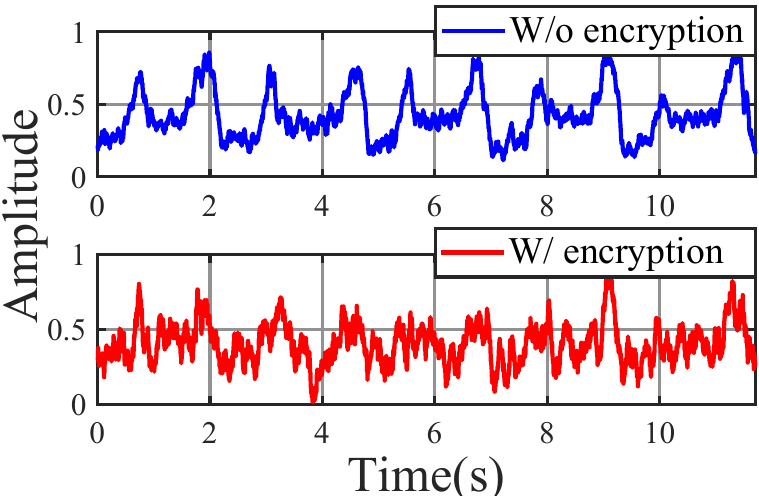}
			\label{sfig:amplitude}
			\vspace{-2.8ex}
		\end{minipage}
	}
	\subfloat[Phase pattern.]{
		\begin{minipage}[b]{0.47\linewidth}
			\centering
			\includegraphics[width = \textwidth]{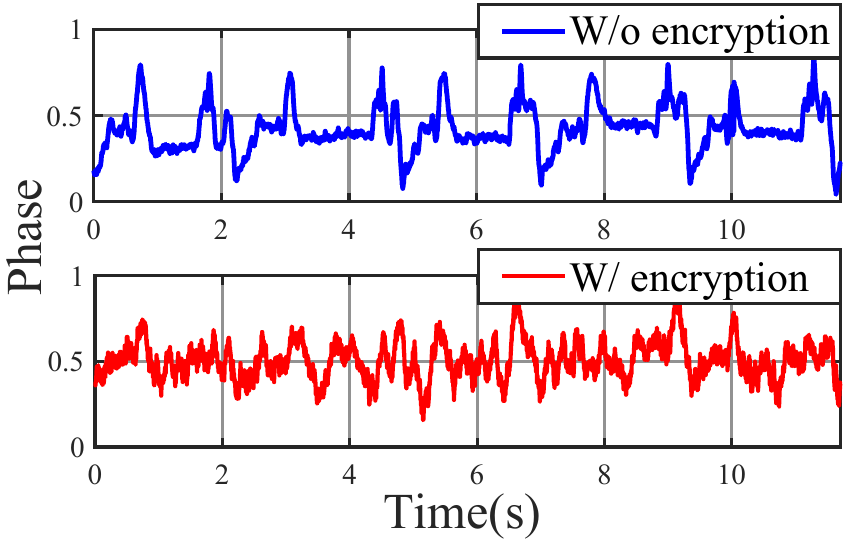}
			\label{sfig:phase}
			\vspace{-2.8ex}
		\end{minipage}
	}
	\\ \vspace{-1ex}
	% \subfloat[Doppler frequency shift.]{
	\subfloat[Time-frequency analysis on phase.]{
		\begin{minipage}[b]{0.88\linewidth}
			\centering
			\includegraphics[width = \textwidth]{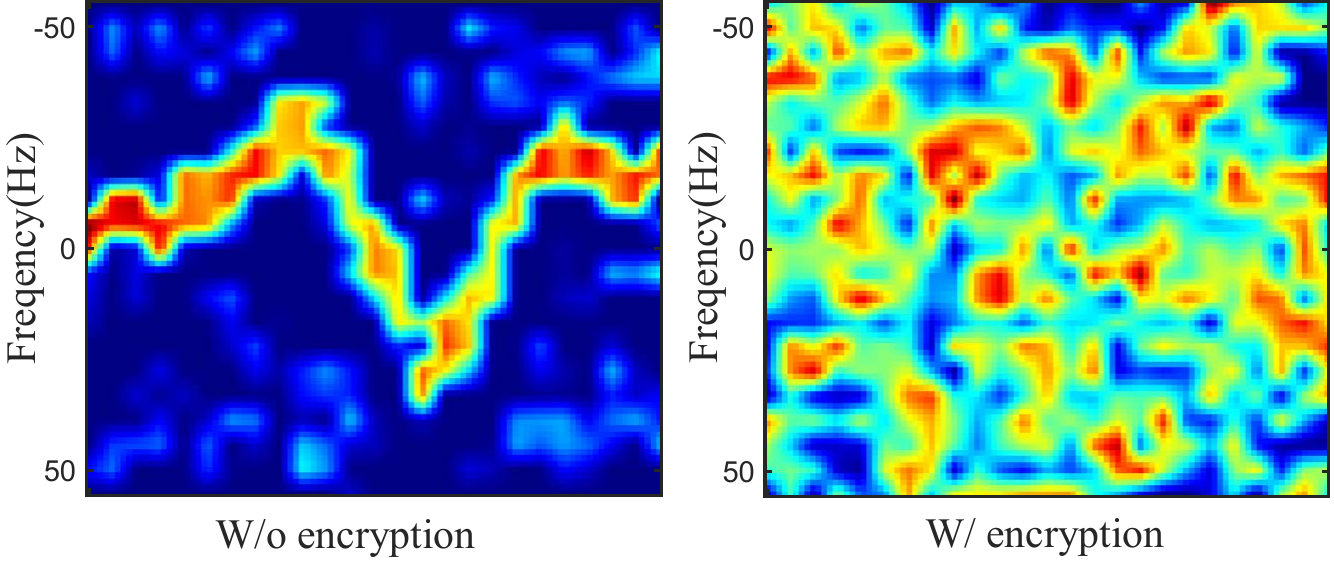}
			\label{sfig:dfs}
			\vspace{-2.5ex}
		\end{minipage}
	}
	\caption{CSI sensing results of ``zigzag'' hand waving. Results with/without MIMO encryption appear in forms of (a) amplitude, (b) phase, and (c) time-frequency analysis.}
	\label{fig:teve}
	%\vspace{-1.5ex}
\end{figure}

We set up an experiment using two combined WARP nodes with 8 antennas to act as Alice, and another WARP node with 1 antenna as Eve. The distance between them is set to 3.5~\!m, while one victim waves a hand along a ``zigzag'' path at 0.5~\!m distance from Alice. We extract CSI features from one randomly selected subcarrier (the 32-th in this case) for illustration purpose.
% to represent hand gestures. 
%We select the CSI of the 32-th subcarrier possessing the smallest phase measurement error~\cite{AWL-CoNEXT17} to extract the relevant features. 
The amplitude and phase variations of this subcarrier induced by the hand gesture with or without applying $\bm{\Psi}$ are respectively shown in Figures~\ref{sfig:amplitude} and~\ref{sfig:phase}; we further perform time-frequency analysis on the phases to derive Figure~\ref{sfig:dfs}. Apparently, while the CSI can clearly ``depict'' the zigzag hand gesture (with two peaks indicating the turning points of the motion path), such critical information gets totally deprived of from the CSI after applying $\bm{\Psi}$.

\subsection{Communications for Legitimate Users} \label{ssec:plu}
\begin{figure}[t]
	\setlength\abovecaptionskip{8pt}
	\vspace{-1ex}
	\centering
	\subfloat[Encrypted LTS.]{  %Scrambled LTS.
		\begin{minipage}[b]{0.47\linewidth}
			\centering
			\includegraphics[width = \textwidth]{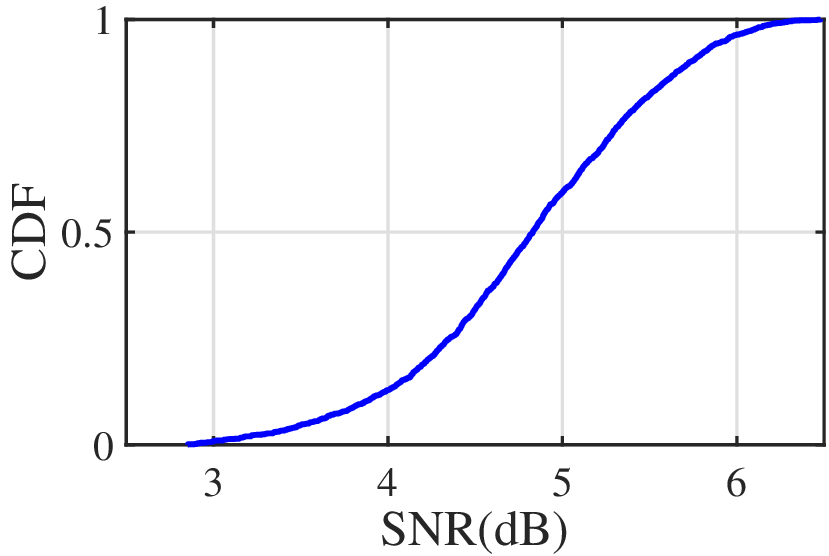}
			\label{sfig:scrambled_snr}
			\vspace{-2ex}
		\end{minipage}
	}
	\hfill
	\subfloat[Affected payload.]{
		\begin{minipage}[b]{0.47\linewidth}
			\centering
			\includegraphics[width = \textwidth]{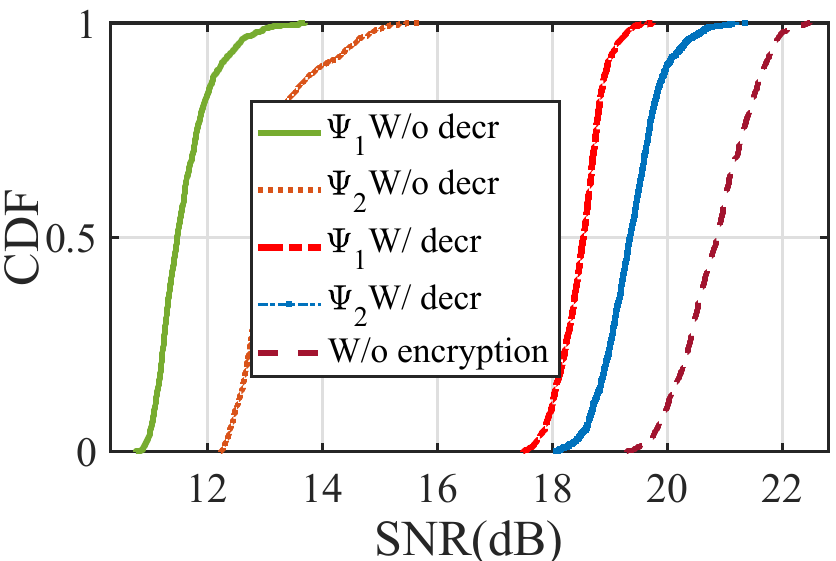}
			\label{sfig:decoded_snr}
			\vspace{-2ex}
		\end{minipage}
	}
	\caption{Channel encryption significantly degrades the Rx quality for LTS indicated by SNR (a) and affects the Rx quality of the payload (b). Though decrypting LTS may counteract the loss in payload SNR (b), the extent of compensation vary given different encryption matrices $\bm{\Psi}$'s.}
	\label{fig:scrambled_decoded_snr}
	\vspace{-1.5ex}
\end{figure}

\vspace{-1.5ex}
Applying the encryption matrix $\bm{\Psi}$ is a double-sided sword: since it scrambles the CSIs to all potential receivers, it hurts both eavesdroppers and legitimate users (both $\mathrm{Bob^C}$ and $\mathrm{Bob^S}$). We hereby use a simple experiment to showcase the degradation of communication quality, judged by the commonly adopted SNR (signal-to-noise ratio) metric.
To effectively measure this metric, we fix the antenna gain of the 8-antenna WARP node to 3~\!dBi and the transmission power of radio-frequency (RF) front-end to 20~\!dBm. We let the node (Alice) transmit 1000 data packets with encrypted LTS, while another 1-antenna node (Bob) observes the SNR distribution of the Rx LTS signals. As depicted in Figure~\ref{sfig:scrambled_snr}, 
the average SNR is only 4.81~\!dB, less than the required threshold: Wi-Fi network normally demands at least a 5~\!dB SNR, below which even the error rate of LTS (using BPSK) can be drastically increased~\cite{SNR-MobiCom16}. Meanwhile, the payload (no $\bm{\Psi}$ applied) also suffers as shown by the two leftmost curves in Figure~\ref{sfig:decoded_snr}, because correct CSIs (estimated from LTS's) are crucial to communication qualities.

Fortunately, initiating the MIMO encryption by an AP (thus our source-defined concept) provides a natural leverage to handle this problem: legitimate users can decrypt the CSIs with $\bm{\Psi}$ securely conveyed to them. This is in stark contrast to prior channel obfuscation approaches lacking of ability to distinguish between Eves and Bobs~\cite{PhyCloak-NSDI16,IRShield-SP22}.
Nevertheless, owning $\bm{\Psi}$ does not directly lead to the recovery of true CSIs, because the encryption may cause higher CSI estimation errors due to the twisted waveform (to be elaborated in Section~\ref{ssec:optimization}). 
We choose two arbitrary $\bm{\Psi}$'s to share between Alice and Bob; observations made on the SNR distributions of payload in Figure~\ref{sfig:decoded_snr} are: i) though significantly elevated compared with no LTS decryption, the average SNRs are still below that of normal Wi-Fi communications, and ii) the SNR distributions corresponding to the two arbitrary $\bm{\Psi}$'s bear discernible differences, indicating a chance of further improving the SNR by selecting the optimal $\bm{\Psi}$'s. In summary, 
% these results demonstrate that, 
while optimizing $\bm{\Psi}$ is crucial to protecting legitimate users, our source-defined MIMO encryption scheme offers a handle to approach the optimization problem involving multiple objectives: namely thwarting eavesdroppers while still preserving the service quality for legitimate users.
%
%In other words, these results demonstrate that, while optimizing $\bm{\Psi}$ is crucial for protecting legitimate users, the source-defined channel scrambling scheme does offer us a handle to execute this optimization.

%
\vspace{-1ex}
\subsection{Efficient Sensing for Legitimate Users} \label{ssec:effsense}
\vspace{-1.5ex}
Although Bobs can perform both communication and sensing tasks by first decrypting the received CSI, there might exist more efficient ways to execute the sensing tasks. In particular, a straightforward implementation of HGR under MIMO encryption with $\bm{\Psi}$ is to first recover the true CSI vector $[\mathcal{h}_{m}]$ and then feed it into a trained neural model (e.g.,~\cite{Widar3-MobiSys19,RFNet-SenSys20}). 
%Nevertheless, this process involves the inversion of $\bm{\Psi}$ and thus results in additional time costs that cannot be ignored for resource-limited Bobs. 
Nevertheless, this process involves the inversion of $\bm{\Psi}$ and additional matrix computations, thus resulting in extra time costs that cannot be ignored for resource-limited Bobs. 
Fortunately, as sensing tasks often demand a lower signal granularity than communication tasks, explicitly inverting $\bm{\Psi}$ can be avoided. Basically, whereas communications have to recover symbols represented by every Rx frame,
% $[\mathcal{h}_{n,m,q}]\bm{\Psi}\mathcal{s}_m$, 
HGR only requires differentiating a countable amount of gestures given multiple Rx frames. 
Therefore, a light-weight implementation could rely on a neural model to concurrently handle both $\bm{\Psi}$-related operations and HGR.
% at the same time.
% Therefore, a more light-weight implementation could rely on a neural model to handle both (partial) inversion and HGR at the same time.

\begin{figure}[t]
	\setlength\abovecaptionskip{8pt}
	\vspace{-1.5ex}
	\raggedleft
	\!\!\!\!\!\!\!\!
	\subfloat[Trained w/o decrypt.]{
		\begin{minipage}[b]{0.32\columnwidth}
			\centering
			\includegraphics[width = \textwidth]{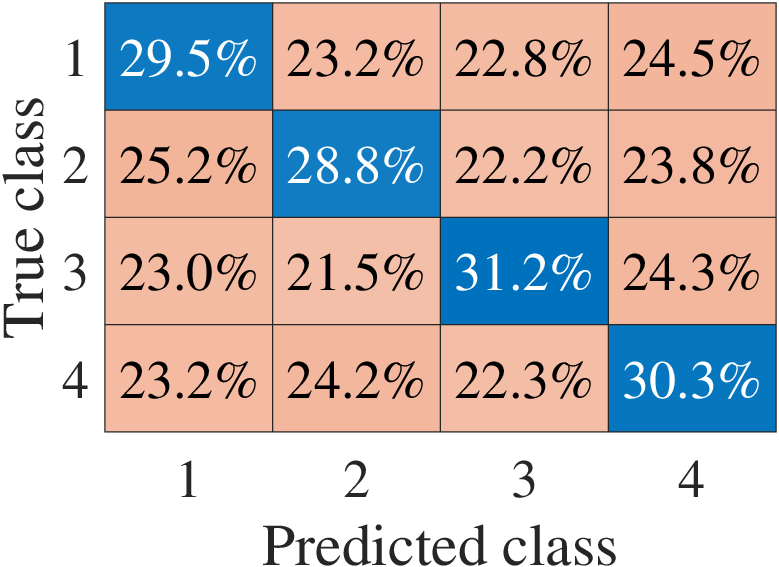}
			\label{sfig:scrambled_acc}
			\vspace{-2.2ex}
		\end{minipage}
	}
	\subfloat[Trained w/ decrypt.]{
		\begin{minipage}[b]{0.32\columnwidth}
			\centering
			\includegraphics[width = \textwidth]{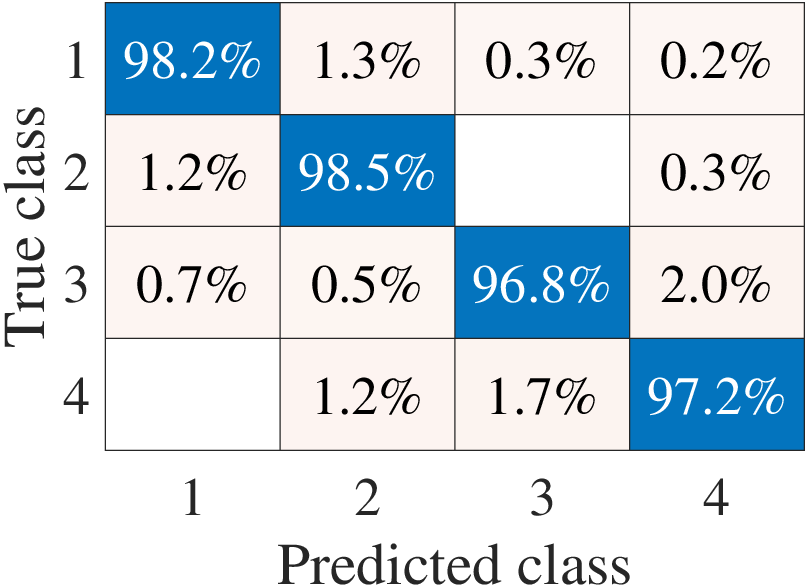}
			\label{sfig:decoded_acc}
			\vspace{-2.2ex}
		\end{minipage}
	}
	\subfloat[Retrain w/o decrypt.]{
		\begin{minipage}[b]{0.32\columnwidth}
			\centering
			\includegraphics[width = \textwidth]{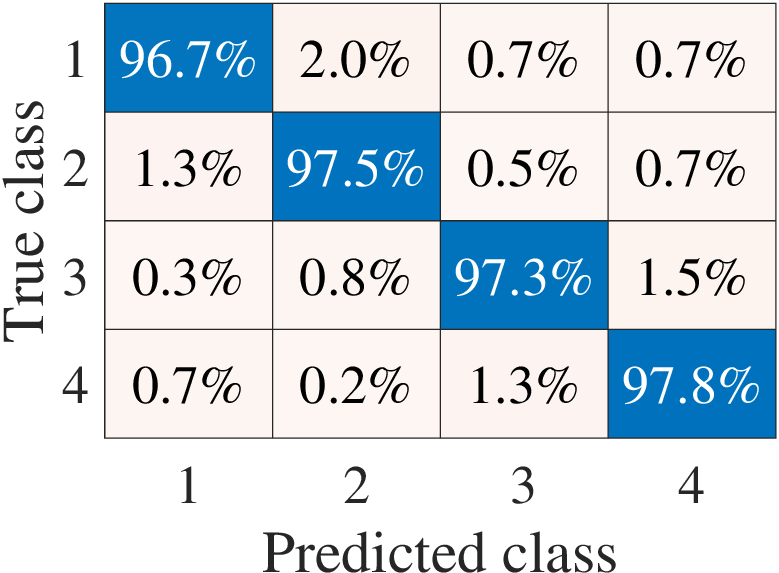}
			\label{sfig:learned_acc}
			\vspace{-2.2ex}
		\end{minipage}
	}
	\caption{HGR performance under different training and inputs: (a) a \textit{trained} model with encrypted CSIs, (b) the \textit{trained} model with decrypted CSIs, and (c) a \textit{retrained} model with encrypted CSIs.}
	\label{fig:train_retrained}
	\vspace{-1ex}
\end{figure}

We hereby briefly verify the feasibility of the aforementioned two HGR strategies. We first collect a dataset with 12,000 CSI samples (each sample contains 2~\!s recording of CSI vector) of four common gestures (i.e., zigzag, circle, push, and clap) under both encrypted and decrypted cases, 
% \rev{There are 80\% of samples used for training a gesture recognition model while the remaining part for testing the HGR performance.} 
80\% of which are used for training an HGR model while the remaining part for testing the HGR performance. 
We adopt the HGR model in~\cite{Widar3-MobiSys19} as the basis, but we either directly use the trained model with both encrypted and decryped CSIs as inputs, or retrain the model using the encrypted CSIs. 
As expected, Figures~\ref{sfig:decoded_acc} and~\ref{sfig:learned_acc} demonstrate that the models taking $\bm{\Psi}$ into account (by either decrypting or retraining) perform much better than that with encrypted CSIs as input in Figure~\ref{sfig:scrambled_acc}. Moreover, the two well-performed strategies achieved very similar average accuracy of around 97\%, but obtaining the decrypted CSIs requires an extra time consumption of 0.83~\!ms
% an extra time consumption of 76.2~\!ms
on the Snapdragon 8 Gen-1 CPU of a Xiaomi 12 smartphone. In reality, $\bm{\Psi}$ can be updated (similar to the security key update~\cite{Keyupdate-1,Keyupdate-CCS21}), so retraining an HGR model upon each update is highly impractical; 
% an efficient strategy should 
incurring a minimal cost in updating $\bm{\Psi}$ is desirable.

\vspace{-.5ex}
\section{\name: Derivation and Design} \label{sec:sys}
\vspace{-1ex}
This section presents the theoretical derivations and technical workflow of \sname.
% in resisting Wi-Fi based eavesdropping attacks and retaining performance for legitimate users.
%
% \subsection{System Overview}
% \label{ssec:sysoverview}
%
% As briefly discussed in Sections~\ref{ssec:attackm} and~\ref{sec:motiv}, \sname aims to thwart eavesdroppers~(Eves) while preserving the sensing and communication capabilities for legitimate users~(Bobs). To achieve these objectives, \sname essentially applies $\bm{\Psi}$ to scramble Wi-Fi LTS and it securely shares $\bm{\Psi}$ only with Bobs but not Eves, so that Eves are deprived of sensing capability but Bobs get to retain both sensing and communications capabilities. As the design of \sname is substantially based on common MIMO settings, we hereby refine the model introduced in Section~\ref{ssec:basic}, by dividing the $Q$ Tx-Rx pairs into respective Tx and Rx antennas, i.e., $Q =  Q^{\mathrm{Tx}} \times Q^{\mathrm{Rx}}$. To offer sufficient diversity for Alice to perform channel scrambling, $Q^{\mathrm{Tx}}$ should be an adequate quantity (e.g., TP-Link AX6600~\cite{Router-TP} and Huawei AirEngine~\cite{Router-HW} have 8 and 16 antennas, respectively). As for Bobs or Eves, our derivation in Section~\ref{ssec:scrambling} shall prove that, unless $Q^{\mathrm{Rx}}$ is huge (e.g., hundreds or even more), there is a natural asymmetry in diversity between Alice and Eve, which serves as the fundamental basis or the security and validity of \sname. 
%
\newrev{
% While applying $\bm{\Psi}$ achieves the basic objectives of \sname, Sections~\ref{ssec:plu} and~\ref{ssec:effsense} also motivate a few aspects for further system optimization, prominently including selecting optimal $\bm{\Psi}$ to strike an adequate balance among three conflicting objectives and improving HGR efficiency for Bobs. To this end, 
We first introduce a new metric to measure the Wi-Fi sensing quality for HGR (based on the well-known SNR metric for communications) in Section~\ref{ssec:psnr}.
Then we present the core design of \sname in terms of spatial-temporal MIMO encryption in Section~\ref{ssec:scrambling}, followed by the security analysis in Section~\ref{ssec:secana} and 
$\bm{\Psi}$ fine-tuning via a multi-objective optimization framework in Section~\ref{ssec:optimization}.
% that involves verifying communication and sensing abilities, 
Finally, we propose 
an adaptive inference architecture (for Bob$^{\mathrm{S}}$) to accommodate encrypted CSIs with minor communication overhead in Section~\ref{sec:ssCSI}, before summarizing the workflow of \name in Section~\ref{ssec:together}.
}

\vspace{-.5ex}
\subsection{Quantifying Motion Sensing Quality} \label{ssec:psnr}
\vspace{-1ex}
\newrev{To have a security measure for the physical plaintext, we leverage SNR to quantify \textit{relative entropy}~\cite{SIG-018,entroy-snr}.} Though SNR is a well-known indicator for signal quality in communications, we need to refine it so as to suit the sensing task. Since Wi-Fi sensing (in particular HGR in our case) mainly focuses on motions, we modify the SNR definition to quantify the ratio between Wi-Fi (motion) sensing signal and other non-signal components. Based on Eqn.~\eqref{eq:csi_subc}, we have the \textit{sensing SNR} represented as follows:
\begin{equation} \label{eq:ssnr}
    \eta_{\text{S}} = \frac{1}{M} {\textstyle{\sum_{m = 1}^{M}}} \frac{| \mathcal{h}_{m}^{\mathrm{D}} |^2}{| \mathcal{h}_{m}^{\mathrm{S}}|^2 + \sigma^2},
\end{equation}
%
%\begin{align} \label{eq:ssnr}
%	\eta_{\text{S}} = \frac{1}{N M} \sum_{n = 1}^{N} \sum_{m = 1}^{M} \frac{| \mathcal{h}_{n,m}^{\mathrm{D}} |^2}{| \mathcal{h}_{n,m}^{\mathrm{S}}|^2 + \sigma^2},
%\end{align}
%
where $\sigma^2 $ is the variance of Gaussian white noise. We denote the dynamic reflection channels caused by human motion as 
\rev{$\mathcal{h}_{m}^{\mathrm{D}} = \sum_{k \in K_{\mathrm{D}}} \alpha_{m,k}^{\mathrm{D}} e^{-j 2 \pi f_{\mathrm{c}} \tau_k^{\mathrm{D}}}$}
%$\mathcal{h}_{n,m}^{\mathrm{D}} = \sum_{k \in K_{D}} \alpha_{n,m,k}^{\mathrm{D}} e^{-j 2 \pi (f_{\text{c}} + n\Delta f ) \tau_k^{\mathrm{D}}}$, 
and the static reflection channels from the surrounding objects (e.g., walls and furniture) as \rev{$\mathcal{h}_{m}^{\mathrm{S}} = \sum_{k \in K_{\mathrm{S}}} \alpha_{m,k}^{\mathrm{S}} e^{-j 2 \pi f_{\mathrm{c}} \tau_k^{\mathrm{S}} }$.}
%$\mathcal{h}_{n,m}^{\mathrm{S}} = \sum_{k \in K_{S}} \alpha_{n,m,k}^{\mathrm{S}} e^{-j 2 \pi (f_{\text{c}} + n \Delta f ) \tau_k^{\mathrm{S}} }$. 
%
According to Eqn.~\eqref{eq:ssnr}, a large $\eta_{\text{S}}$ indicates a stronger dynamic reflection signals over non-signal components, potentially resulting in better sensing performance. 

\begin{figure}[b]
	\setlength\abovecaptionskip{6pt}
	\vspace{-1.5ex}
	\centering
	\subfloat[SDNR.]{
		\begin{minipage}[b]{0.45\linewidth}
			\centering
			\includegraphics[width = \textwidth]{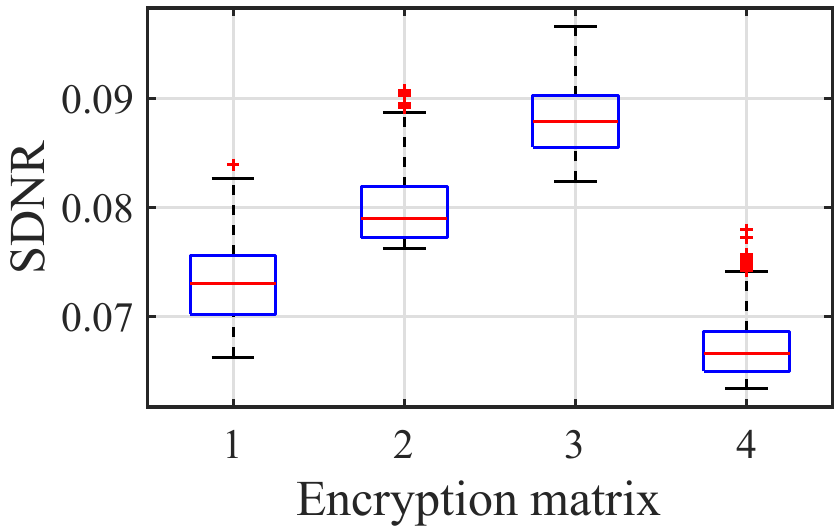}
			\label{sfig:sdnr}
			\vspace{-3ex}
		\end{minipage}
	}
	\vspace{.1ex}
	\subfloat[Envelope similarity.]{
		\begin{minipage}[b]{0.45\linewidth}
			\centering
			\includegraphics[width = \textwidth]{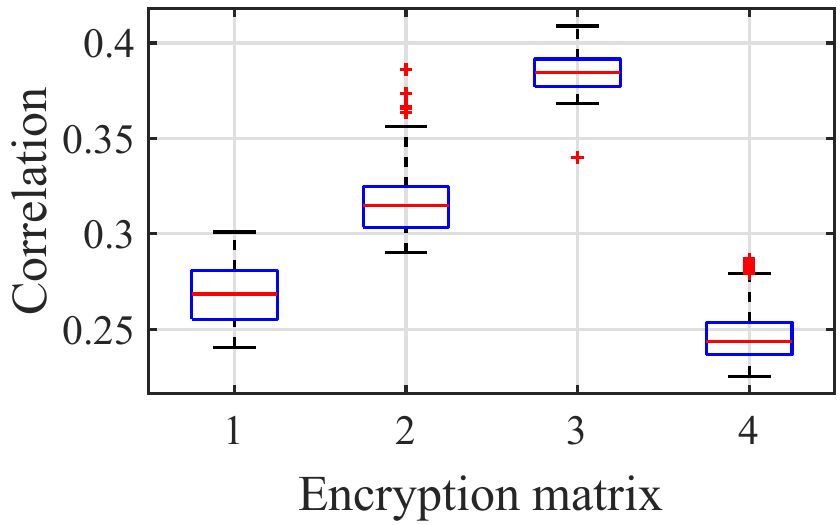}
			\label{sfig:sndr_sim}
			\vspace{-3ex}
		\end{minipage}
	}
	\caption{Four distinct $\bm{\Psi}$'s lead to
	% Variation occurred in signal distortion-to-noise ratio 
	(a) \newrev{SDNRs of encrypted CSIs} and (b) CSI envelope similarities different from non-encrypted ones; (a) and (b) are very similar in trend.}
	\label{fig:scramble_sdnr_sim}
	\vspace{-.5ex}
\end{figure}

%As our design involves channel scrambling and 
As \newrev{signal distortion caused by MIMO encryption is} not considered in Eqn.~\eqref{eq:ssnr}, we hereby introduce a novel \textit{signal-to-distortion-plus-noise ratio}~(SDNR): 
\begin{align} \label{eq:sdnr}
	\eta_{\mathrm{SD}} = \frac{1}{M} {\textstyle{\sum_{m = 1}^{M}}} \frac{| \hat{\mathcal{h}}^{\mathrm{D}}_{m}|^2}{| \hat{\mathcal{h}}^{\mathrm{}}_{m}  -    \mathcal{h}_{m}^{\mathrm{}}|^2 +| 
    \mathcal{h}_{m}^{\mathrm{S}}|^2 + \sigma^2}
\end{align}
%
%\begin{align} \label{eq:sdnr}
	%\eta_{\mathrm{SD}} = \frac{1}{N M} \sum_{n = 1}^{N} \sum_{m = 1}^{M} %\frac{| \hat{\mathcal{h}}^{\mathrm{D}}_{n,m}|^2}{| %\hat{\mathcal{h}}^{\mathrm{}}_{n,m}  -    %\mathcal{h}_{n,m}^{\mathrm{}}|^2 + | %\mathcal{h}_{n,m}^{\mathrm{S}}|^2 + \sigma^2}
%\end{align}
%
where 
%$ \varsigma =  \sigma^2  + | \mathcal{h}_{n,m}^{\mathrm{S}}|^2$ and 
%$ \hat{ \mathcal{h} }_{n,m} $
\rev{$ \hat{ \mathcal{h} }_{m} $ is the encrypted CSI,
%scrambled CSI,
% at the $n$-th OFDM subcarrier of the $m$-th packet. The term 
and $| \hat{\mathcal{h}}^{\mathrm{}}_{m}  -    \mathcal{h}_{m}^{\mathrm{}}|^2$ }
%$| \hat{\mathcal{h}}^{\mathrm{}}_{n,m}  -    \mathcal{h}_{n,m}^{\mathrm{}}|^2$ 
% in Eqn.~\eqref{eq:sdnr} 
represents the distortion caused by \newrev{MIMO encryption:
% scrambling.
% that larger difference between $\hat{\mathcal{h}}^{\mathrm{}}_{n,m}$ and $ \mathcal{h}_{n,m}$ results in lower $\eta_{\mathrm{SD}}$. 
% more analysis on $\hat{\mathcal{h}}^{\mathrm{}}_{n,m}$ will be described in Section~\ref{ssec:naive_scrambling}. 
%
% Subsequently, we explore the effectiveness of SDNR in representing signal distortion induced by encrypting, hence 
it also indicates a gain in entropy.} 
% scrambling. 
We conduct experiments with the same settings adopted in Section~\ref{ssec:thw-eve}, and we calculate the $\eta_{\mathrm{SD}}$ of original and encrypted CSIs (via 4 encryption matrices $\bm{\Psi}$'s to be elaborated in Section~\ref{ssec:scrambling}), while leveraging cosine similarity to compare their signal envelopes. The results in Figure~\ref{fig:scramble_sdnr_sim} reveal two critical points: i) $\eta_{\text{SD}}$ and signal similarity share the same variation trend, hence confirming SDNR's effectiveness, ii) encryption indeed works but to different extents determined by $\bm{\Psi}$, so we can further optimize $\bm{\Psi}$'s to \newrev{fine-tune the encryption strength} in Section~\ref{ssec:optimization}.
%scrambling strength in Section~\ref{ssec:optimization}.

\vspace{-.5ex}
\subsection{Spatial-Temporal Encryption}
\vspace{-1ex}
%\subsection{Spatial-Temporal Scrambling}
\label{ssec:scrambling}
According to 
% the motivating example in 
Section~\ref{ssec:thw-eve}, applying an encryption matrix $\bm{\Psi}$ to Wi-Fi LTS can largely thwart the Eve's intention to infer privacy-sensitive information via Wi-Fi sensing. To better characterize our source-defined channel encryption,
% (necessary to further optimizations)
we hereby consider only one Wi-Fi NIC (either Bob or Eve) when presenting \sname's 
% spatial-temporal 
encryption operations; the derivations can be readily extended to cover multiple NICs due to Wi-Fi's broadcast and time-division nature. How to differentiate Bob and Eve shall be discussed later, as the encryption operation deems them equally.
%scrambling operation deems them equally.

\begin{figure}[b]
    \setlength\abovecaptionskip{8pt}
    \vspace{-1ex}
	\centering
	\includegraphics[width = 0.85\columnwidth]{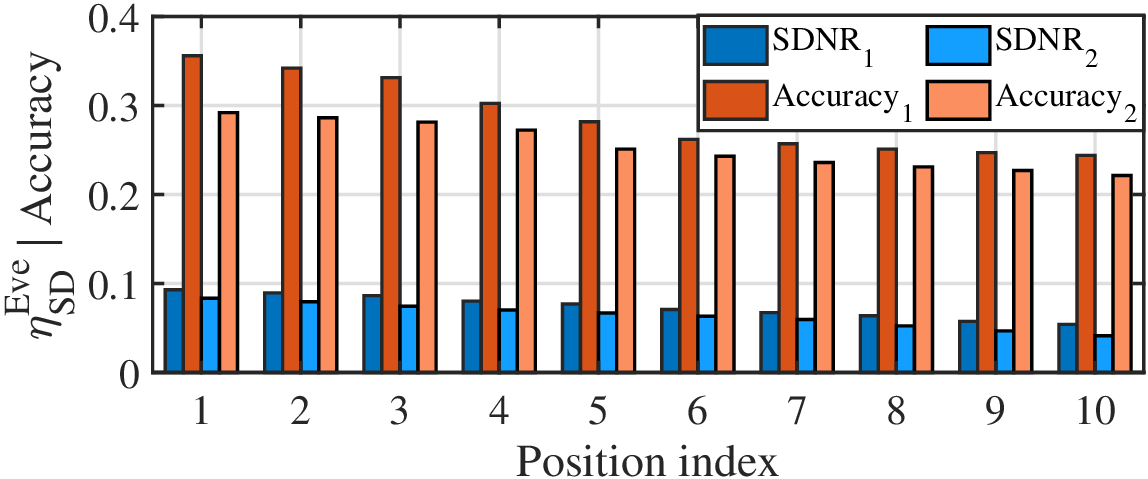}
	\caption{Hand gesture recognition accuracy under two $\bm{\Psi}$'s and varying SDNR $\eta_{\mathrm{SD}}$ measured at ten positions.} 
	\label{sfig:ssnr_accuracy}
	\vspace{-.5ex}
\end{figure}

\emph{MIMO-Driven Spatial-Temporal Encryption.}
%\emph{MIMO-Driven Spatial Scrambling.}
%
Let us formally present the source-defined encryption scheme in terms of the interaction between channels (or CSIs) and the encryption matrix $\bm{\Psi}$. Essentially, the distorted CSIs observed by a single Rx antenna (of Bob or Eve) can be written as:
\begin{align} \label{eq:mimoforAll}
% \hat{\bm{{H}}} =  \left[\bm{{H}}_1, \cdots, \bm{{H}}_q, \cdots, \bm{{H}}_{Q^{\mathrm{Tx}}} \right] \bm{\Psi} 
\hat{\bm{\mathcal{h}}} =  \left[\bm{\mathcal{h}}_1, \cdots, \bm{\mathcal{h}}_q, \cdots, \bm{\mathcal{h}}_{Q^{\mathrm{Tx}}} \right] \bm{\Psi} 
\end{align}
% where $\bm{\Psi} = [\psi_{\ell,n}]$ is a block-column matrix with each block $\Psi_q$ being an $N \times N$ square matrix. 
where \newrev{$\bm{\Psi} = [\Psi_1, \cdots, \Psi_q, \cdots, \Psi_{Q^{\mathrm{Tx}}}]^{\mathrm{T}}$: $\Psi_q = \mathrm{diag}(\mathcal{d}_m)_q$ is an $M \times M$ beamformer matrix applied to the $q$-th Tx antenna and $\mathcal{d}_m$ is a complex value distinct to the $m$-th packet (CSI). We have $\Psi_q = I$ (identify matrix) or $\bm{0}$ without encryption, depending on whether the $q$-th antenna is used to transmit signals or not.
Essentially, encrypting a time sequence of $M$ CSIs leads to a \textit{key length} of $M$, while the (implied) summation in Eqn.~\eqref{eq:mimoforAll}, indicating a ``(in the air) spatial mix'' of all $Q^{\mathrm{Tx}}$ transmissions, is similar to the MixColumns operation unique to AES~\cite{heron2009advanced}, which makes the AES secure against the differential and linear cryptanalysis.}
Without proper \newrev{decryption}, the degradation in the sensing performance of Eve can be characterized as, given the SDNR metric defined in Eqn.~\eqref{eq:sdnr}:
\vspace{-1ex}
\begin{equation} \label{eq:naMIMO}
	\eta_{\mathrm{SD}}^{\mathrm{Eve}} = \frac{1}{Q^{\mathrm{Tx}}M} 
        % \sum_{q = m = 1}^{Q^{\mathrm{Tx}},M} 
	% \sum_{m = 1}^{M} 
	%\frac{ \left\Vert \rev{\bm{h}_q^{\mathrm{D}} } \Psi_q \right\Vert^2_2   }{ \left\Vert \rev{ \bm{h}_q} \!-\! \rev{\bm{h}_q } \Psi_q \right\Vert_2^2 +\left\Vert \rev{ \bm{h}_q^{\mathrm{S}} } \Psi_q \right  \Vert^2_2 +  \sigma^2 },
        \frac{ \|\hat{\bm{\mathcal{h}}}^{\mathrm{D}} \|^2_2 } { \| \bm{\mathcal{h}} - \hat{\bm{\mathcal{h}}} \|_2^2 +\| \bm{\mathcal{h}}^{\mathrm{S}} \|^2_2 +  \sigma^2 },
\end{equation}
%
%\begin{scriptsize} 
%\begin{align} \label{eq:naMIMO}
%	\!\!\!\!\eta_{\mathrm{SD}}^{\mathrm{Eve}} = \frac{1}{Q^{\mathrm{Tx}}NM} \sum_{q = n = m = 1}^{Q^{\mathrm{Tx}},N,M} 
	% \sum_{m = 1}^{M} 
%	\frac{ \left\Vert \bm{H}_q^{\mathrm{D}} \Psi_q \right\Vert^2_2   }{ \left\Vert \bm{H}_q - \bm{H}_q \Psi_q \right\Vert_2^2 +\left\Vert \bm{H}_q^{\mathrm{S}} \Psi_q \right  \Vert^2_2 +  \sigma^2 }, 
%\end{align}
%\end{scriptsize}
%
where the operation $\Vert \cdot \Vert_2$ denotes the $l^2$ norm. 
To verify the relation between $\eta_{\text{SD}}$ (affected by \newrev{encryption}) and HGR sensing accuracy, we repeat the earlier experiments but adjust the distance between Alice and Eve from 2.5 to 5.2~\!m with a step size of 0.3~\!m. The victim then performs gestures at each Alice-Eve distance while the corresponding $\eta_{\mathrm{SD}}$ and HGR accuracy are measured. Figure~\ref{sfig:ssnr_accuracy} presents the average $\eta_{\mathrm{SD}}$ and HGR accuracy at the ten distinct Alice-Eve distances; it demonstrates a strong correlation between them, hence 
% shares the consistently increasing pattern, and demonstrates deep correlation. 
% Therefore, we have briefly 
proving the effectiveness of $\eta_{\mathrm{SD}}$ in qualifying the sensing quality of Wi-Fi-based HGR, as well as the strength of $\bm{\Psi}$ in thwarting potential eavesdropping.
%

\begin{figure}[t]
    \setlength\abovecaptionskip{8pt}
    %\vspace{-.3ex}
	\centering
	\includegraphics[width = 0.88\columnwidth]{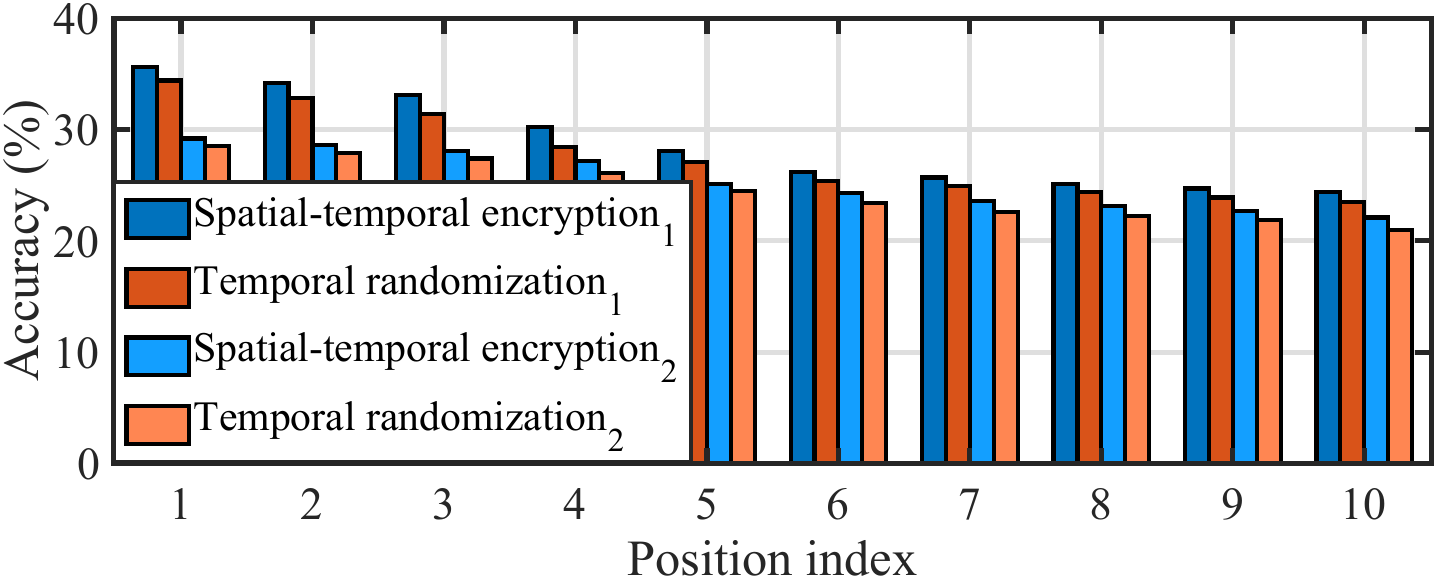}
	\caption{Hand gesture recognition accuracy under 
 \rev{spatial-temporal encryption and temporal randomization}.} 
 %two scrambling cases.} 
	\label{sfig:sampling_scrambling}
	\vspace{-1ex}
\end{figure}

%\emph{Temporal Scrambling via Sampling Rate Randomization.}
%\emph{Temporal \rev{Encryption} via Packet Interval Randomization.}
\emph{\newrev{Temporal Randomization.}}
%
% Apart from \rev{encrypting} CSIs with respect to both subcarriers and antennas 
\newrev{As the spatial-temporary encryption enforced via $\bm{\Psi}$ assumes regular time intervals, the encyption strength can be further enhanced by randomizing the temporal dimension,} 
% temporal dimension of CSIs can also be \rev{encrypted} to thwart eavesdropping, 
because neural models are often trained under the assumption of a known and constant input interval. To perform temporal \newrev{randomization}, we let the packet intervals take irregular values $ \hat{t}_m = (m +\beta_m)  \Delta t$ where $\Delta t $ is the original (constant) interval and $\beta_m \in \left[ -0.5, 0.5 \right] $ is the \newrev{randomization} ratio for the $m$-th packet. Consequently, the irregular interval time vector $[\hat{t}_m]$ (unknown to Eve) forces Eve to use a pre-trained neural model to handle the encrypted CSIs with further tampered Doppler frequency shift patterns.
% induced by human gestures. 
We specifically evaluate whether the packet interval randomization can \newrev{enhance the encryption strength of $\bm{\Psi}$,
% to further enhance the strength of channel \rev{encryption}. We 
reusing} the previous experiment setting but under two cases: i) $\bm{\Psi}$ only, and ii) adding temporal randomization. Then we leverage a model trained with authentic CSIs (see Section~\ref{ssec:effsense}) to perform HGR for the two cases. As shown in Figure~\ref{sfig:sampling_scrambling},
% adding packet interval 
\newrev{temporal} randomization can further degrade HGR accuracy compared with that of $\bm{\Psi}$ only, especially at a closer distance. 

\subsection{Security Analysis} \label{ssec:secana}
\vspace{-1ex}
\newrev{
This analysis focuses on two major threats: the observation diversity of Eves 
%(i.e., the amount of eavesdropping antennas) 
and the side-channel KPA. We postpone extended discussions to Appendix~\ref{apx:entropy}.

\emph{Observation Diversity.}
As Eve is totally unaware of the complex-valued parameters involved in $\bm{\Psi}$ whose quantity is $Q^{\mathrm{Tx}}\times M$, it may attempt to gain more ``observations'' via different antennas (including antennas from colluded Eves) to somewhat remove $\bm{\Psi}$. This diversity gain works under normal situation as the channel $\mathcal{h}_{m}$ is the only unknown fact in $\mathcal{h}_{m} = \mathcal{y}_{m} / \mathcal{s}_{m}$ for Rx decoding. However, \name encrypts the channels with $\bm{\Psi}$ as indicated by Eqn.~\eqref{eq:mimoforAll}. As a result, 
% for $Q^{\mathrm{Tx}} = 8$, 
every new antenna Eve adds to its possession introduces $Q^{\mathrm{Tx}}$ new unknown parameters (the channels between AP's antennas and that of Eve's), while $\bm{\Psi}$ remains unknown. Consequently, we may surely conclude that Eve earns no diversity gain by adding more antennas. 
% It is worth noting that 
%
% Nonetheless, Eve's $Q^{\mathrm{Rx}}$ needs to at least match the above quantities, which is highly impossible as common Wi-Fi NICs own only 2 antennas~\cite{AntennaNumber-TMC} and the inherent stealthiness of Eves prevents them from carrying devices with ostentatious features (e.g.,~\cite{BigStation-SIGCOMM13,Argos-MobiCom12}). Moreover, the independence among multiple receptions cannot be guaranteed even if Eve manages to compress many antennas into a compact device. 
%
% Given the same parameters used by the earlier case study, a received LTS vector $\hat{\bm{ {H}}} \bm{\mathcal{s}}_m$ contains $8\times64\times64$ unknown complex elements, apart from the 8 unknown channels. 
% For Bob or Eve, each Rx only can establish one equation for solving unknown elements. To remove the element freedom degree and thereby correctly estimate CSIs, the Wi-Fi NIC requires to 
%
% Therefore, unless Eve constructs $8\times64\times64$ independent equations by at least $8\times64\times64$ antennas, it has no chance to recover the channel CSI.

\emph{Side-channel KPA.}
Although our attack model in Section~\ref{ssec:attackm} forbids full-scale KPAs, one may wonder if a few gesture samples serve as ``side-channel'' to break the encryption; a digital version of this attack was proposed by~\cite{KPA-NDSS14} leveraging partial plaintext such as protocol headers. In fact, this side-channel KPA has been thwarted by a \textit{moving target defense}~\cite{Robin} that changes its (orthogonal) blinding patterns via physical antenna manipulations.
% only \needrev{in the scale of multiple frames (second level).}
% [Zhe] These part is not correct, since Robin changes the channel state at a per symbol or per frame rate via multiple antennas, but not second level. The description is located in last sentence in page 1: ROBin leverages a pattern reconfigurable antenna to vary the channel state at a per symbol or per frame rate, resulting in an artificially created fastchanging wireless channel unsuitable for the known-plaintext attack, for which can be viewed as one of the proactive/dynamic defense (or moving target defense) mechanisms.
Because \name adopts a cryptographic key whose beamformer patterns allow for much diversified choices from full space,
% and are digitally controlled by beamformers, 
side-channel KPA should not be a concern. In fact, this attack barely causes loss in conditional entropy, since $\eta_{\mathrm{SD}}^{\mathrm{Eve}}$ in Eqn.~\eqref{eq:naMIMO}, evaluated in statistical mean sense over a large amount of samples, cannot be significantly altered due to the exposure of a few samples; 
we shall further elaborate on this issue in both Section~\ref{sec:limfur} and Appendix~\ref{apx:entropy}.
% this is also attributed to the fundamental difference between physical and digital plaintext.
% Moreover, our physical plaintext does not accommodate the same side-channel as exploited by attacks under digital plaintext: a digital symbol is scrambled by the same blinding scheme~\cite{Robin}, whereas a physical symbol (a gesture) is encrypted by a sequence of time-variant MIMO beamformers.
%
Unfortunately, the combination of physical encryption and physical plaintext also makes common security analysis on encryption schemes unsuitable: 
% On one hand, it is not clear entropy can be defined for a physical symbol (a time-variant process). On the other hand, 
the encryption process is very different since complex operations are involved (i.e., beamformers comprise complex numbers) instead of digital substitution and transposition.
%[Zhe] I think "permuatation" is more suitable, since AES leverages substitution–permutation network to encrypt plaintext.
% reference: https://en.wikipedia.org/wiki/Substitution%E2%80%93permutation_network
}

\subsection{Balancing Security and Performance}
\label{ssec:optimization}
\begin{figure}[b]
	\setlength\abovecaptionskip{8pt}
	\vspace{-1ex}
	\centering
	\subfloat[Raw CSI.]{
		\begin{minipage}[b]{0.23\linewidth}
			\centering
			\includegraphics[width = \textwidth]{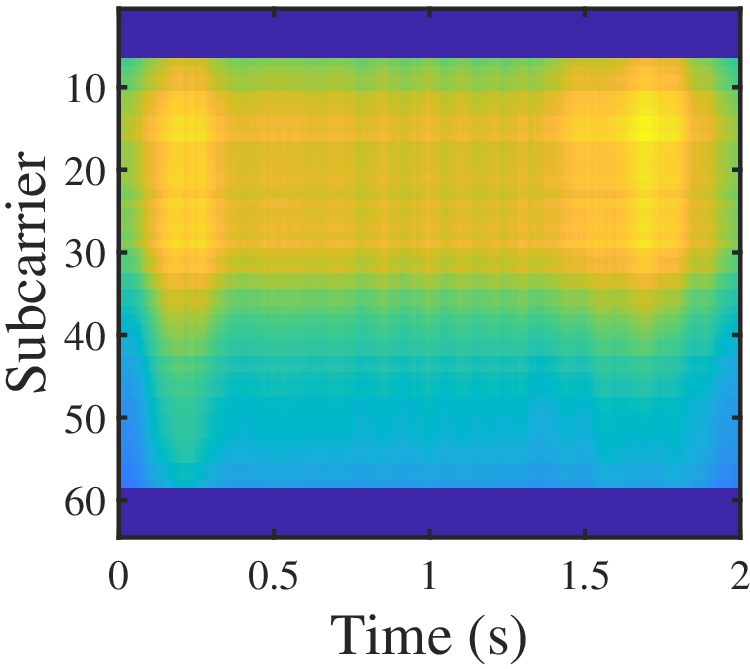}
			\label{sfig:rawcsi}
			\vspace{-3ex}
		\end{minipage}
	}
	\subfloat[$\bm{\Psi}_1$.]{
		\begin{minipage}[b]{0.23\linewidth}
			\centering
			\includegraphics[width = \textwidth]{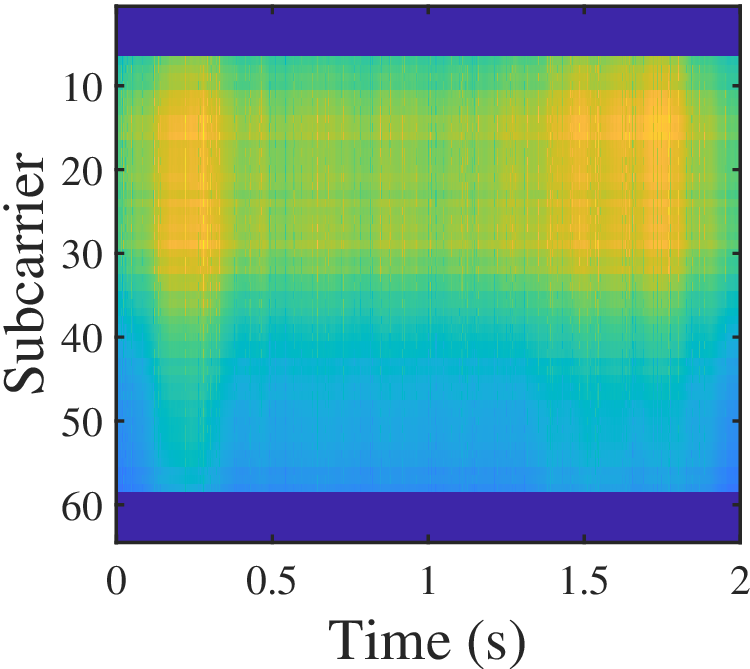}
			\label{sfig:desc1}
			\vspace{-3ex}
		\end{minipage}
	}
	\subfloat[$\bm{\Psi}_2$.]{
		\begin{minipage}[b]{0.23\linewidth}
			\centering
			\includegraphics[width = \textwidth]{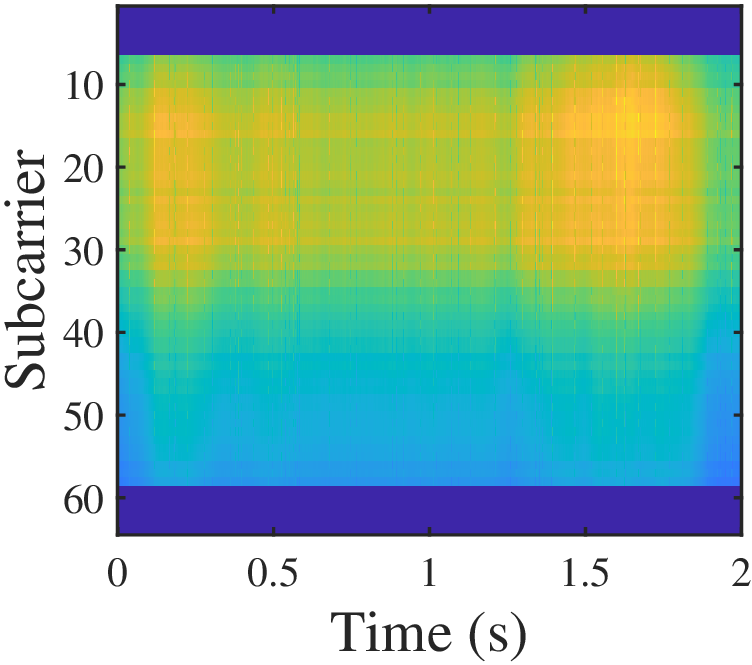}
			\label{sfig:desc2}
			\vspace{-3ex}
		\end{minipage}
	}
	\subfloat[$\bm{\Psi}_3$.]{
		\begin{minipage}[b]{0.23\linewidth}
			\centering
			\includegraphics[width = \textwidth]{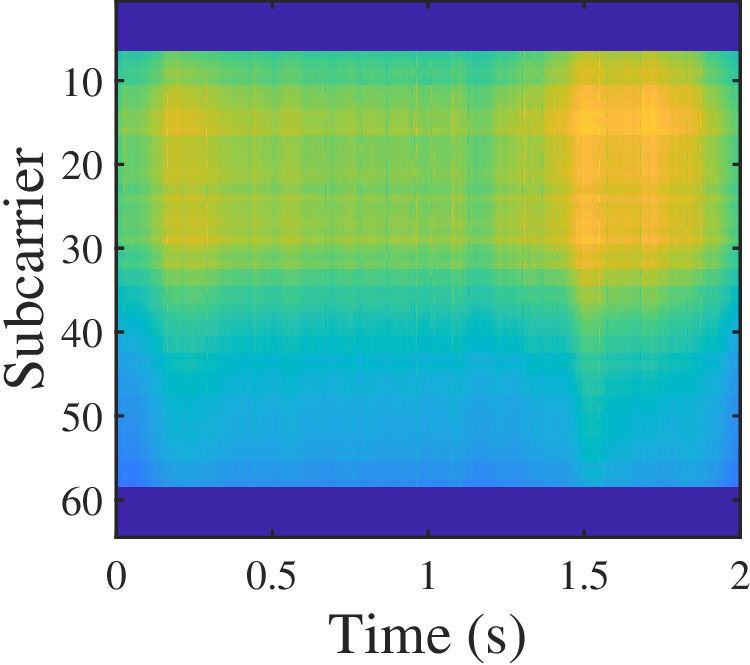}
			\label{sfig:desc3}
			\vspace{-3ex}
		\end{minipage}
	}
		\caption{CSI distortion persists even with the inverse of $\bm{\Psi}$ for decryption. We hereby compare a raw CSI sample (a) with three encrypted and then decrypted versions (corresponding to three distinct $\bm{\Psi}$'s) (b)-(d).}
 %\caption{CSI distortion persists even with the inverse of $\bm{\Psi}$ for descrambling. We hereby compare a raw CSI sample (a) with three scrambled and then descrambled versions (corresponding to three distinct $\bm{\Psi}$'s) (b)-(d).}
	\label{fig:csi_distortion}
	\vspace{-.5ex}
\end{figure}

\vspace{-1ex}
Whereas Section~\ref{ssec:scrambling} indicates that the encryption of \sname deems Bobs and Eves equally, their inherent difference can still be signified: Alice securely (re-)distributes the encryption matrices $\bm{\Psi}$'s to Bobs (who have registered to the Wi-Fi network), which can be readily achieved via key distribution mechanisms~\cite{menezes2018handbook}.
Recalling that arbitrarily applying $\bm{\Psi}$'s may degrade the communication quality for Bobs (see Figure~\ref{fig:scrambled_decoded_snr} in the Section~\ref{ssec:plu}), 
% (Bob$^{\mathrm{C}}$ hereafter), 
since the tolerance range for CSI estimation errors~\cite{li2006effects} can be violated by scrambling, which we further illustrate using Figure~\ref{fig:csi_distortion}.
Because the same issue exists for Bobs performing sensing too,
% Moreover, as motivated in Section~\ref{ssec:effsense}, there exist more efficient solution for Bobs using the sensing service (Bob$^{\mathrm{S}}$ hereafter). 
% Therefore, 
% it is necessary to propose a security and performance balance optimization method.
we hereby propose a \textit{multi-objective optimization framework} to strike a balance among i) risk of eavesdropping, ii) communication quality, and iii) sensing accuracy. In forming the optimization, Alice relies on the standard feedback~\cite{ieee80211ax} from Bobs to obtain their CSIs, while estimating those for Eves by averaging the received feedback.

%\emph{Optimizing Spatial Scrambling.}
%\paragraph{Optimizing Spatial Scrambling} 
%
Since a good tradeoff among thwarting eavesdropping, enhancing (legitimate) sensing accuracy, and preserving communication quality cannot be achieved by a single objective, 
%optimization problem cannot ensure their tradeoff. Therefore, 
we model this tradeoff via a multi-objective optimization problem with the encryption matrix $\bm{\Psi}$ as the variable:
\begin{align} \label{eq:mooOpt}
	\max_{\bm{ \Psi }} & \left[ -\eta_{\mathrm{SD}}^{\mathrm{Eve}},~~~\eta_{\mathrm{C}}^{\mathrm{Bob}},~~~\eta_{\mathrm{SD}}^{\mathrm{Bob}} \right] \\
	\mathrm{s.t.}~~& \Vert \bm{\Psi}   \Vert_2^2 ~~\le~ M  Q^{\mathrm{Tx}} , \label{eq:mooOpt:const1} \\
	& \eta_{\mathrm{C}}^{\mathrm{Bob}} \ge \epsilon_{\mathrm{C}}, ~~~\eta_{\mathrm{SD}}^{\mathrm{Bob}} \ge \epsilon_{\mathrm{SD}}, \label{eq:mooOpt:const2}
\end{align}
where the constraint~\eqref{eq:mooOpt:const1} bounds the total power from above, $\epsilon_{\mathrm{C}}$ and $\epsilon_{\mathrm{SD}}$ in the constraint~\eqref{eq:mooOpt:const2} act as the lower-bounds for the performance of Bob$^{\mathrm{C}}$ (legitimate communication users) and Bob$^{\mathrm{S}}$ (legitimate sensing users), respectively. 
The symbol $\eta_{\mathrm{C}}^{\mathrm{Bob}} = \| \hat{\bm{\mathcal{h}}}  \|_2^2 / \sigma^2 $ is the communication SNR of Bob$^{\mathrm{C}}$, and $\eta_{\mathrm{SD}}^{\mathrm{Bob}} $ is the SDNR of Bob$^{\mathrm{S}}$: a revision to Eqn.~\eqref{eq:naMIMO} as decryption is applied here:
\begin{align} \label{eq:sdnrBob}
	\eta_{\mathrm{SD}}^{\mathrm{Bob}} = \frac{1}{Q^{\mathrm{Tx}}M} 
        % \sum_{q = n = m = 1}^{Q^{\mathrm{Tx}},N,M} 
	% \sum_{m = 1}^{M} 
	\frac{ \Vert  \tilde{\bm{\mathcal{h}}}^{\mathrm{D}} \Vert^2_2   }{   \Vert {\bm{\Psi}}_q^{-1}  \sigma	 \Vert^{2}_2   +  \Vert  \tilde{\bm{\mathcal{h}}}^{\mathrm{S}}  \Vert^2_2 +  \sigma^2 }, 
\end{align}
where $ \bm{\Psi}^{-1} $ is the inverse of $\bm{\Psi}$,
%~\cite{ben2003generalized}. 
$ \tilde{\bm{\mathcal{h}}}^{\mathrm{D}}$  and $ \tilde{\bm{\mathcal{h}}}^{\mathrm{S}} $ are the dynamic and static reflection channels from Alice's Tx antennas to Bob's Rx antenna after decryption.
% by $ {\Psi}_q^{-1} $. 
The term $ \Vert \bm{\Psi}^{-1}  \sigma	 \Vert^{2} $ indicates that there still exists a \textit{residue distortion}, as we have illustrated in Figure~\ref{fig:csi_distortion}. Another observation to be drawn from Figure~\ref{fig:csi_distortion} is that, comparing with the original CSI, different $\bm{\Psi}$'s lead to distinct residue distortions, hence the need for optimization over $\bm{\Psi}$.
% for better robustness, it is urge to figure out the suitable  

%
To locate a good tradeoff on the Pareto optimal frontier of the above optimization problem, we apply the scalarization trick via a set of weights~\cite{cvx}:
\begin{align} \label{eq:moo_sum}
	\max_{\bm{ \Psi }} &~~  \omega_1 \eta_{\mathrm{C}}^{\mathrm{Bob}}  +\omega_2 \eta_{\mathrm{SD}}^{\mathrm{Bob}}   -\omega_3  \eta_{\mathrm{SD}}^{\mathrm{Eve}}   \\
	\mathrm{s.t.} &~~ \eqref{eq:mooOpt:const1},~~ \eqref{eq:mooOpt:const2},~~ \textstyle{\sum_i}  \omega_i = 1,  \label{eq:moo_sum:const3}
\end{align}
where $\omega_i \in [0, 1]$ is the scalar weight applied to an objective in the original problem~\eqref{eq:mooOpt}. To solve the problem~\eqref{eq:moo_sum}, we leverage ADMM~\cite{boyd2011distributed} to obtain local optimal solutions of this nonlinear non-convex optimization. 
\newrev{We omit the optimization for temporal randomization as this part does not affect the performance of Bobs. 

\vspace{1ex}
\noindent\emph{Remarks}: We deliberately put the optimization after the security analysis, because optimizing $\bm{\Psi}$ does not violate any arguments made in Section~\ref{ssec:secana}, hence it has virtually no impact on the encryption strength of $\bm{\Psi}$.
}
%
%\emph{Temporal Scrambling Optimization.}
%\paragraph{Temporal Scrambling Optimization} 
%
% Similar to $\bm{\Psi}$, we may also optimize over the irregular packet interval vector $[\hat{t}_m]$. Different from the spatial scrambling, the temporal scrambling does not affect the communication SNR $\eta_{\mathrm{SD}}^{\mathrm{Bob}}$. Also, as far as Alice may securely convey $[\hat{t}_m]$ to $\mathrm{Bob^S}$, the sensing performance for $\mathrm{Bob^S}$ should not be degraded at all. Therefore, the optimization can be performed only against Eve in terms of reducing its HGR accuracy, i.e., a single-objective optimization. Moreover, the outcome of this optimization can be readily derived as the most ``random'' time sequence, which, given the nature of Tx times (e.g., its positivity and constant interval), is known to be following an exponential distribution between consecutive $\hat{t}_m$'s.
%

\vspace{-1ex}
\subsection{Sensing under Encrypted CSIs} \label{sec:ssCSI}
\vspace{-1ex}
According to Section~\ref{ssec:effsense} and~\ref{ssec:optimization}, Bobs can perform both communication and sensing after receiving the securely distributed $\bm{\Psi}$ from Alice. Nonetheless, whereas this is necessary for Bob$^{\mathrm{C}}$ and its overhead can be amortized over the large volume of data traffic, \newrev{direct decryption for Bob$^{\mathrm{S}}$ appears to be neither efficient nor necessary, because learning HGR can be conducted in the encrypted domain. 
% distributing $\bm{\Psi}$ to Bob$^{\mathrm{S}}$ incurs non-negligible overhead, especially when \rev{the numbers of packets $M$} and Tx antennas $Q^{\mathrm{Tx}}$ are large. Fortunately, 
In particular, as} sensing tasks often demand a lower signal granularity
than communication tasks (as explained in Section~\ref{ssec:effsense}), we may build a deep sub-model to filter the encrypted CSIs so that they can correctly drive a given HGR model to perform classification. Though this sub-model pre-distributed to Bobs upon their registration still demands a \textit{decryption key} to act on a specific $\bm{\Psi}$, this key should incur a much lower overhead than $\bm{\Psi}$.

\begin{figure}[b]
	\setlength\abovecaptionskip{8pt}
	\vspace{-1ex}
	\centering
	\includegraphics[width = 0.95\columnwidth]{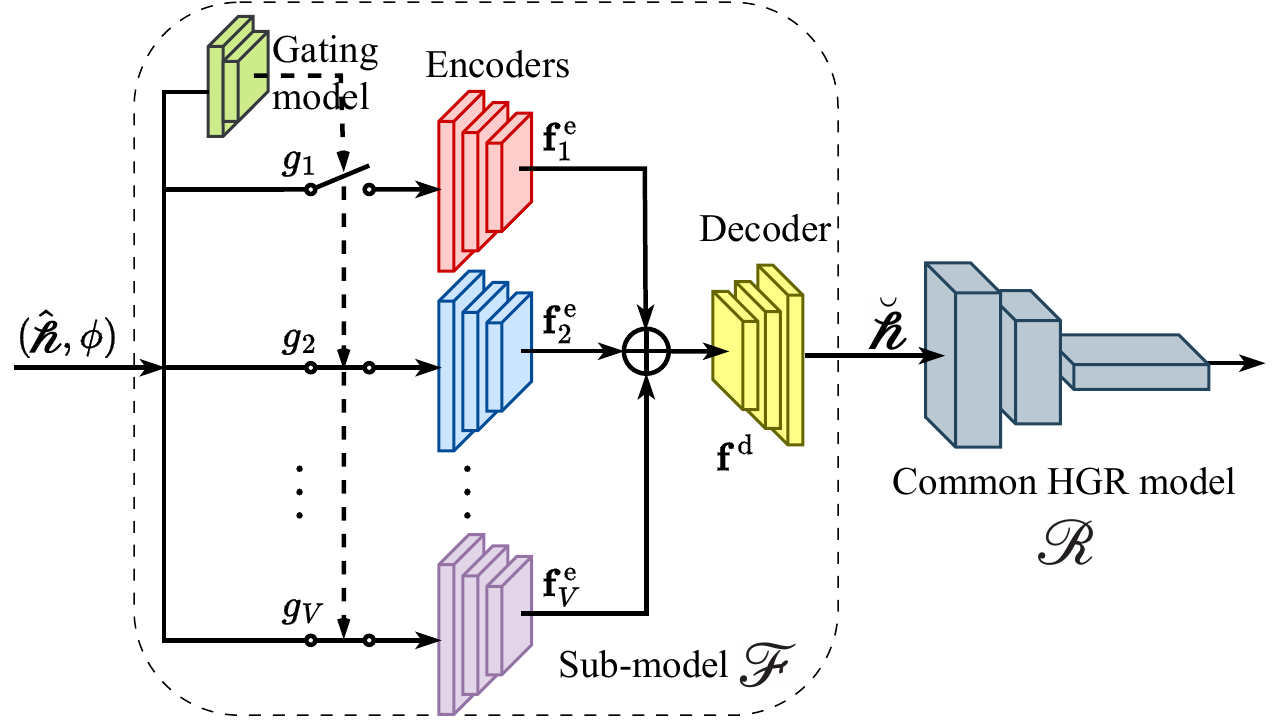}
	\caption{The ``one-fits-all'' HGR pipeline for \sname; it involves a dynamically adaptable pre-processing sub-model $\mathcal{F}$ and a common HGR model $\mathcal{R}$.}
	\label{fig:ill_HGRmodel}
	\vspace{-.5ex}
\end{figure}

\vspace{.5ex}
\emph{HGR Pipeline Given Encrypted CSIs.}
% \paragraph{HGR Pipeline Given Scrambled CSIs}
%
Our HGR pipeline under \sname framework comprises two major components: i) a plug-and-play sub-model to pre-process the encrypted CSIs, and ii) a commonly adopted HGR neural networks (e.g., Widar3~\cite{Widar3-MobiSys19}). As shown in Figure~\ref{fig:ill_HGRmodel}, these two components are denoted by $\mathcal{F}$ and $\mathcal{R}$, respectively. The output of this HGR pipeline is represented as \rev{$\bm{z} = \mathcal{R} \circ \mathcal{F}  ( [\hat{\bm{\mathcal{h}}}], \phi ) $} where 
% $[\hat{\bm{H}}_q]$ is the \rev{encrypted} CSI tensor and 
$\phi$ represents the \textit{key} generated by hashing the \rev{encryption} matrix $\bm{\Psi}$ that has produced \rev{$\hat{\bm{\mathcal{h}}}$}. Consequently, instead of distributing $\bm{\Psi}$, Alice needs only to securely share the key $\phi$ with Bobs, and $\phi$ is only several hundreds of bits at most given a normal cryptographic hash function (e.g., SHA-512~\cite{SHA}). 
The sub-model $\mathcal{F}$ to pre-process the \rev{encrypted} CSIs is constructed and trained to adapt to \rev{encrypted} CSIs produced by different $\bm{\Psi}$'s; it outputs \textit{surrogate CSI}s \rev{$\breve{\bm{\mathcal{h}}}$} to properly drive the HGR neural network $\mathcal{R}$ for classification. It is worth noting that $\mathcal{F}$ does not perform $\bm{\Psi}$-related matrix operation; 
%matrix inversion to a $\bm{\Psi}$, 
otherwise it would have to bear a high complexity in both model parameters and training. This is possible because Bob$^{\mathrm{S}}$ only demands the feature map generated by $\mathcal{R}$ (given an input $\hat{\bm{\mathcal{h}}}$) is well-aligned with the decision boundaries of its classifier; this is much easier than the symbol decoding by Bob$^{\mathrm{C}}$ to be completed only within a very short period of Wi-Fi signals (i.e., \rev{a few readings of $[\hat{\mathcal{h}}_{m}]$ for a given $m$}). %(i.e., a few readings of $[\hat{\mathcal{h}}_{n,m}]$ for a given $m$).

\vspace{.5ex}
\emph{Keyed Dynamic Neural Network.}
To accommodate the CSIs encrypted with distinct encryption matrix $\bm{\Psi}$, a naive solution is to train multiple versions of (static) neural networks $[\mathcal{F}_1, \mathcal{F}_2, \cdots ]$, according to different keys and CSIs produced by $\bm{\Psi}$'s, and to switch to the corresponding version when processing certain input. However, this naive solution introduces large overhead in terms of storage and training time. Therefore, we borrow the idea from \textit{dynamic neural networks}~\cite{DNN-TPAMI} that uses a small set of $V$ parallel neural networks, each controlled by a gate, to form the sub-model $\mathcal{F}$. When in action, the controllable gates may allow $\mathcal{F}$ to have up to $2^V$ different structures, hence dynamically adapting $\mathcal{F}$ to suit a particularly encrypted CSI input. In order to speed up the training, we deliberately add the key $\phi$; it delivers extra information to control the gates.
\begin{figure}[t]
	\setlength\abovecaptionskip{8pt}
	\vspace{-1ex}
	\centering
	\subfloat[Raw $\bm{\mathcal{h}}$.]{
		\begin{minipage}[b]{0.30\linewidth}
			\centering
			\includegraphics[width = \textwidth]{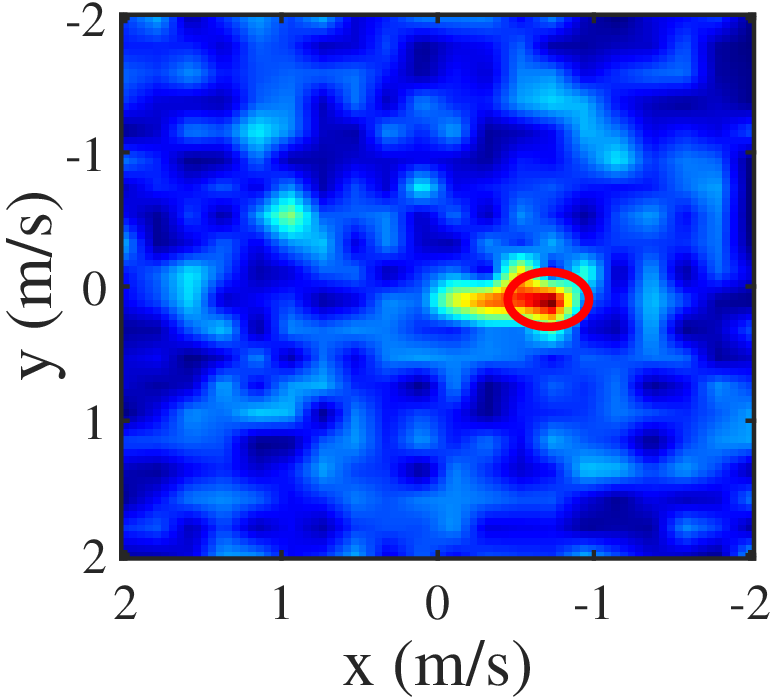}
			\label{sfig:rawBVP}
			\vspace{-3ex}
		\end{minipage}
	}
	%\vspace{.1ex}
	%
	%\subfloat[Scrambled $\hat{\bm{H}}_q$.]{
     \subfloat[Encrypted $\hat{\bm{\mathcal{h}}}$.]{
		\begin{minipage}[b]{0.30\linewidth}
			\centering
			\includegraphics[width = \textwidth]{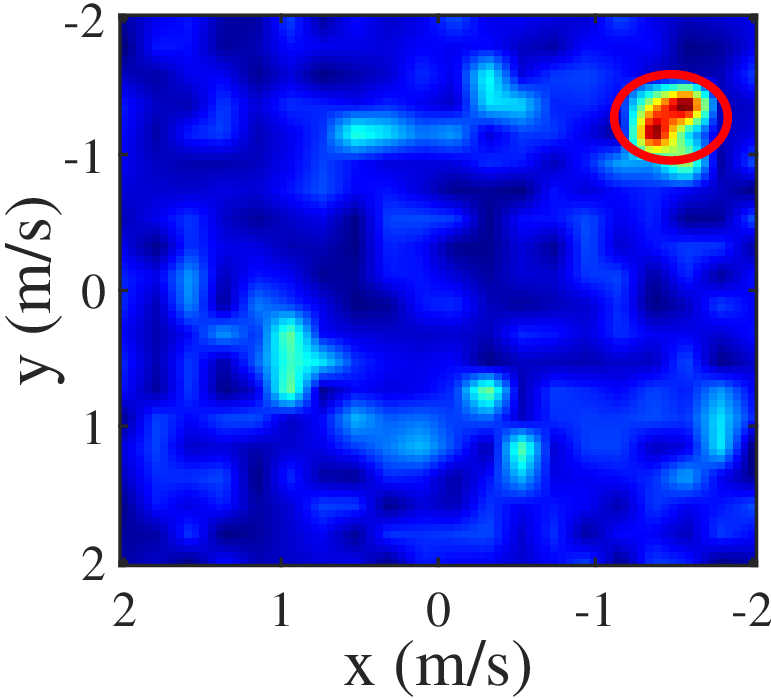}
			\label{sfig:scraBVP}
			\vspace{-3ex}
		\end{minipage}
	}
	%\vspace{.1ex}
	%
    \subfloat[Surrogate $\breve{\bm{\mathcal{h}}}$.]{
		\begin{minipage}[b]{0.30\linewidth}
			\centering
			\includegraphics[width = \textwidth]{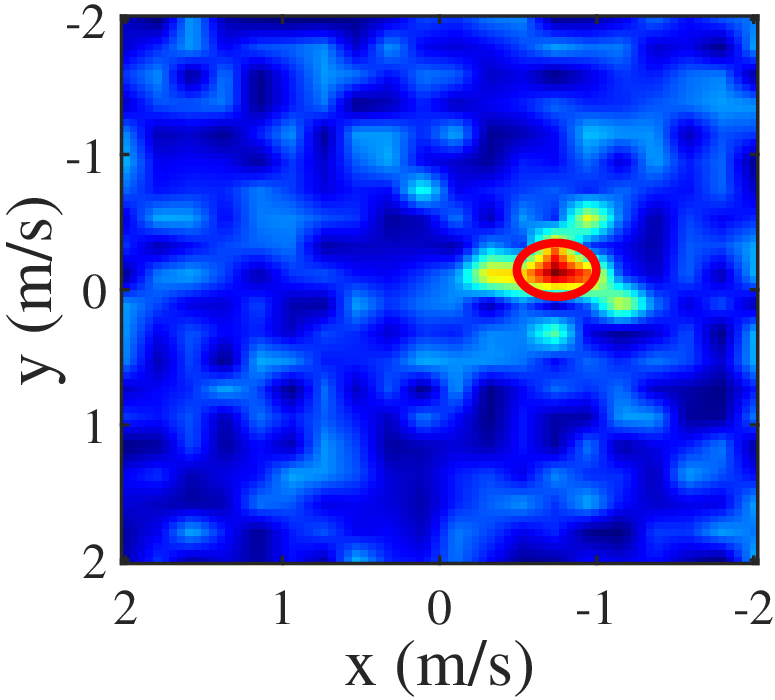}
			\label{sfig:SurrBVP}
			\vspace{-3ex}
		\end{minipage}
	}
	%\vspace{.1ex}
	%
	\caption{Feature maps of body-coordinate velocity profile (BVP) for Widar3 under different CSI inputs: (a) raw $\bm{\mathcal{h}}$, (b) encrypted $\hat{\bm{\mathcal{h}}}$, and (c) surrogate $\breve{\bm{\mathcal{h}}}$.}
	\label{fig:sensingBVP}
	\vspace{-1ex}
\end{figure}

Figure~\ref{fig:ill_HGRmodel} illustrates our keyed dynamic neural network, where $V$ encoders $\mathbf{f}^{\mathrm{e}}_i$ are controlled by a gating model $\mathbf{g}$ to obtain adaptability and only one decoder $\mathbf{f}^{\mathrm{d}}$ is used to generate the surrogate CSIs $\breve{\bm{\mathcal{h}}}$. Consequently, the sub-model $ \mathcal{F} $ can be represented as \rev{
$ \mathbf{f}^{\mathrm{d}} \circ ( \sum_{i = 1}^{V}  g_{i}  \circ \mathbf{f}^{\mathrm{e}}_i ) ( \bm{\hat{\bm{\mathcal{h}}}} , \phi )  $ }where $g_i  \in \{ 0, 1 \}$ is the binary output of $\mathbf{g}$ for controlling the activation of $i$-th encoder function $\mathbf{f}^{\mathrm{e}}_i $. We refer to Section~\ref{sec:imple} for detailed parameter settings for $ \mathcal{F} $.
During the training phase, we fix the weights for the pre-trained HGR model $\mathcal{R}$ and only update the weights of $\mathcal{F}$ via the gradient initiated by the cross-entropy loss function: 
\begin{align} \label{eq:deCNN}
	\min_{\mathcal{F}} &  - \frac{1}{ |S| }  \sum_{j \in S}^{} \sum_{c \in C}^{}  y_{j,c} \log \left( [\mathcal{R } \circ \mathcal{F} \left(\bm{\hat{\bm{\mathcal{h}}}}, \phi \right) ]_c \right)
\end{align}
where $S$ and $C$ are the set of training samples and classes, respectively. A trained $ \mathcal{F} $, when facing a specific input pair $(\bm{\hat{\bm{\mathcal{h}}}} , \phi)$, shall be able to dynamically adapted to it via the combinatorial structure activated by $\mathbf{g}$ and hence to produce the correct surrogate CSIs $\breve{\bm{\mathcal{h}}}$. In Figure~\ref{fig:sensingBVP}, we plot the feature maps generated by the Widar3 HGR model~\cite{Widar3-MobiSys19}, given different CSI inputs: apparently, while the encrypted CSIs lead to very different feature maps from that of raw CSIs, surrogate CSIs may almost perfectly reproduce the feature maps of raw CSIs.

\vspace{-1ex}
\subsection{Putting All Together} \label{ssec:together}
\vspace{-1ex}
According to 
% our system overview in Section~\ref{ssec:sysoverview} and 
the illustration of Figure~\ref{fig:teaser}, a trusted party Alice controls Wi-Fi APs of an indoor space where privacy sensitivity is high. It applies encryption matrix $\bm{\Psi}$ to encrypt the channels (hence CSIs), so that any malicious eavesdroppers (Eves) obtaining the encrypted CSIs $\hat{\bm{\mathcal{h}}}$ cannot leverage them to infer hand gestures of a nearby victim, as proven in Section~\ref{ssec:secana}. In the meantime, the Wi-Fi network should also support legitimate users (Bobs) separated into two groups: Bob$^{\mathrm{C}}$ and Bob$^{\mathrm{S}}$ respectively use communication and sensing services. Upon Bobs' registration with Alice, they become eligible to obtain the information about the encryption matrix $\bm{\Psi}$ (either direct or indirect as $\phi$)
% periodic updates 
from Alice 
% in terms of the encryption matrix 
for decrypting $\hat{\bm{\mathcal{h}}}$. 
Since the MIMO encryption may still affect the performance of Bobs even after decryption, Alice optimizes $\bm{\Psi}$'s (see Section~\ref{ssec:optimization}), leveraging a novel sensing metric defined in Section~\ref{ssec:psnr}, so as to strike a good balance between thwarting Eves and protecting Bobs.
\newrev{
While a key needs to be $\bm{\Psi}$ itself for Bob$^{\mathrm{C}}$, Bob$^{\mathrm{S}}$ may enjoy a substantially improved efficiency with a hashed $\bm{\Psi}$ of only several hundreds of bits, as explained in Section~\ref{sec:ssCSI}.
% for the sake of conciseness of the paper body.
}

\section{\MakeUppercase{Implementation}}\label{sec:imple}
\begin{figure}[b]
	\setlength\abovecaptionskip{8pt}
	\vspace{-1ex}
	\centering
	\subfloat[WARP for Alice and Bob/Eve.]{
		\begin{minipage}[b]{0.47\linewidth}
			\centering
			\includegraphics[width = .92\textwidth]{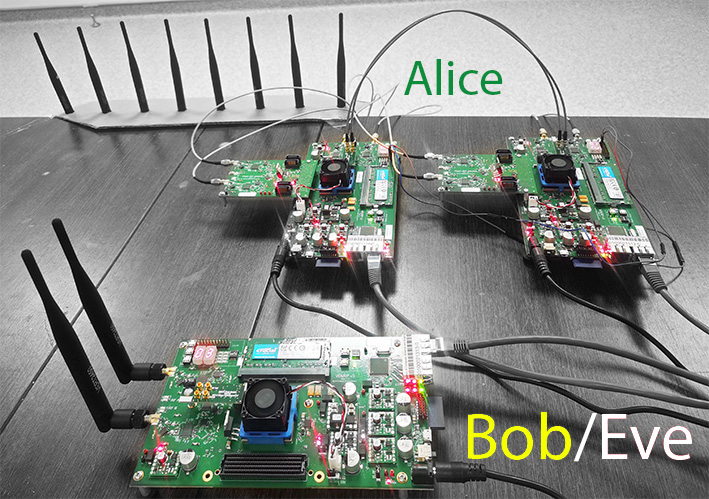}
			\label{sfig:warp}
			\vspace{0ex}
            %\vspace{-3ex}
		\end{minipage}
	}
	\hfill
	\subfloat[Experiment layout.]{
		\begin{minipage}[b]{0.47\linewidth}
			\centering
			\includegraphics[width = .92\textwidth]{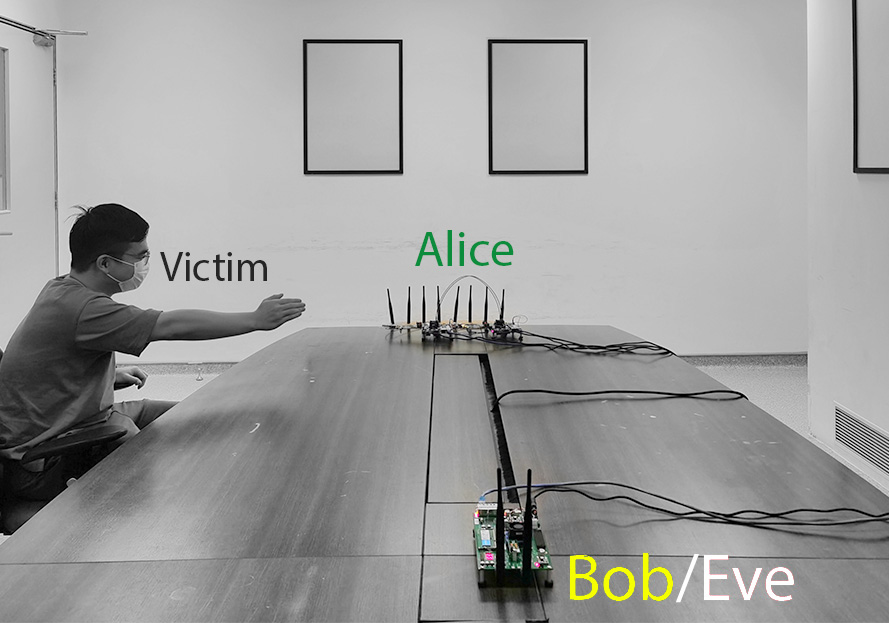}
			\label{sfig:scenario}
			\vspace{0ex}
           %\vspace{-3ex}
		\end{minipage}
	}
	\caption{\sname implementation: (a) WARP configurations and (b) experiment layout in an office area.}
	\label{fig:experiement_setting}
	\vspace{-1ex}
\end{figure}

\emph{Hardware.} 
We utilize the widely-adopted SDR platform WARP v3~\cite{WARP-web} to build our \sname prototype shown in Figure~\ref{sfig:warp}.  
\sname operates at 2.4\!~GHz with 20\!~MHz bandwidth, and AP (Alice) transmits packets
% and its AP (Alice) continuously transmits Wi-Fi packets according to the 802.11ac PHY specification~\cite{802_11}, 
at a default rate of 1,000 packets per second according to Widar3~\cite{Widar3-MobiSys19} and WiSee~\cite{WiSee-MobiCom13}.
Whereas we let Bob and Eve directly take the basic version of WARP node with 2 antennas, Alice is equipped with 8 antennas by leveraging FMC module Mango FMC-RF-2X245~\cite{fmc} to extend the number of antennas in a single WARP node to 4 and then synchronizing two independent nodes via the clock module CM-MMCX~\cite{mmcx}. All WARP nodes are connected to a PC server supported by an Intel Core i7-12700H CPU, 32GB RAM, and a GeForce RTX 3070Ti graphics card via a 1\!~GHz Ethernet switch, where the software of \sname (described in the following) operates. 
 
\emph{Software.} 
%\paragraph{Software Implementations.} 
%
We implement the software framework of \sname in low level C language and Python~3.7 running on the PC server, with the neural models built upon PyTorch 1.11.0. As shown in Figure~\ref{fig:ill_HGRmodel}, we utilize Widar3 and WiSee as the common HGR models; they respectively depend on the body-coordinate velocity profile (BVP) and Doppler frequency shift (DFS) features. For our sensing sub-model $\mathcal{F}$, its gating model $\mathbf{g}$ uses a standard residual block~\cite{He_2016_CVPR} followed by 4 full-connection layers, with $V = 8$ for our current design. Each of the 8 encoders consists of 4 consecutive convolutional layers, with a kernel size 3, stride 1, padding size 1, and their output channels being 32, 64, 64 and 128, respectively. For the decoder, 5 consecutive transposed convolutional layers constructed by ConvTranspose2d~\cite{ConvTranspose2d} are utilized, their kernel size is set to 3, as well as stridden to 1 (but 2 for the fourth layer), and the output channels of these layers are set to 64, 32, 32, 16 and 16, respectively. Finally, the output of each layer is followed by an activation function ReLU~\cite{ReLU-ICML10} and a normalization operator of BatchNorm2d~\cite{BatchNor-ICML15}. 
During the training phase, all weights of $\mathcal{F}$ are initialized by the Xavier uniform initializer~\cite{Xavier}, the random parameter dropout ratio is set to 0.5, and the batch size is 128. Meanwhile, RMSprop optimizer~\cite{RMSprop} is utilized to update parameters, with a learning rate is of 0.001. Upon inference, we deploy the keyed dynamic neural network~(introduced in Section~\ref{sec:ssCSI}) in every Bob, but only the common HGR model on Eve.
% according to the Section~\ref{ssec:attackm}.

\section{\MakeUppercase{Evaluations}} \label{sec:eval}
In this section, we first present the overall performance of \sname in thwarting Eve, protecting legitimate sensing of $\mathrm{Bob^S}$ and communication of $\mathrm{Bob^C}$. Then the impact of various parameter settings on our prototype is evaluated.

\subsection{Experiment Setup} \label{ssec:expsetup}
%
% To obtain CSI samples constructing our model, 
We recruit 6 volunteers (3 females and 3 males) as victims for participating in data collection. We first ask them to watch the example video for standardizing their gesture-performing behaviors. Then each victim performs eight common gestures (i.e., push-pull, clap, slide, tap, pinch-spread, drawing a circle, square, and zigzag) in a large meeting room with an area of about 32~\!m$^2$ as shown in Figure~\ref{sfig:scenario}. To ease the exposition, these gestures are abbreviated as PP, CL, SL, TA, PS, DC, DS, and DZ respectively. \newrev{Given the passive nature of Wi-Fi sensing (see Section~\ref{ssec:attackm}), we may leverage only one WARP node to emulate a multi-user scenario with several Bobs and Eves, by separately conducting either 
% communications with Alice or 
legitimate sensing or eavesdropping on the hand gestures of one target/victim. Of course, this does not emulate multi-Eve collusion cases, but such cases have been either forbidden or proven ineffective in Sections~\ref{ssec:attackm} and~\ref{ssec:secana}.}
We collect two datasets spanning totally 42.6~\!hours, corresponding to without/with enabling \sname's MIMO encryption: \textsf{Data-1} contains 25920 samples (8 gestures $\times$ 540 instances $\times$ 6 objects) while \textsf{Data-2} owns 8640 samples (8 gestures $\times$ 180 instances $\times$ 6 objects). The data collection process has strictly followed the standard procedures required by our IRB.

For adapting Widar3 and WiSee to our environment while optimizing parameters of $\mathcal{F}$, we leverage two-thirds of samples in \textsf{Data-1} to retrain them and the remaining part for testing gesture recognition result. Moreover, \sname employs two common metrics, namely accuracy and bit error ratio (BER), to quantify sensing and communication performance, where $\mathrm{accuracy} = \frac{N^{\mathrm{cor}}}{N^{\mathrm{all}}}$  measures the ratio of correctly recognized gestures, with $N^{\mathrm{cor}}$ and $N^{\mathrm{all}}$ respectively denoting the number of correctly recognized and of all gestures, and $\mathrm{BER} = \frac{N^{\mathrm{err}}}{N^{\mathrm{all}}}$ indicates the ratio of incorrectly decoded bits, with $N^{\mathrm{err}}$ and $N^{\mathrm{all}}$ being the number of error bits and of totally transmitted bits, respectively.

\subsection{\sname Performance} \label{ssec:performance}

\subsubsection{Thwarting Eve}\label{sssec:scramblingPer}

\begin{figure}[t]
	\setlength\abovecaptionskip{8pt}
	\vspace{-1.5ex}
	\centering
	\subfloat[LoS and Widar3.]{
		\begin{minipage}[b]{0.49\linewidth}
			\centering
			\includegraphics[width = \textwidth]{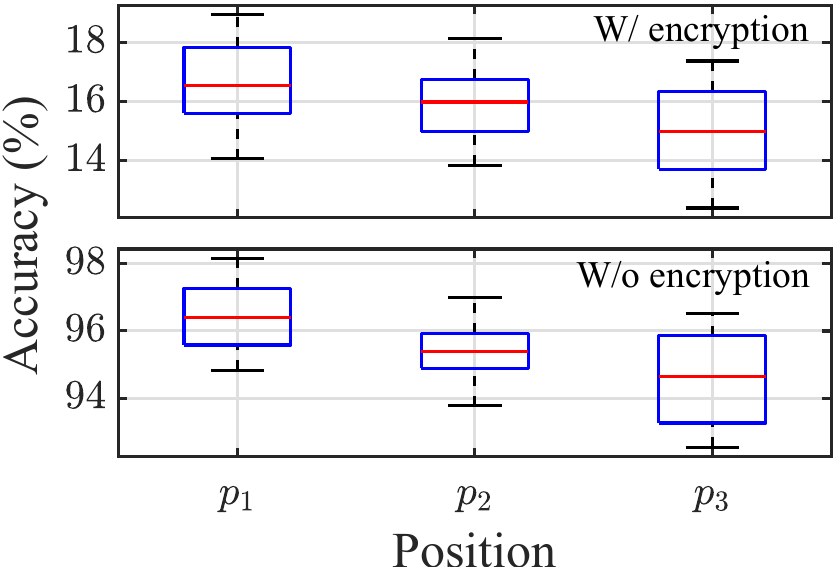}
			\label{sfig:acc_widar_los}
			\vspace{-2.4ex}
		\end{minipage}
	}
	\subfloat[NLoS and Widar3.]{
		\begin{minipage}[b]{0.49\linewidth}
			\centering
			\includegraphics[width = \textwidth]{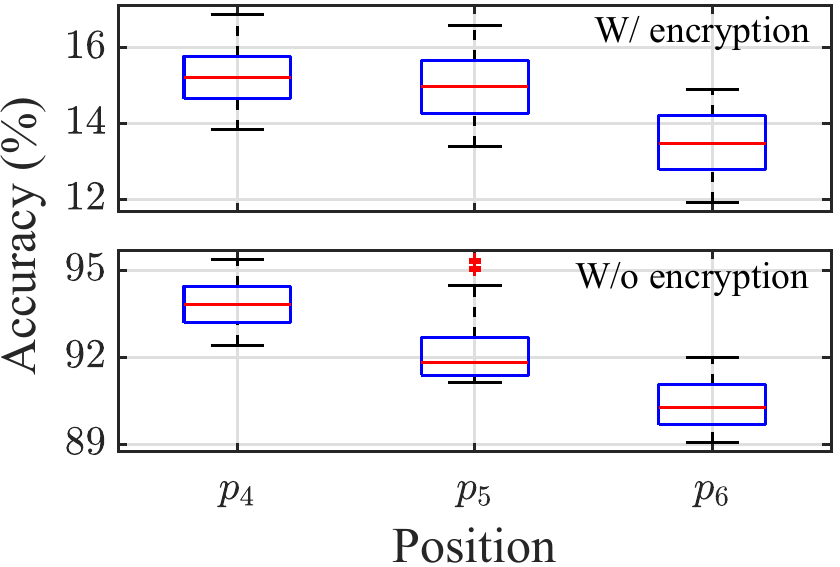}
			\label{sfig:acc_widar_nlos}
			\vspace{-2.4ex}
		\end{minipage}
	}
	\\ \vspace{-0.1ex}
	\subfloat[LoS and WiSee.]{
		\begin{minipage}[b]{0.49\linewidth}
			\centering
			\includegraphics[width = \textwidth]{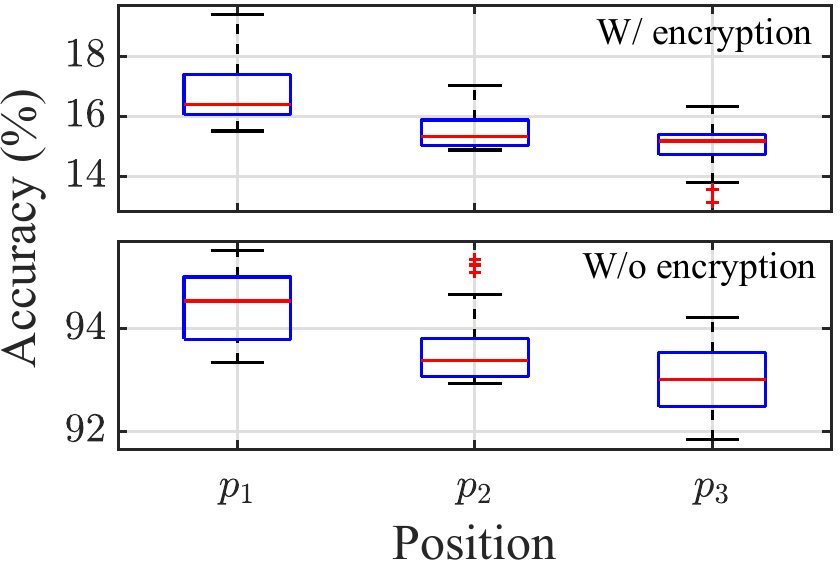}
			\label{sfig:acc_wisee_los}
			\vspace{-2.4ex}
		\end{minipage}
	}
	\subfloat[NLoS and WiSee.]{
		\begin{minipage}[b]{0.49\linewidth}
			\centering
			\includegraphics[width = \textwidth]{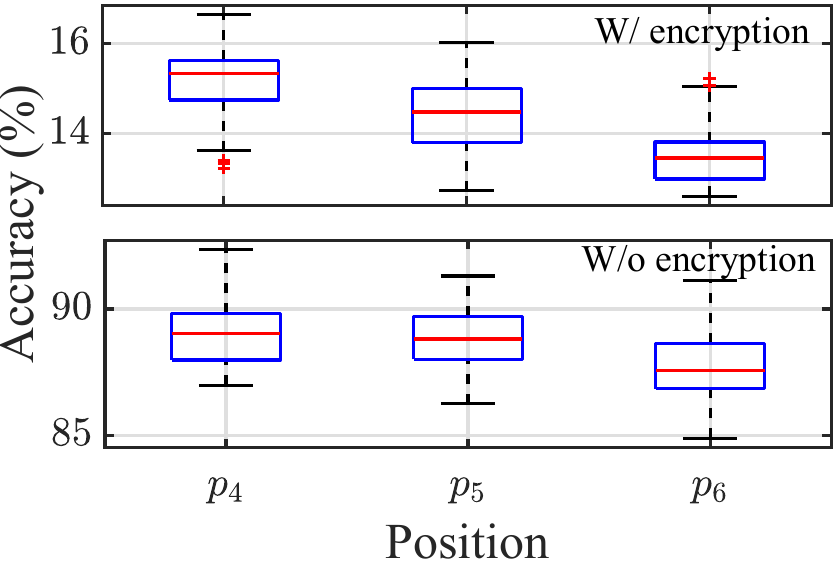}
			\label{sfig:acc_wisee_nlos}
			\vspace{-2.4ex}
		\end{minipage}
	}
	\caption{Eve's eavesdropping performance evaluated by Widar3 under (a) LoS ad (b) NLoS, when turning on/off spatial-temporal encryption; the results under similar settings for WiSee are further shown in (c) and (d).}
	\label{fig:overall_overall_scrambling}
    \vspace{-1.5ex}
\end{figure}

To evaluate the performance of \sname on thwarting malicious eavesdropping via \rev{\textit{spatial-temporal encryption}}, we place Eve at six positions: the first three (i.e., $p_1$, $p_2$, and $p_3$) having LoS path with Alice, at a distance of 1.5~\!m, 3~\!m and 4.5~\!m respectively; the others ($p_4$, $p_5$, and $p_6$) are co-located with previous ones but changed to NLoS type by adding obstacles in-between. The victim performs different hand gestures at 0.5~\!m from Alice. As shown in Figure~\ref{sfig:acc_widar_los}, with encryption, Eve's eavesdropping (HGR) accuracy on average is less than 16\% in three LoS positions: compared with eavesdropping without encryption, \sname has successfully reduced Eve's HGR accuracy by about 80\%.
\begin{figure}[b]
	\setlength\abovecaptionskip{8pt}
	\vspace{-2ex}
	\centering
	\subfloat[Widar3.]{
		\begin{minipage}[b]{0.49\linewidth}
			\centering
			\includegraphics[width = \textwidth]{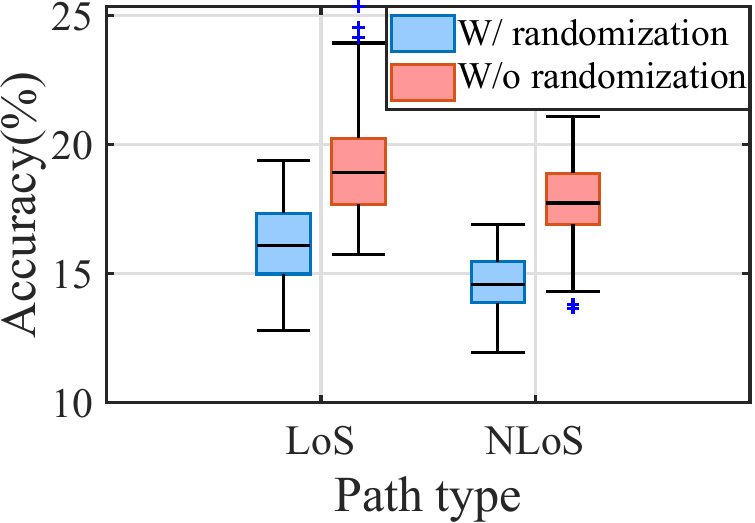}
			\label{sfig:temporals1}
			\vspace{-2.5ex}
		\end{minipage}
	}
	\subfloat[WiSee.]{
		\begin{minipage}[b]{0.49\linewidth}
			\centering
			\includegraphics[width = \textwidth]{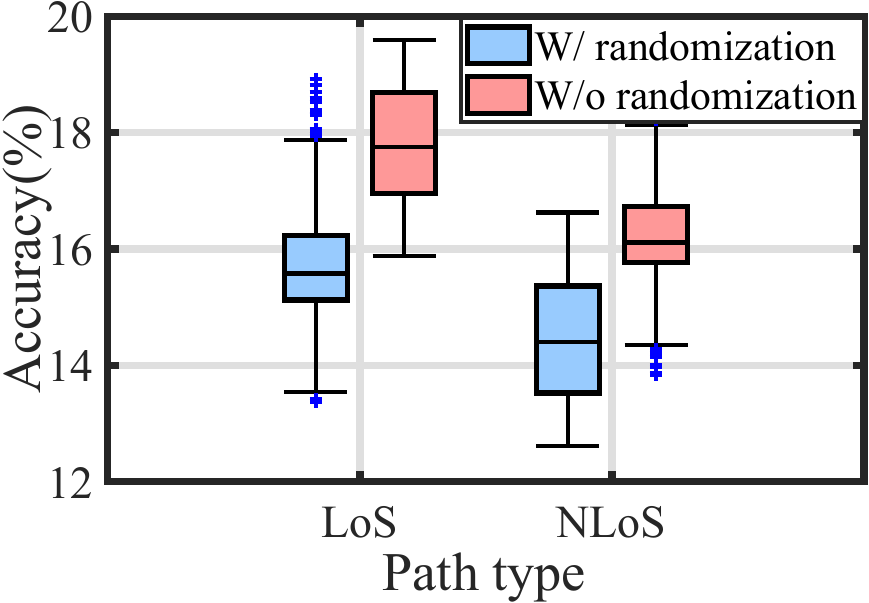}
			\label{sfig:temporals2}
			\vspace{-2.5ex}
		\end{minipage}
	}
	\caption{Eve's gesture recognition accuracy when using (a) Widar3 and (b) WiSee respectively, with/without enabling the temporal randomization.}
	\label{fig:temporalper}
	\vspace{-.5ex}
\end{figure}
A similar trend for the NLoS setting is further demonstrated by Figure~\ref{sfig:acc_widar_nlos}, while both results also confirm that Eve's eavesdropping performance can also be worsen by either a longer distance or certain obstacles, consistent with our study in Section~\ref{ssec:psnr}. Finally, we leverage another well-known HGR model WiSee (also adopted by PhyCloak~\cite{PhyCloak-NSDI16}) to reconfirm the excellent performance of \sname for thwarting Eve in Figure~\ref{sfig:acc_wisee_los} and~\ref{sfig:acc_wisee_nlos}. \newrev{The results clearly indicate that the encryption matrix $\bm{\Psi}$, endowed with diversified choices from full space (as described in Section~\ref{ssec:secana}), can effectively encrypt a wide range of channels to thwart Eve's attacks. }

We also evaluate \sname's temporal randomization performance against Eve; the results are reported in Figure~\ref{fig:temporalper}. Essentially, temporal randomization further reduces Eve’s eavesdropping accuracy by 4.87\% for Widar3 and 6.01\% for WiSee; it also lessens the accuracy difference between LoS and NLoS to only 0.32\%. The effectiveness of temporal randomization can be attributed to the common HGR model often pre-trained based on fixed and regular input (CSI) sample intervals. 

%\vspace{-1ex}
\subsubsection{Protecting Bob$^{\mathrm{S}}$}\label{sssec:legalsensing}
Apart from effectively thwarting Eve, \sname also needs to retain the sensing ability of Bob$\mathrm{^S}$. Therefore, we hereby verify this performance aspect of \sname by comparing the Bob$^{\mathrm{S}}$'s HGR accuracy at the same six positions taken by Eve earlier to ensure a fair comparison. The common HGR models adopted are again Widar3 and WiSee, but we consider two settings: i) non-encrypted CSIs fed to these models and ii) encrypted CSIs fed to $\mathcal{F}$ before 
%our ``one-fits-all'' HGR pipeline with 
using these models as classifiers.
\begin{figure}[b]
	\setlength\abovecaptionskip{8pt}
	\vspace{-1.5ex}
	\centering
	\subfloat[HGR accuracy delivered by Widar3.]{
		\begin{minipage}[b]{0.9\linewidth}
			\centering
			\includegraphics[width = \textwidth]{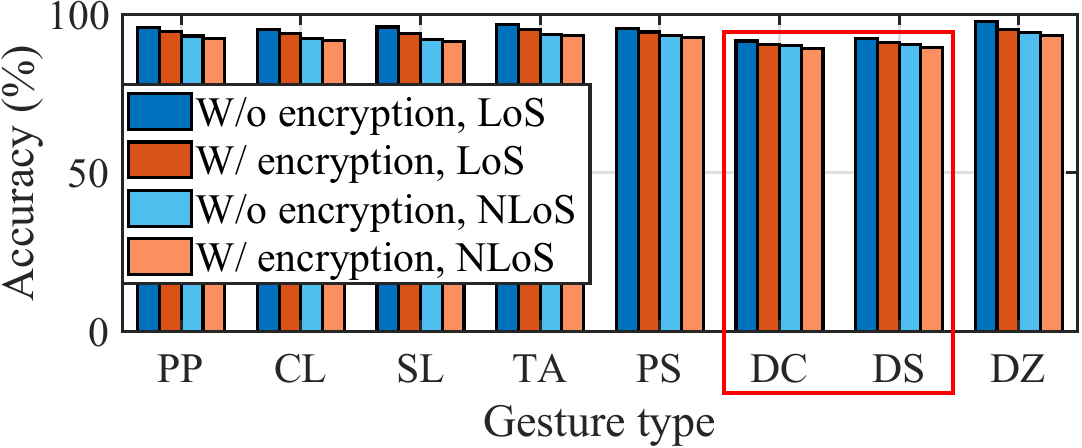}
			\label{sfig:acc_widar_sensing}
			\vspace{-3ex}
		\end{minipage}
	}
	\\
	\subfloat[HGR accuracy delivered by WiSee.]{
		\begin{minipage}[b]{0.9\linewidth}
			\centering
			\includegraphics[width = \textwidth]{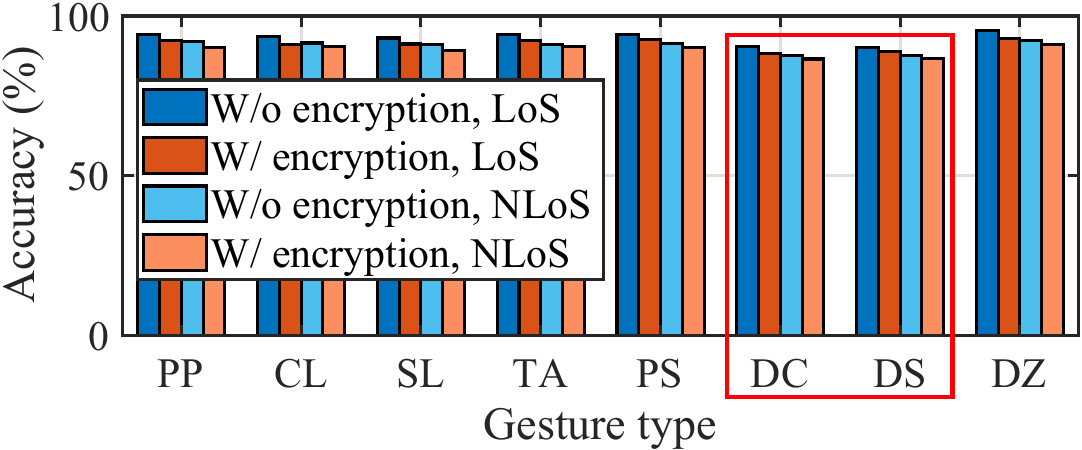}
			\label{sfig:acc_wisee_sensing}
			\vspace{-3ex}
		\end{minipage}
	}
	\caption{HGR accuracy of Bob$\mathrm{^S}$ given i) non-encrypted CSIs and ii) encrypted ones filtered by $\mathcal{F}$ before Widar3 (a) and WiSee (b), under both LoS and NLoS settings.}
	\label{fig:overall_sensing}
	\vspace{-.5ex}
\end{figure}

Figures~\ref{sfig:acc_widar_sensing} and~\ref{sfig:acc_wisee_sensing} present the average accuracy of HGR by Widar3 and WiSee, respectively. 
Apparently, the HGR accuracy for Bob$^{\mathrm{S}}$ only drops slightly when the encryption function is turned on, proving \sname's efficacy on this performance aspect. Among all gestures, drawing circle (DC) and square (DS) have relatively lower recognition accuracy since differentiating them is indeed a challenge. Also, the average accuracy 90.62\% of WiSee is lower than Widar3's 92.74\%, suggesting that the BVP feature employed by Widar3 may be more robust to environment interference than the DFS of WiSee. Therefore, extracting environment-robust (cross-domain) features to construct HGR models can make it more practical.

\subsubsection{Protecting Bob$^{\mathrm{C}}$}\label{sssec:legalcommunication}
\begin{figure}[t]
	\setlength\abovecaptionskip{8pt}
	\vspace{-2ex}
	\centering
	\subfloat[QAM-32 modulation.]{
	        \begin{minipage}[b]{0.49\linewidth}
			\centering
			\includegraphics[width = \textwidth]{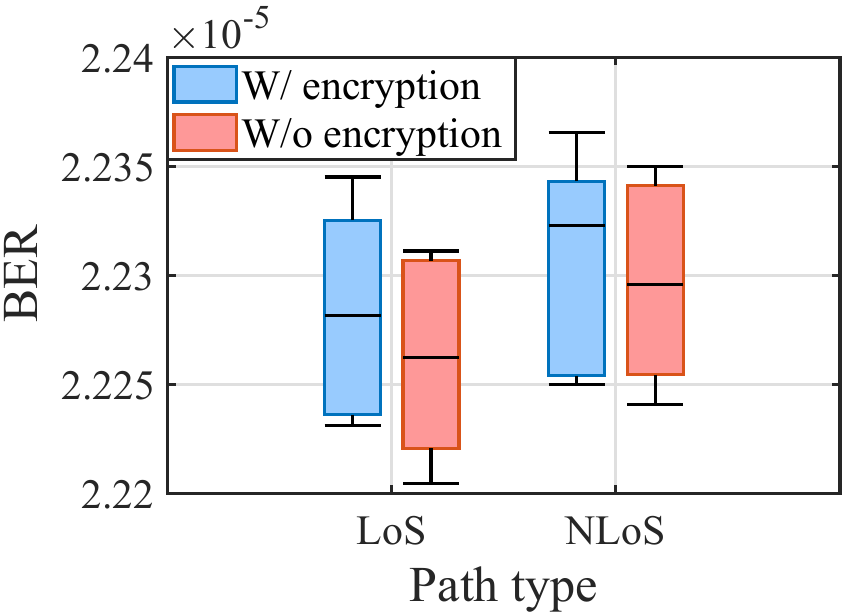}
			\label{sfig:com_16}
			\vspace{-2.5ex}
		\end{minipage}
	}
	\subfloat[QAM-64 modulation.]{
		\begin{minipage}[b]{0.49\linewidth}
			\centering
			\includegraphics[width = \textwidth]{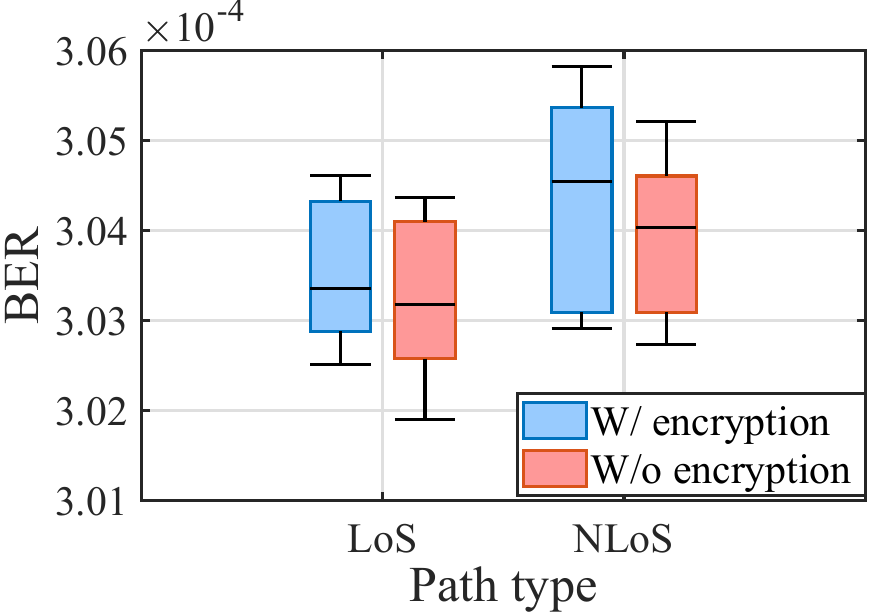}
			\label{sfig:com_nlos}
			\vspace{-2.5ex}
		\end{minipage}
	}
	\caption{BERs of payload data modulated by QAM-32 (a) and QAM-64 (b), respectively, under the two cases of enabling and disabling \sname\ encryption.}
	\label{fig:overall_communication}
	\vspace{-.5ex}
\end{figure}

In this section, we evaluate the communication performance of \sname in protecting Bob$^{\mathrm{C}}$, using BER of payload as the metric. We randomly choose 28 positions for Bob$^{\mathrm{C}}$ to measure its BER. At each position, we let Alice communicate with  Bob$^{\mathrm{C}}$ for 30 minutes, using two common modulations~(QAM-32 and QAM-64) for payload, and Bob$^{\mathrm{S}}$ also performs HGR from time to time.\footnote{These two users share the same device in our experiments, yet they can be either separated or co-located in reality.} The boxplots in Figure~\ref{fig:overall_communication} demonstrate that \sname only affects the communication performance to a negligible extent: the median BER increases only in the digit of $10^{-6}$, and the BER variances are also increased at the same order, regardless of whether it is LoS or NLoS setting and what modulation scheme is adopted. 
\sname achieves this excellent performance (for both Bob$^{\mathrm{C}}$ and Bob$^{\mathrm{S}}$) thanks to our multi-objective optimization framework in Section~\ref{ssec:optimization} that successfully reaches the best tradeoff among conflicting objectives; this is in stark contrast to existing proposals that simply evaluate special cases to ``prove'' their channel scrambling barely affecting communications~\cite{PhyCloak-NSDI16,IRShield-SP22}.

\subsection{Impact of Parameter Settings}
We hereby study the impact of several system parameters on \sname.

%\subsubsection{Obfuscating Matrix Optimization}
\subsubsection{Encryption Matrix Optimization}
\label{sssec:matrix_opti}
\begin{figure}[t]
	\setlength\abovecaptionskip{8pt}
	\vspace{-1.5ex}
	\centering
	\subfloat[Eve.]{
	        \begin{minipage}[b]{0.32\linewidth}
			\centering
			\includegraphics[width = \textwidth]{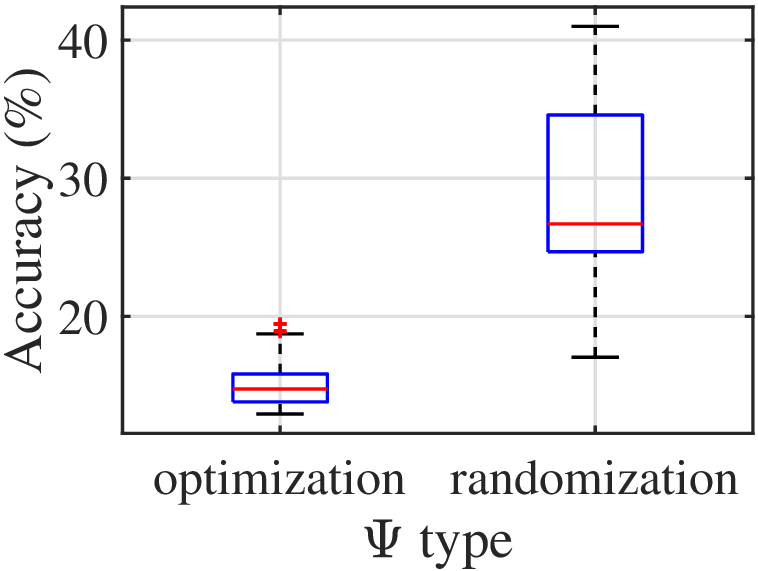}
			\label{sfig:permat_scra}
			\vspace{-2.5ex}
		\end{minipage}
	}
	\subfloat[Bob$^{\mathrm{S}}$.]{
		\begin{minipage}[b]{0.32\linewidth}
			\centering
			\includegraphics[width = \textwidth]{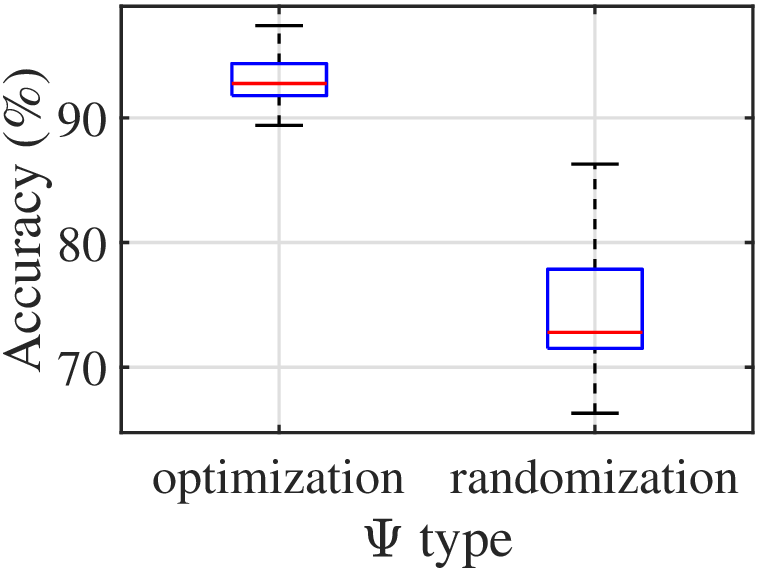}
			\label{sfig:permat_sensing}
			\vspace{-2.5ex}
		\end{minipage}
	}
	\subfloat[Bob$^{\mathrm{C}}$.]{
		\begin{minipage}[b]{0.32\linewidth}
			\centering
			\includegraphics[width = \textwidth]{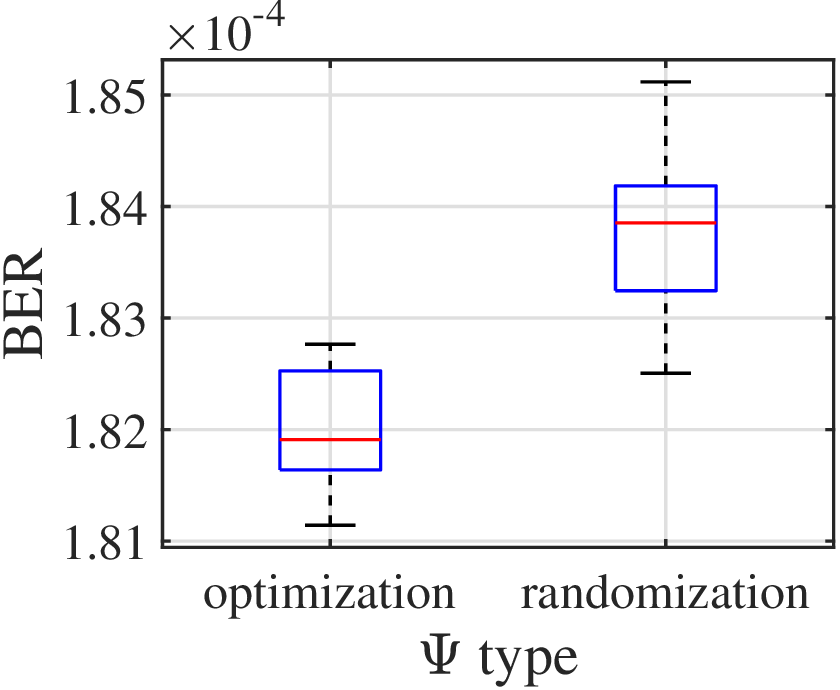}
			\label{sfig:permat_comm}
			\vspace{-2.5ex}
		\end{minipage}
	}
	\caption{{\sname's performance for (a) eavesdropping, (b) legitimate sensing, and (c) data communication under optimized or randomized encryption matrices.}}
    %obfuscating matrices.}}
	\label{fig:per_optMat}
	\vspace{-1.5ex}
\end{figure}

According to Section~\ref{ssec:optimization}, we leverage a multi-objective optimization framework to identify the optimal $\bm{\Psi}$'s realizing the best tradeoffs among the performances of Eve, Bob$^{\mathrm{S}}$ and Bob$^{\mathrm{C}}$. We randomly generate five hundred $\bm{\Psi}$'s, and use them as the initial solution. We further obtain the same number of optimized $\bm{\Psi}$'s. The results in Figure~\ref{fig:per_optMat} clearly demonstrate that, compared with random $\bm{\Psi}$'s, the optimized ones have much better performances in terms of worsening the sensing accuracy of Eve, improving that of Bob$^{\mathrm{S}}$, and  reducing the BER for Bob$^{\mathrm{C}}$. This firmly indicates that \sname can successfully identify the best tradeoffs on the Pareto optimal frontier of the multi-objective optimization.  

\subsubsection{Sub-model for Bob$^{\mathrm{S}}$'s Sensing}\label{sssec:submodel}
\begin{figure}[b]
	\setlength\abovecaptionskip{8pt}
	\vspace{-2ex}
	\centering
	\subfloat[CSI correlation.]{
	        \begin{minipage}[b]{0.45\linewidth}
			\centering
			\includegraphics[width = \textwidth]{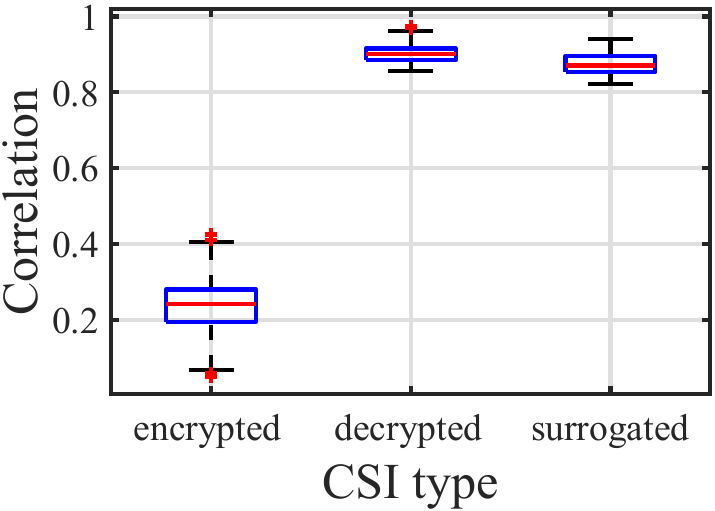}
			\label{sfig:submodel}
			\vspace{-2.5ex}
		\end{minipage}
	}
	\subfloat[Encoder quantity $V$.]{
		\begin{minipage}[b]{0.45\linewidth}
			\centering
			\includegraphics[width = \textwidth]{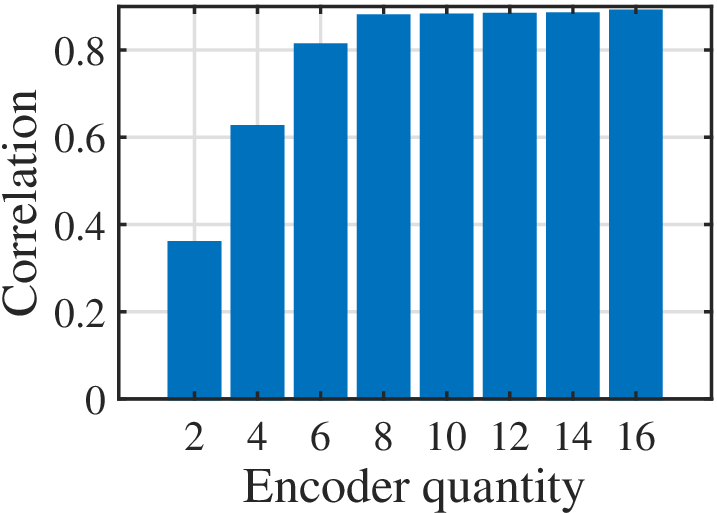}
			\label{sfig:decoderquantity}
			\vspace{-2.5ex}
		\end{minipage}
	}
	\\
        \vspace{-1ex}
	\subfloat[Training data size.]{
		\begin{minipage}[b]{0.8\linewidth}
			\centering
			\includegraphics[width = \textwidth]{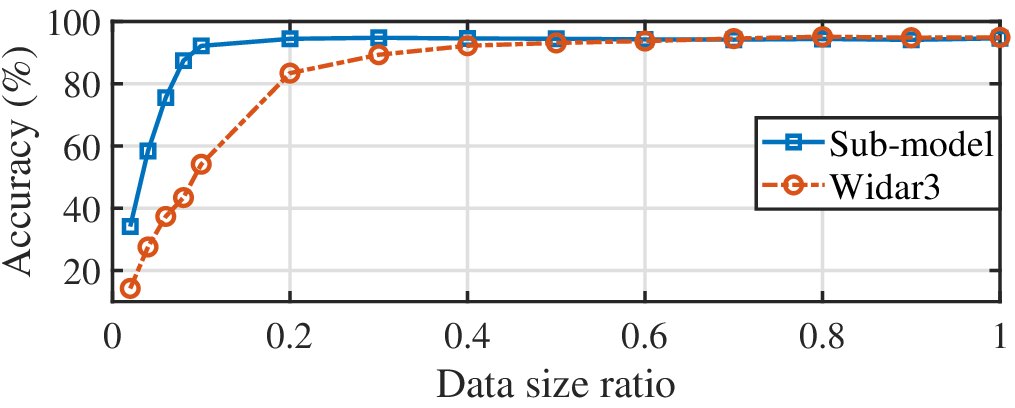}
			\label{sfig:datasize}
			\vspace{-2.5ex}
		\end{minipage}
	}
	\caption{The power of surrogating: (a) correlations between three types of feature maps (encrypted, decrypted and surrogated and the
     %type feature maps (scrambled, descrambled and surrogated) and 
     the raw ones, (b) the impact of encoder quantity $V$, and (c) the impact of training data size on sub-model performance.}
	\label{fig:submodel_per}
	\vspace{-.5ex}
\end{figure}

To verify the effectiveness of our design in Section~\ref{sec:ssCSI}, we calculate the cosine similarity of the feature maps respectively induced by encrypted, descrypted, and surrogate CSIs with those induced by raw CSIs.  For a fairness comparison, we use the same Data-1~(introduced in Section~\ref{ssec:expsetup}) to obtain the results. As shown in Figure~\ref{sfig:submodel}, the average correlation between surrogate and raw feature maps is around 90\%, only 1.5\% lower than that of the descrypted ones, clearly explaining why our sub-model can still perform accurate HGR shown in Figure~\ref{fig:overall_sensing}. 

We also study how the parameter $V$ of the encoders quantity affects the sub-model. We adjust $V$ from 2 to 16, and then plot the corresponding feature map correlations in Figure~\ref{sfig:decoderquantity}. Apparently, the performance saturates after $V = 8$, so we set the default $V$ as 8 in our prototype. These results also corroborate our statement in Section~\ref{sec:ssCSI} that, instead of a naive solution using a large number of static models, our sub-model only needs a small value of $V$, even for a large quantity of $\bm{\Psi}$'s. Moreover, training data size determines whether our sub-model can fully understand feature structures. To study its impact, we train the sub-model with increasing fraction of data, and compare the resulting accuracy with that of the Widar3 HGR model (during pre-training phase) in Figure~\ref{sfig:datasize}. Apparently, training our sub-model converges much faster (at least five times) than pre-training Widar3.

\subsubsection{Antenna Quantity}\label{sssec:antqua}
\begin{figure}[t]
	\setlength\abovecaptionskip{8pt}
	\vspace{-1.5ex}
	\centering
	\subfloat[Antennas of Alice.]{
	        \begin{minipage}[b]{0.47\linewidth}
			\centering
			\includegraphics[width = \textwidth]{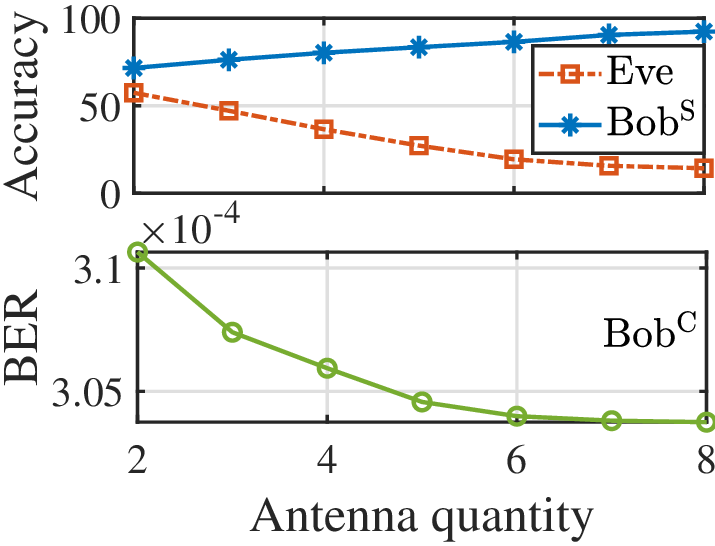}
			\label{sfig:antennaTX}
			\vspace{-2.5ex}
		\end{minipage}
	}
	\subfloat[Antennas of Eve.]{
		\begin{minipage}[b]{0.47\linewidth}
			\centering
			\includegraphics[width = \textwidth]{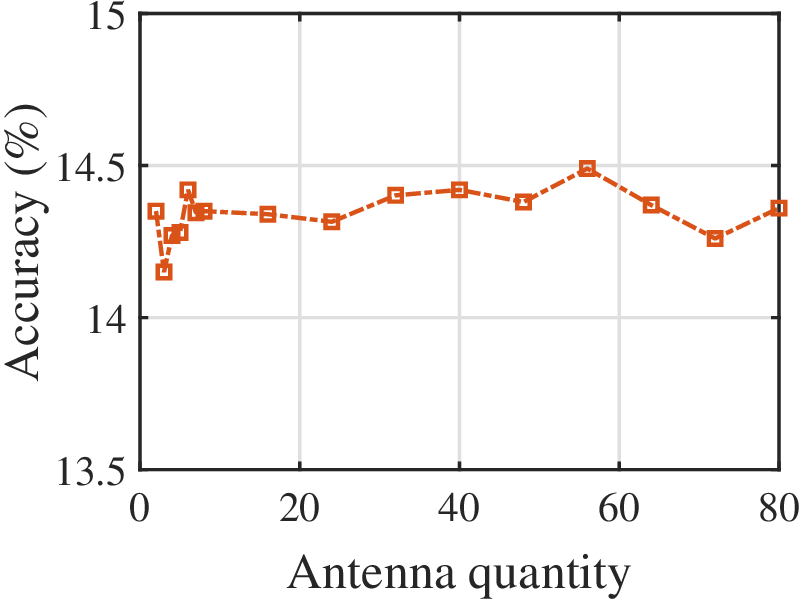}
			\label{sfig:antennaRX}
			\vspace{-2.5ex}
		\end{minipage}
	}
	\caption{Antenna matters? (a) \sname's improved performance when increasing antenna quantity for Alice, and (b) increasing that quantity for Eve fails to work.}
	\label{fig:antenna}
	\vspace{-1ex}
\end{figure}

For Alice, a large number of antennas allows it to improve both the diversity in $\bm{\Psi}$ (thwarting Eve) and the signal power distribution (protecting Bobs). Figure~\ref{sfig:antennaTX} presents \sname's improved performance (on both sides) when equipping Alice with an increasing number of antennas. Moreover, though we have argued in Section~\ref{ssec:attackm} that Eves often cannot have too many physical antennas, \newrev{they may leverage virtual antennas (by changing positions at different times)~\cite{Virtual_Ant1} to gain higher diversity.} Fortunately, the results in Figure~\ref{sfig:antennaRX} evidently confirm that, with up to 80 virtual antennas, Eve has no chance to crack $\bm{\Psi}$'s and thus to improve its HGR accuracy, \newrev{even with higher diversity in both time and space, as proven in Section~\ref{ssec:secana}.}

\subsubsection{Cost Reduction in Legitimate Sensing}\label{sssec:timecost}
For Bob$^{\mathrm{S}}$, \sname achieves cost reduction in two aspects: i) it cuts the transmission volume by employing a much shorter decryption key to replace $\bm{\Psi}$ thanks to the lower signal granularity demanded by gesture sensing, and ii) it substantially lessens the computation time cost by directly leveraging encrypted CSIs for gesture recognition, avoiding additional $\bm{\Psi}$-related matrix operations. We hereby conduct a simple case study based on our experiment parameters to concretely demonstrate these two aspects.

\newrev{On one hand, given the settings of Alice's 8 antenna transmitting 1,500 packets (for a 1.5-second gesture) to Bob$^{\mathrm{S}}$, the encryption matrix $\bm{\Psi}$ contains $8 \times 1500$ complex numbers (roughly 4 bytes each) and the total data volume is 48,000 bytes, since $\bm{\Psi}$ covers spatial-temporal dimensions as
% is a diagonal matrix 
introduced in Section~\ref{ssec:scrambling}. However, \sname only relies on a decryption key $\phi$ to activate specific encoders for processing the encrypted CSIs, 
%Given the settings of 8 Tx antennas for Alice, the \rev{encryption} matrix $\bm{\Psi}$ contains $8\times64\times 64$ complex numbers (roughly 4 bytes each), and the total data volume is 131,072 bytes. However, \sname only relies on a \rev{decryption} key $\phi$ to activate specific encoders for processing the \rev{encrypted} CSIs,
%
% rather than for decoding data bits one by one in Bob$^{\mathrm{C}}$. Therefore, we can hash obfuscating matrix $\bm{\Psi}$ to 
%
so the transmission volume is reduced to the 2048-bit decryption key, less than 0.54\%~of the original volume. Although either $\bm{\Psi}$ or $\phi$ needs only to be updated periodically (e.g., every several weeks or even months) as explained in Sections~\ref{ssec:secana}, \ref{ssec:optimization}, and~\ref{sec:ssCSI}, such a reduction is still significant.
% transmit one time but can handle scrambled CSIs in one longish key-update period of 12 hours. 

%Moreover, 
On the other hand, each $\bm{\Psi}$-related decryption (matrix) operation 
needs an extra 0.83~\!ms as indicated in Section~\ref{ssec:effsense}, so the extra time cost for a 1.5-second gesture is reduced from $1500 \times 0.83 = 1245$~\!ms (given the packet rate of 1,000 packets per second) to zero;} here only the extra time cost matters because the average inference times of the normal HGR models with and without the sub-model $\mathcal{F}$ are very close to each other.
% of one gesture for $\bm{\Psi}$-dependent way and our HGR pipeline are very close, that is 72~\!ms and 85~\!ms respectively. 
To sum up, our plug-and-play sub-model can provide evident cost reduction in both transmission volume and computation time.

\section{\MakeUppercase{Limitations and Discussions}} \label{sec:limfur}
%

%\vspace{-.5ex}
\noindent\emph{New User Registration.} 
We have so far implicitly assumed that Bobs have registered to Alice before \sname activating its encryption, but what about newly joining Bobs? After activating MIMO encryption, unregistered Bobs are treated as the Eves, and hence face difficulty in further registration (albeit still possible as a Bob enjoy non-zero communication throughput under encryption). To address this issue under Wi-Fi standard, the Alice may sporadically broadcast Wi-Fi's beacon frames without encryption, hence providing opportunity for newly arriving Bobs to register. Since the beacon frames are relatively short~\cite{Beacon} and 
% the temporal randomization can also thwart Eve but not affect data decoding of unregistered Bobs 
\newrev{Eves are not interested in registration for the sake of its stealthiness (see Section~\ref{ssec:attackm}),} accommodating newly joining Bobs should be readily solvable for \sname.

%\vspace{.5ex}
% \noindent\emph{Eve's Antenna Quantity.} One may argue that confining the antenna number of Eves and Bobs to 2 is a limitation for \sname. On one hand, we have explained in Sections~\ref{ssec:attackm} and~\ref{ssec:scrambling} that this is the inevitable consequence of Eve's stealthiness and hence adopting mobile devices. On the other hand, we have proven that increasing antennas quantity does not bring any eavesdropping gain for Eves in Sections~\ref{ssec:scrambling} and~\ref{sssec:antqua}.  For Bobs with more than 2 antennas~(i.e. 4 or 8), we just extend CSI tensor from $[{\bm{H}}_q] = [\mathcal{h}_{n,m,q}]$ in Eqn.~\eqref{eq:sdnrBob} to $[{\bm{H}}_q] = [\mathcal{h}_{n,m,r,q}]$ (where $r$ represents the index of Bob's antenna), and still perform the multi-objective optimization presented in Section~\ref{ssec:optimization} to obtain the optimal $\bm{\Psi}$. Essentially, increasing the antennas of Bob endows a higher degree of freedom to its CSI tensor, hence potentially allowing it to gain more opportunities in improving its own performance without affecting others.

\vspace{.5ex}
\newrev{
\noindent\emph{Analysis on Physical Encryption.}
As discussed in Section~\ref{ssec:secana}, \name excels in applying digitally controlled MIMO beamformers to physically encrypt physical plaintext (e.g., hand gestures), as opposed to existing (seemingly similar) techniques of \textit{orthogonal blinding}~\cite{Strobe, KPA-NDSS14, Robin} that only physically obfuscate digital plaintext. The resulting benefit is twofold: full-scale KPA is too obvious due to its physical presence and hence forbidden on one hand (see Section~\ref{ssec:attackm}), while on the other hand, side-channel KPA barely affects the encryption strength as the conditional entropy of physical ciphertext is evaluated in statistical sense (instead of the combinatorial sense for digital ciphertext), as the entropy is now closely related to SNR~\cite{SIG-018,entroy-snr}. Applying physical encryption based on complex multiplication (as opposed to the permutation used for digital encryption) also allows for further optimization on the choice of keys, which is a concept absent from conventional cryptography. This makes us believe that we may extend such encryption schemes to secure digital plaintext too. Nonetheless, it is unclear how to perform formal security analysis on physical encryption: to the best of our knowledge, convention attacks to encryption schemes, such as brute-force attack, known-plaintext attack, and differential cryptanalysis, may not be readily migrated to this new case. Consequently, we would leave this problem open for future exploration.
}

%\vspace{-.5ex}
\section{\MakeUppercase{Conclusion}} \label{sec:conclusion}
%\vspace{-1ex}
%
In this paper, we have designed and implemented \sname as a MIMO-driven privacy-preserving system for supporting multi-user Wi-Fi sensing and communication applications. In particular, \newrev{we have proposed a source-defined channel encryption to thwart stealthy Wi-Fi eavesdropping attack
% and protect legitimate sensing/communication, 
and seriously analyzed its security. To strike an adequate balance between thwarting attacks and preserving legitimate services}, we have explored a multi-objective optimization framework to guide the channel encryption design. Finally, we have innovated in a keyed dynamic neural network to produce surrogate CSIs for legitimate sensing users and hence to reduce the key distribution load for them. Our extensive experiments with a WARP-based \sname prototype have evidently demonstrated the efficacy of \sname in securing sensing and protecting entitled abilities of legitimate users.

\section*{Acknowledgement}
We are grateful to anonymous reviewers for their constructive suggestions. 
This research is support by National Research Foundation, Singapore and Infocomm Media Development Authority under its Future Communications Research \& Development Programme grant FCP-NTU-RG-2022-015, as well as MoE Tier 1 grant RG16/22. Zhe Chen is the corresponding author.

\balance
\bibliographystyle{IEEEtran} 
\bibliography{reference}

\newpage
\appendices

\section{Extended Security Analysis} \label{apx:entropy}
\newrev{
Given limited space for the main texts, we spare some detailed analysis to this appendix, focusing on issues not fully discussed in Sections~\ref{ssec:secana} and~\ref{sec:limfur}, including whether certain side-channel may substantially reduce the entropy of the encryption to the extend of breaking it. 

First of all, as \name applies a formal cryptographic encryption with a sequence of scrambling patterns and the sequence length as the key length, it is fundamentally different from the orthogonal blinding schemes appeared in~\cite{Strobe, Robin} in mainly two aspects. On one hand, as legitimate users (Bobs) can obtain a key for CSI decryption, the encryption keys can be chosen within a full space, rather than forced to stay in orthogonal to Bobs' channel for blinding Eves. On the other hand, as \name's encryption key often has a length of around or even over 1,000 scrambling patterns (as shown in Section~\ref{sssec:timecost}) that get switched from one to another in millisecond scale, the numerical analysis of how entropy decays in time~\cite{Robin} does not apply anymore. In short, \name's physical encryption is much stronger than existing orthogonal blinding.

Second, one may wonder if a period of no activity may leak the information about \name's encryption key. In fact, such a period can be deemed as the static background already counted in Eqn.~\eqref{eq:naMIMO}: as it is statistically independent of gestures, it does not matter if the CSIs are encrypted or not, and the related term in the denominator of Eqn.~\eqref{eq:naMIMO} simply takes the unencrypted form. Therefore, eavesdropping no-activity period does not concern the strength of \name's protection at all.
Moreover, a no-activity period bears no function of ``pilot signal'' that leverages known
information to probe unknown one (e.g., known LTS to probe the channel). As \name encrypts the
channel, both the encryption key and the channel are unknown. As we have discussed in Sections~\ref{ssec:secana} and \ref{sssec:antqua},
this situation makes adding observation diversity (either temporal by observing longer or
spatial by adding antennas) totally fruitless. Note that channel conditions may still vary and become independent beyond coherence time~\cite{tse2005fundamentals}, so no activity does not suggest static CSIs.

Third, there exists a major difference in the interpretation of \textit{entropy} (commonly used for security analysis) between digital and physical plaintexts: whereas the former can be computed in absolute sense given the discrete distribution of digial symbols (e.g., entropy may become zero if the encryption key is exposed), the latter is characterized in relative sense against the noise~\cite{SIG-018} and hence get related to SNR, which may never become zero but can be maximized if the signal is scrambled to be under the noise floor. Consequently, unlike digital encryption (upon digital plaintext) for which exposing a few pairs of unencrypted and encrypted symbols can substantially decrease the entropy of encryption, the same exposure causes negligible loss in (relative) entropy for physical encryption, because SNR is computed in statistical sense by averaging across many observations/samples. Therefore, as argued in Section~\ref{ssec:secana}, side-channel KPA most likely fails to break \name; as far as we understand, the only chance one may break \name is via a full-scale KPA to re-train the HGR model under encrypted CSIs, but this is forbidden by our attack model in Section~\ref{ssec:attackm}.

Last but not least, we have omitted the analysis on the sub-model $\mathcal{F}$ introduced in Section~\ref{sec:ssCSI}, for a slightly different reason as that explained in Section~\ref{sec:limfur} for physical encryption. Although we suspect that hashing $\bm{\Psi}$ to produce $\phi$ should not affect the encryption strength, our current proposal in Section~\ref{sec:ssCSI} still requires $\mathcal{F}$ to be distributed only to Bob$^{\mathrm{S}}$, albeit no apparent reason against also exposing $\mathcal{F}$ to Eves(as the case for the common HGR models). In general, the security analysis on physically encrypted physical plaintext is somehow ``interfered'' by the existence of a deep learning model acting as a ``translator'' between physical plaintexts and their digital labels (or interpretations). Whereas the security issue of such models in terms of being tricked to generate wrong translations (a.k.a, adversarial examples) is well understood, whether one can breach a keyed model like our $\mathcal{F}$ so as to obtained the encrypted translations remains a largely unexplored question.
}

\section{Extended Experiments}
% vspace{-2ex}

\begin{figure}[t]
	\setlength\abovecaptionskip{8pt}
	%\vspace{-.5ex}
	\centering
	\subfloat[Obfuscating eavesdropping.]{
		\begin{minipage}[b]{0.8\linewidth}
			\centering
			\includegraphics[width = \textwidth]{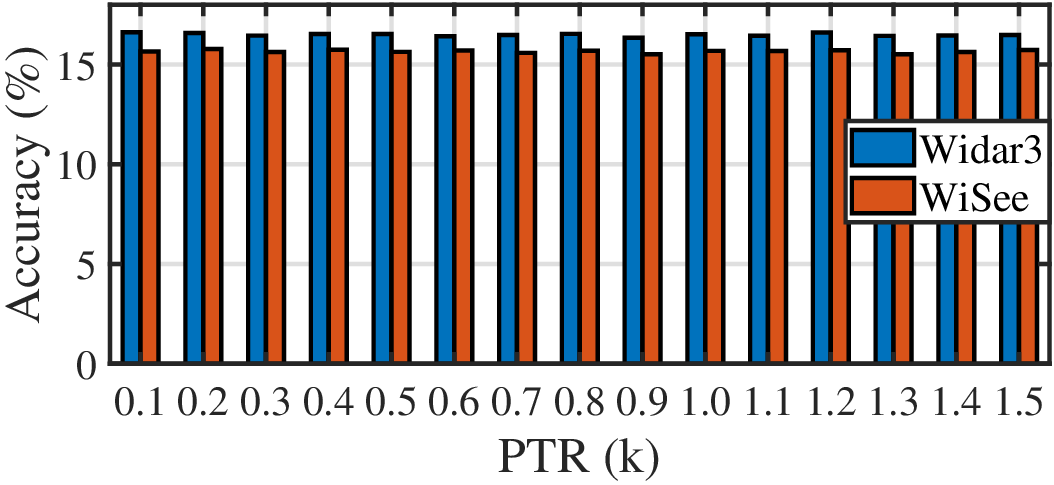}
			\label{sfig:app_ptr1}
			\vspace{-3ex}
		\end{minipage}
	}
        \vspace{-2ex}
	\\
	\subfloat[Legitimate sensing.]{
		\begin{minipage}[b]{0.8\linewidth}
			\centering
			\includegraphics[width = \textwidth]{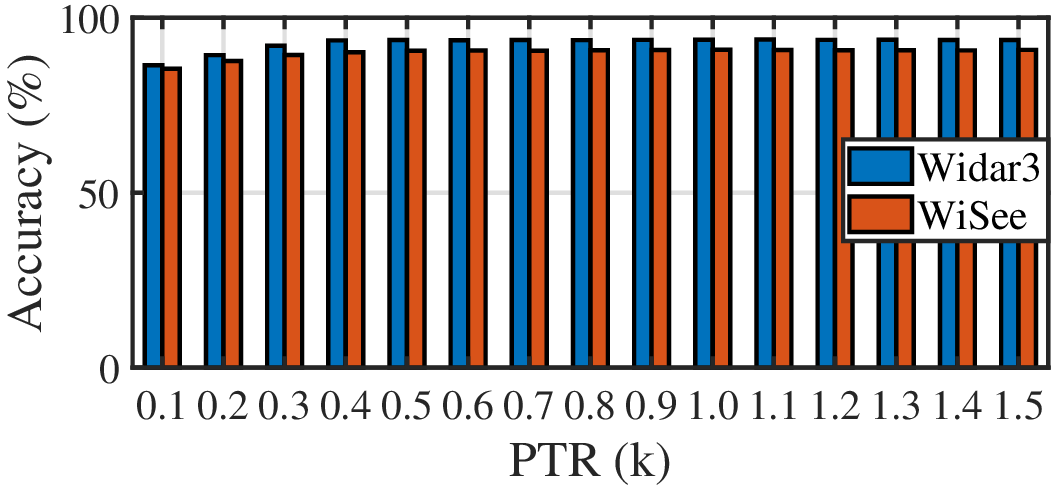}
			\label{sfig:app_ptr2}
			\vspace{-3ex}
		\end{minipage}
	}
         \vspace{-2ex}
	\\
	\subfloat[Legitimate communication.]{
		\begin{minipage}[b]{0.78\linewidth}
			\centering
			\includegraphics[width = \textwidth]{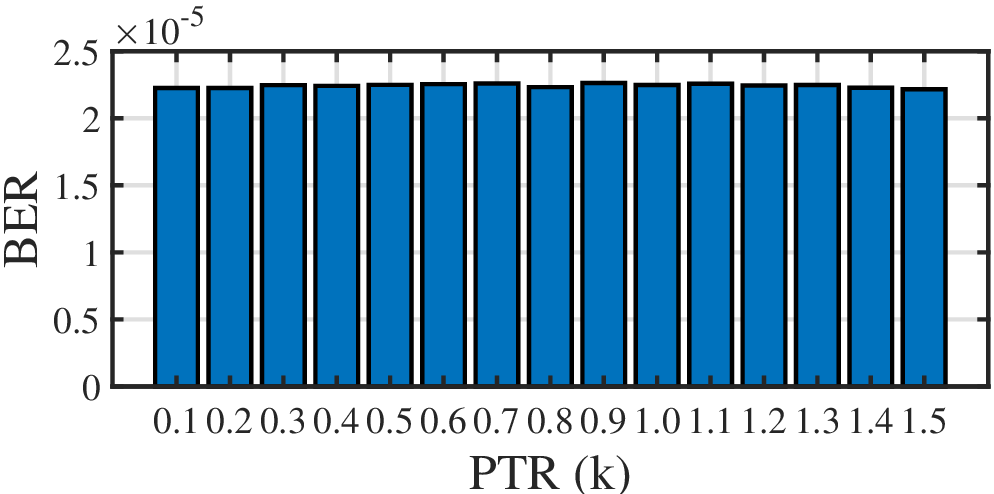}
			\label{sfig:app_ptr3}
			\vspace{-3ex}
		\end{minipage}
	}
	\caption{\sname's performance in (a) thwarting Eve, while (b) protecting Bob$^{\mathrm{S}}$ and (c) Bob$^{\mathrm{C}}$, under distinct packet transmission rates.}
	\label{fig:app_ptr}
	\vspace{-1.5ex}
\end{figure}

\noindent\emph{The Impact of Packet Transmission Rate.}
Packet transmission rate (PTR) can directly impose effects on signal density and hence sensing tasks, \newrev{and it also affects how fast $\bm{\Psi}$ switches its scrambling patterns from one to the next.} \sname sets the default PTR to 1,000 packets per second referring to~\cite{Widar3-MobiSys19, WiSee-MobiCom13,oneshot_sensing}, to provide sufficient channel information density for gesture recognition. However, the LTS-based CSI estimation is designed for data communication services, not specifically for sensing. Therefore, PTR is determined by communication service-related parameter settings, such as Wi-Fi protocols and wireless channel states. \rev{Subsequently, we adjust PTR to the range from 100 to 1,500 pkt/s} %Subsequently, we adjust PTR by down-sample original CSI samples to the range from 100 to 1,500 (i.e., 1.5~\!k)
for fully evaluate \sname under distinct communication environments. As depicted in Figure~\ref{sfig:app_ptr2}, 
the legitimate sensing accuracy owns a slight improvement when the PTR (less than 0.4~\!k) increases, and then stabilizes around 93.5\% of Widar3 and 90.8\% of WiSee. Moreover, thwarting eavesdropping in Figure~\ref{sfig:app_ptr1} and protecting legitimate communication in Figure~\ref{sfig:app_ptr3} keep a stable and satisfactory performance in the full PTR range. In general, these experiments lead to two conclusions: \sname can be adequately compatible with distinct PTR value settings, thus PTR imposes negligible effect on our privacy-preserving mechanism for Wi-Fi sensing; the reason why malicious eavesdropping can be successfully obfuscated is driven by \sname rather than signal granularity.

\begin{figure}[b]
	\setlength\abovecaptionskip{8pt}
	\vspace{-2.5ex}
	\centering
	\subfloat[Widar3.]{
        \begin{minipage}[b]{0.45\linewidth}
    	\centering
    	\includegraphics[width = \textwidth]{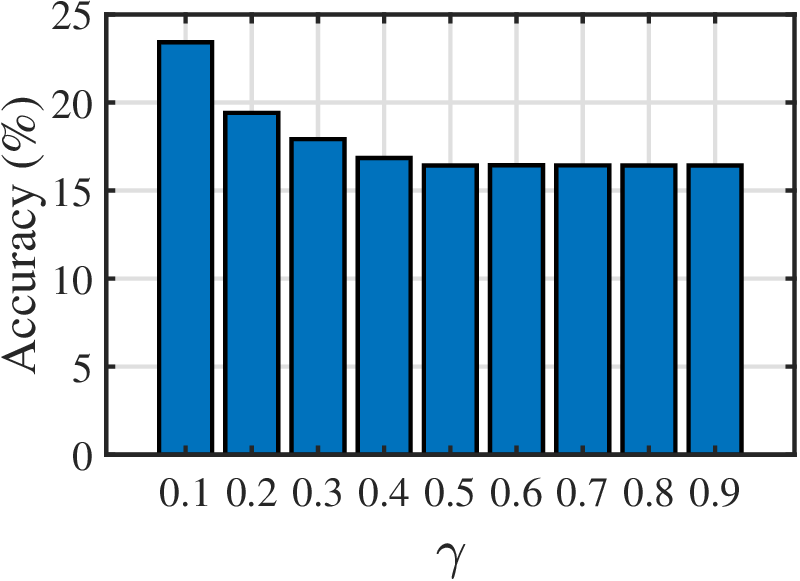}
    	\label{sfig:packetInter1}
    	\vspace{-2.5ex}
	\end{minipage}
	}
        \hfill
	\subfloat[WiSee.]{
		\begin{minipage}[b]{0.45\linewidth}
			\centering
			\includegraphics[width = \textwidth]{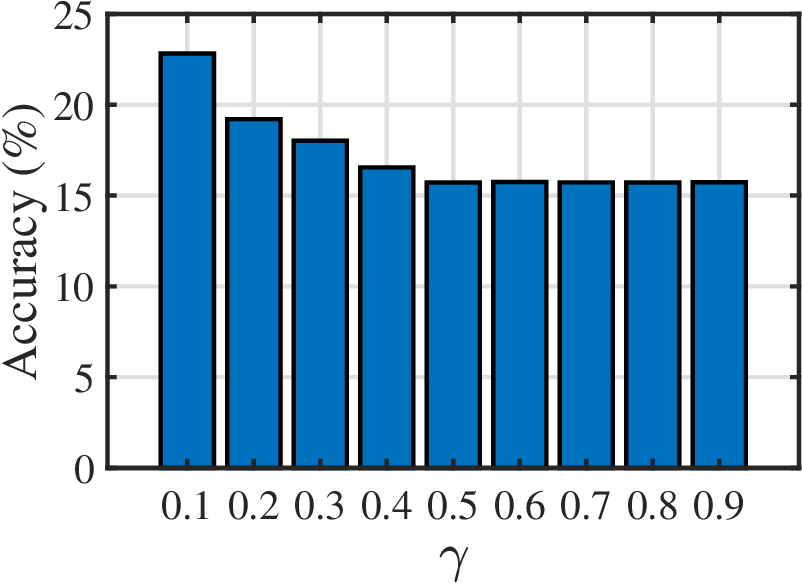}
			\label{sfig:packetInter2}
			\vspace{-2.5ex}
		\end{minipage}
	}
        \vspace{-.5ex}
	\caption{Gesture recognition accuracy of Eve offered by (a) Widar3 and (b) WiSee, under distinct value ranges of the randomization ratio.}
	\label{fig:packetInterAll}
	\vspace{-.5ex}
\end{figure}

\vspace{1ex}
\noindent\emph{The Impact of Random Packet Interval.}
\sname adds temporal randomization to fixed packet transmission interval to achieve temporal randomization, which complements the spatial-temporal encryption for further obfuscating the correspondence between CSI features and gestures. In Section~\ref{sssec:scramblingPer}, we evaluate the effectiveness of temporal randomization as the randomized ``jitter'' ratio $\beta_m$ falls within $\left[ -0.5, 0.5 \right]$. Since the $\beta_m$'s value boundary $\gamma$ (i.e., $\beta_m \in \left[-\gamma, \gamma \right]$) affects the effectiveness of temporal randomization, we thus study the performance in thwarting Eve under distinct $\gamma$ values. Figure~\ref{fig:packetInterAll} presents Eve's HGR accuracy when the $\gamma$ is adjusted from 0.1 to 0.9. It is clear that, for $\gamma$ no less than 0.5, Eve's HGR accuracy is minimized to around 15\% and remains stable in both Widar3 and WiSee models. 
% In particular, for Wi-Fi communications, data packets always undergo uncertain propagation delays, thus this experiment also indicates the compatibility of \sname with common communication cases.
This also corroborates our earlier choice of $\gamma = 0.5$.

\begin{figure}[h]
	\setlength\abovecaptionskip{8pt}
 	% \setlength\belowcaptionskip{12pt}
	%\vspace{-2ex}
	\centering
	\includegraphics[width = 0.6\columnwidth]{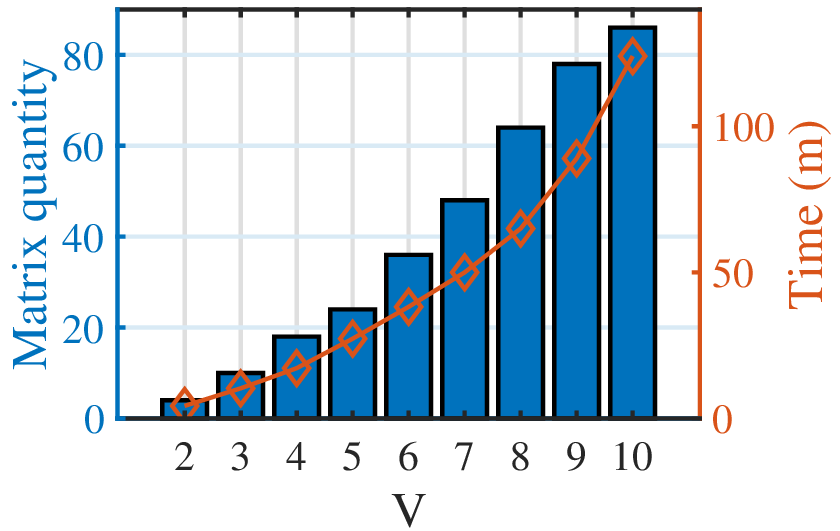}
	\caption{The encryption matrix quantity pre-processed by a sub-model and corresponding model training time cost.}
% 	, as adjusting $V$ from 2 to 10.}
	\label{fig:submodelNumber}
	% \vspace{-2.5ex}
\end{figure}

\vspace{1ex}
\noindent\emph{The Impact of Encoder Quantity.}
We study how the different number of encoders impacts on \sname.
For a specific encoder quantity $V$, the amount of encryption matrices $\bm{\Psi}$ that the sub-model can pre-process is limited. 
% Therefore, for enlarging the sub-model with greater pre-processing adaptability, \rev{it is straightforward to increase the encoder quantity $V$. } 
In this section, with offering satisfactory HGR accuracy, we record the maximum quantity of $\bm{\Psi}$ being pre-processed when $V$ is adjusted from 2 to 10. As shown in Figure~\ref{fig:submodelNumber}, when $V$ increases, the maximum quantity of $\bm{\Psi}$ varies from 4 to 86; meanwhile, the average time spent on offline model training increases apparently. As $V$ is set as 8, the model training time is around one hour while supporting pre-process 64 encryption matrices. This might not meet our expectation~(i.e., $2^8 = 256$) in Section~\ref{sec:ssCSI}, since the representation capacity of our implemented model~(specified in Section~\ref{sec:imple}) can be limited. Though a careful design of the encoders may improve the overall representation capacity of the sub-model, this is out of scope of our paper.
Moreover, as we analyze in Section~\ref{ssec:secana}, one $\bm{\Psi}$ may thwart Eve for a rather long time period, 
% \sname can still effectively obfuscate Eve and thus 
so 64 $\bm{\Psi}$s is quite sufficient for \sname's privacy-preservation purpose.

% for secure sensing in each descrambling key update period. 
% Consequently, \sname acquiescently sets $V$ as 8~(introdued in Section~\ref{sec:imple}) for constructing our ``one-fits-all'' HGR pipeline.

% This case ensures the sub-model's adaptability and its time cost for model training is acceptable. However, as described in Section~\ref{ssec:ssCSI}, $V$ equal to 8 should provide a sub-model with $2^8$ different structures and thus pre-process the same number of $\bm{\Psi}$, but the actual result is 64. The reason for this difference is that multiple similar sub-model structures can only handle one identical $\bm{\Psi}$. Moreover, the security analysis in Section~\ref{ssec:scrambling} states that even if owning only one $\bm{\Psi}$, \sname can still effectively obfuscate Eve and thus 64 $\bm{\Psi}$s is enough for secure sensing. Consequently, \sname acquiescently sets $V$ as 8 for constructing our ``one-fits-all'' HGR pipeline.

\balance
\section{Future Work}

\noindent\emph{Towards General Sensing Application.}
Although we take gesture recognition in this paper as an example to evaluate the performance of \sname in thwarting Eve while protecting legitimate sensing and communication, \sname can be easily adapted to other similar sensing applications, such as activity tracking~\cite{activity_rec} and motion recognition~\cite{WiDeo-NSDI15}. 
%
% Taking gesture recognition as an example, we evaluate the performance of \sname in thwarting Eve while protecting legitimate sensing and communication. In addition, \sname can easily adapt to other common sensing applications, such as activity tracking~\cite{activity_rec} and motion recognition~\cite{WiDeo-NSDI15}. 
Its scalability stems from the following facts. First, Wi-Fi-based sensing applications share a common characteristic which they all extract features from CSI phase and amplitude via deep neural network. Consequently, \sname can encrypt them to prevent malicious eavesdropping, no matter what the sensing applications we encounter. Second, for legitimate sensing, users only need to integrate our plug-and-play sub-model to any base recognition model. Third, for communication services, \sname can readily recover the LTS with the help of encryption matrices and then decode the payload data. 
In the future, we shall deploy \sname into a neural network acceleration ASIC to explore the high-speed inference capability for more sensing applications. 

\vspace{1ex}
\noindent\emph{Compatibility with Wi-Fi Framework.}
%
%\rev{
Our proposed \sname is compatible with current Wi-Fi standards, and potentially integrated into COTS Wi-Fi chip via firmware modification. Though in this paper, we utilize the WARP SDR to build the system prototype of \sname while evaluating its effectiveness in securing sensing and protecting communication, the WARP is widely leveraged by many researchers to verify their innovative Wi-Fi related ideas including both communication and sensing areas. We also implement our prototype strictly follows current Wi-Fi standards as the other researchers. More importantly, our source-defined encryption mechanism only involves simple encryption matrix operations. It does not need to redesign the hardware components, but only the embedded software~(a.k.a firmware) in the existing COTS Wi-Fi chip. To sum up, \name is a promising Wi-Fi privacy-preserving mechanism providing an innovative idea for future chip design with secure sensing considerations. 

%}
% The implementability of our proposed privacy-preserving mechanism is mainly reflected in the compatibility with current Wi-Fi standards and COTS. 

% On the one hand, \sname utilizes the WARP3 SDR to build a system prototype while evaluating its effectiveness in securing sensing and protecting communication.} The WARP3 is specially leveraged to help researchers evaluate innovative Wi-Fi sensing ideas and also includes our scrambling mechanism, thus \sname prototype strictly follows current Wi-Fi standards. On the other hand, our source-defined scrambling mechanism only involves simple obfuscating matrix operations, without reconstructing the existing Wi-Fi COTS. Therefore, \sname is compatible with Wi-Fi chips (e.g., Broadcom BCM4339~\cite{broadcom}) via controlling firmware. To sum up, \sname is a promising Wi-Fi privacy-preserving mechanism providing an innovative idea for future chip design with secure sensing considerations.

\vspace{1ex}
\noindent\emph{Privacy-preserving for The Active Wi-Fi Sensing.}
Different from the conventional passive Wi-Fi sensing discussed in this paper, active Wi-Fi sensing means designing special radar-like hardware that transmits Wi-Fi signals actively~\cite{Octopus-MobiCom21, isacot}. In this way, Eve does not need to eavesdrop the signals from Alice, but generates/transmits them directly and then analyzes the reflection signals from victims to learn their activities. \newrev{Recently, RF-Protect~\cite{shenoy2022rf} leverages similar IRS technology of IRShield~\cite{IRShield-SP22} to protect human behaviors from radar active sensing, yet it faces the same practical problem of IRShield for lack of controllability to differentiate between legitimate sensing users and attackers, as discussed in Section~\ref{ssec:survey}.} Therefore, this problem can be very challenging for source-defined privacy-preserving, and we must redesign our \sname to not only perform encrypting, but also figure out the active signals from the Eve. In the next step, we plan to explore \sname for protecting privacy under active Wi-Fi sensing. 

\newpage

\section*{Meta-Review}

\subsection*{Summary}
The paper proposes a privacy-preserving Wi-Fi gesture recognition system called \sname. So far, the Wi-Fi-based gesture recognition/sensing literature has allowed any Wi-Fi receiver in proximity to legitimate sensing entities also to sense the gestures. The few privacy-preserving human gesture recognition systems are incompatible with Wi-Fi standards or do not support realistic multi-user scenarios.

\sname addresses these drawbacks and designs a multi-user privacy-preserving Wi-Fi-based human gesture recognition system without affecting legitimate communicating entities. The paper's main idea is to scramble the Wi-Fi channels at the transmitter (access point) using a 11secret scrambling matrix‘’ only known to legitimate parties, prohibiting eavesdroppers from estimating CSI changes. The paper presents an exhaustive privacy and performance evaluation of the proposed system through a proof-of-concept implementation.

\subsection*{Scientific Contributions}
\begin{itemize}
% \item Privacy-preserving Wi-Fi gesture recognition/sensing system.
% \vspace{.5ex}
\item % Creative use of physical layer scrambling to prevent eavesdroppers from actively sensing using Wi-Fi signals.
Provides a Valuable Step Forward in an Established Field.
\vspace{.5ex}
\item % Multiple hand gesture recognitions possible due to the physical scrambling.
Establishes a New Research Direction.
\vspace{.5ex}
\end{itemize}

\subsection*{Reasons for Acceptance}
\begin{enumerate}
\item Potentially establish a new direction in privacy-preserving wireless sensing systems.
\vspace{.5ex}
\item Exhaustive experimental evaluation with a strong discussion of the results.
\end{enumerate}

\subsection*{Noteworthy Concerns}
\begin{enumerate} 
\item The need for privacy-preserving hand gesture recognition systems using Wi-Fi is not clear.
\vspace{.5ex}
\item It is not clear how an attacker will be distinguished from a legitimate user.
\vspace{.5ex}
\item There is a possibility that an adversary can decrypt user gestures by comparing the state of the wireless channel with and without user activity. This is a potential threat.
\end{enumerate}

\section*{Response to the Meta-Review} % Optional
\noindent Rebuttal to Noteworthy Concerns:
\vspace{1ex}
\begin{enumerate} 
\item We believe that one thing is rather clear: as hand gesture may reveal keystrokes (in particular password typing), as suggested by a few publications cited at the end of the first paragraph in Section~1, privacy-preserving (or securing) hand gesture recognition (or general Wi-Fi sensing) is imperative.
\vspace{1ex}
\item As we have explained in Section~2.2 of our paper, whether to perform registration is a clear indication to whether a user is legitimate or not. If the concern is on ``how to identify whether or not a group of users performing neural model training for Wi-Fi enabled gesture recognition are attacking \sname via known-plaintext attack (KPA)'', the answer is also clear: as no user is supposed to do so (they are supposed to use only publicly-known neural models as stated in Section~2.2), anyone attempting to do so in public should be deemed as attacking \sname and hence should be ``neuralized''.
\vspace{1ex}
\item Our physical security and encryption, though different in mechanisms from their digital counterpart, work in the same vein: one cannot decrypt digital messages (resp. user gestures) by comparing two states of with or without digital message (resp. the two wireless channel states of with or without user activity). Nevertheless, as a published proposal on security defence, \sname is open to any potential attack proposals against it, and we are ready to counteract such new proposals should they appear.
\end{enumerate}

%\newpage 
%\appendices

\end{document}